\documentclass[onecolumn]{emulateapj}
\usepackage{graphics}

\begin{document}

\title{A Finite Difference Representation of Neutrino Radiation Hydrodynamics
in Spherically Symmetric General Relativistic Space-Time}

\author{Matthias Liebend\"orfer\altaffilmark{1,2,3}, O. E. Bronson Messer\altaffilmark{1,2,6}, Anthony Mezzacappa\altaffilmark{2}, \\ Stephen W. Bruenn\altaffilmark{4}, Christian Y. Cardall\altaffilmark{1,2,6}, and F.-K. Thielemann\altaffilmark{5}} \altaffiltext{1}{Department of Physics and Astronomy, University of Tennessee, Knoxville, Tennessee 37996-1200} \altaffiltext{2}{Physics Division, Oak Ridge National Laboratory, Oak Ridge, Tennessee 37831-6354} \altaffiltext{3}{CITA, University of Toronto, Toronto, Ontario M5S 3H8, Canada }\altaffiltext{4}{Department of Physics, Florida Atlantic University, Boca Raton, Florida 33431-0991} \altaffiltext{5}{Department of Physics and Astronomy, University of Basel, Klingelbergstrasse 82, 4056 Basel, Switzerland} \altaffiltext{6}{Joint Institute for Heavy Ion Research, Oak Ridge National Laboratory, Oak Ridge, Tennessee 37831-6374}

\begin{abstract}
We present an implicit finite difference representation for general
relativistic radiation hydrodynamics in spherical symmetry. Our code,
{\sc agile-boltztran}, solves the Boltzmann transport equation for
the angular and spectral neutrino distribution functions in self-consistent
simulations of stellar core collapse and postbounce evolution. It
implements a dynamically adaptive grid in comoving coordinates. A
comoving frame in the momentum phase space facilitates the evaluation
and tabulation of neutrino-matter interaction cross sections, but
produces a multitude of observer corrections in the transport equation.
Most macroscopically interesting physical quantities are defined by
expectation values of the distribution function. We optimize the finite
differencing of the microscopic transport equation for a consistent
evolution of important expectation values. We test our code in simulations
launched from progenitor stars with 13 solar masses and 40 solar masses.
Half a second after core collapse and bounce, the protoneutron star
in the latter case reaches its maximum mass and collapses further
to form a black hole. When the hydrostatic gravitational contraction
sets in, we find a transient increase in electron flavor neutrino
luminosities due to a change in the accretion rate. The \( \mu  \)-
and \( \tau  \)-neutrino luminosities and rms energies, however,
continue to rise because previously shock-heated material with a non-degenerate
electron gas starts to replace the cool degenerate material at their
production site. We demonstrate this by supplementing the concept
of neutrinospheres with a more detailed statistical description of
the origin of escaping neutrinos. Adhering to our tradition, we compare
the evolution of the 13 solar mass progenitor star to corresponding
simulations with the multi-group flux-limited diffusion approximation,
based on a recently developed flux limiter. We find similar results
in the postbounce phase and validate this {\sc mgfld} approach for
the spherically symmetric case with standard input physics.
\end{abstract}
\keywords{supernovae: general---neutrinos---radiative transfer---hydrodynamics---relativity---methods: numerical}

\section{Introduction}\label{section_neutrino_transport}

A supernova explosion is a dramatic event which includes such a rich
diversity of physics \citep{Bethe_90,Burrows_Young_00,Mezzacappa_Bruenn_00,Janka_Kifonidis_Rampp_01}
that self-consistent numerical simulations on current computer hardware
do not allow to include all relevant pieces at once. After stellar
core collapse, a nascent neutron star is formed at the center of the
event, requiring a description in general relativity. Neutrinos generated
in the compactifying region are tightly coupled to the matter at high
densities, while they leak or stream out at lower densities. The collapsing
stellar core bounces as the equation of state stiffens when nuclear
densities are exceeded, and a shock wave is formed that ploughs outwards
through the still infalling outer layers. Multi-frequency radiation
hydrodynamics must be used to quantify the energy that the neutrinos
deposit in the material behind the shock. This energy deposition has
been considered to be essential for the success or failure of the
supernova explosion \citep{Colgate_White_66,Wilson_85,Bethe_Wilson_85}.
Observational and theoretical evidence suggests that this neutrino
heating drives convection behind the shock. Instabilities in the protoneutron
star, significant rotation, and strong magnetic fields further complicate
the picture. Observations of neutron star kicks, mixing of nuclear
species, inhomogeneous ejecta, and polarization of spectra support
the presence of asymmetries in supernova explosions \citep{Tueller_et_al_91,Strom_et_al_95,Galama_et_al_98,Leonard_et_al_00}.
Motivated by such observations, the neutrino driven explosion mechanism
has been explored in multidimensions \citep{Herant_Benz_Colgate_92,Miller_Wilson_Mayle_93,Herant_et_al_94,Burrows_Hayes_Fryxell_95,Janka_Mueller_96,Mezzacappa_et_al_98b,Fryer_Heger_00,Fryer_Warren_02},
and alternative jet-based explosion scenarios have received new momentum
\citep{Hoeflich_Wheeler_Wang_99,Khokhlov_et_al_99,MacFadyen_Woosley_99,Wheeler_et_al_00}.

Because of their excessive computational demand, multi-dimensional
simulations have had to rely on physically significant approximations.
Simulations are performed that impose the neutrino radiation field
externally or uniformly, that prescribe the neutrino spectrum ab initio,
that artificially seed instabilities, or that insert jets in order
to explore important phenomena of supernova explosions in observational
data. A complementary approach is to model supernovae in spherical
symmetry. Spherically symmetric simulations of stellar core collapse
reach back to the late sixties \citep{Colgate_White_66,May_White_67,Arnett_67,Schwartz_67,Wilson_71}
when computers first became available. Although the assumption of
pure spherical symmetry is also a physically significant simplification,
it can provide an essential contribution to the quality of numerical
supernova models and their interpretation. A spherically symmetric
model can implement full general relativity. Additionally, the reduced
number of computational zones allows an accurate treatment of neutrino
radiation transport and the inclusion of sophisticated microscopic
input physics. During the search for a robust supernova mechanism,
standards in the nuclear input physics \citep{Lattimer_Swesty_91}
and weak interaction physics \citep{Bruenn_85} have been established.
In recognition of the importance of neutrino transport in the supernova,
its numerical treatment has been improved from simple leakage schemes
\citep{VanRiper_Lattimer_81,Baron_Cooperstein_Kahana_85b} to multi-group
flux-limited diffusion ({\sc mgfld}) approximations \citep{Arnett_77,Bowers_Wilson_82,Bruenn_85,Myra_et_al_87}
to investigations of the full Boltzmann transport equation \citep{Mezzacappa_Bruenn_93a,Messer_et_al_98,Yamada_Janka_Suzuki_99,Burrows_et_al_00,Rampp_Janka_02}.
Only recently have self-contained simulations of stellar core collapse
and postbounce evolution with Boltzmann neutrino transport been performed,
in Newtonian gravity and the {\cal O}\( (v/c) \) limit \citep{Rampp_Janka_00,Mezzacappa_et_al_01,Thompson_Burrows_Pinto_03}
and in general relativistic space-time \citep{Liebendoerfer_et_al_01}.
They allow, on the one hand, the exploration of nuclear physics under
extreme conditions, and, on the other hand, provide a solid point
of reference for the construction of accurate multidimensional simulations
\citep{Buras_et_al_03b}.

Basic numerical techniques for radiative transfer have been designed
and continuously improved over the last three decades (see e.g. \citep{Mihalas_Mihalas_84}
and \citep{Lewis_Miller_84} for a review). In most applications, photons
or neutrons are the transported particles. In our application to supernova
dynamics, the transport of neutrinos requires a new combination of
capabilities derived from both fields, photon and neutron transport.

As in \emph{neutron transport} the radiation particles are fermions.
In contrast to photons, the neutrinos assume a Fermi-Dirac distribution
in thermal equilibrium. Weak interactions with nuclear matter can
be suppressed due to Pauli blocking of final states. This makes the
dependence of the collision integral on the distribution function
nonlinear. Several quantities have to be evolved with special care:
An important constraint for the evolution of neutrino and electron
abundances is lepton number conservation. While applications to photon
transport do not require an exact count of photons as particles, the
outcome of a supernova simulation sensitively depends on the deleptonization.
The transfer of lepton number occurs via the reactions \( e^{-}+p^{+}\rightleftharpoons n+\nu _{e} \)
and \( e^{+}+n\rightleftharpoons p+\bar{\nu }_{e} \) where the protons,
\( p \), and neutrons, \( n \), are free or bound in nuclei. The
electron neutrino, \( \nu _{e} \), and antineutrino, \( \bar{\nu }_{e} \),
may escape to infinity or be absorbed at distant locations. These
reactions determine the electron fraction, and hence, the partial
pressure of the electron gas in the fluid. This pressure contributes
importantly to the dynamics if the electron gas is degenerate and
the density lower than nuclear density. Beside of the transport of
lepton number, the transport of energy is a crucial phenomenon in
supernova dynamics. Neutrinos escaping from the accreting material
and the protoneutron star may be absorbed behind the stalled shock.
This energy transfer is believed to add to the thermal pressure behind
the shock such that ultimately a supernova explosion is launched with
the ejection of the outer layers \citep{Colgate_White_66,Bethe_Wilson_85}.
Last, but not least, the evolution of the global energy should be
monitored. Just before collapse, the progenitor star is marginally
bound. During collapse, the binding energy of the nascent protoneutron
star increases dramatically. It is balanced by the internal energy
of the compressed matter and the trapped neutrinos. Only on a longer
time scale is the energy of the radiation field transported away from
the star while the emptied states in the neutrino phase space are
immediately replenished by neutrino emission in the cooling protoneutron
star. Since the gravitational energy, the internal energy, and the
energy in the radiation field rise to the order of \( 10^{53} \)
erg, a detailed analysis of global energy conservation is advised
in order to make the comparatively small explosion energy of \( 10^{51} \)erg
predictable. Whether we choose a kinetic equation for the propagation
of particle number or an equation prescribing the evolution of radiation
intensity, it is a numerical challenge to conserve both lepton number
and energy with the same finite difference representation of the transport
equation. 

The following transport requirements more closely resemble the astrophysical
applications of \emph{photon transport} than neutron transport. Owing
to the protoneutron star or black hole formed at the center of the
event, space-time is curved. The particles follow geodesics in general
relativistic space-time, i.e. their angular distribution is affected
by gravitational bending. Additionally, the particle energies are
subject to a gravitational frequency shift. In contrast to many neutron
transport applications, the fluid is highly dynamic during collapse
and even more so after shock formation. Thus, observer corrections
for Doppler shift and angular aberration have to be implemented as
in photon transport applications. However, there is some freedom in
the choice of where to apply these corrections. In an inertial frame
they complicate the description of interactions in the collision integral.
In a comoving frame they enter the transport terms in the Boltzmann
equation.

We proceed with the latter choice in an \( S_{N} \)-method. Our code,
{\sc agile-boltztran}, emerged from the following components: The
neutrino transport part, {\sc boltztran}, has been developed for the
simulation of stellar core collapse in an implicitly finite differenced
{\cal O}\( (v/c) \) approximation \citep{Mezzacappa_Bruenn_93a,Mezzacappa_Bruenn_93b,Mezzacappa_Bruenn_93c}.
It has been compared to {\sc mgfld} in selected stationary state phases,
and standard test problems for radiative transfer in supernovae have
been performed \citep{Messer_et_al_98,Messer_00}. {\sc agile} is an
implicit general relativistic hydrodynamics code that evolves the
Einstein equations based on conservative finite differencing on an
adaptive grid \citep{Liebendoerfer_Rosswog_Thielemann_02}. In this
paper, we describe how these codes are merged and extended to enable
accurate simulations of the very dynamic postbounce phase. In particular,
we detail the finite differencing of the observer corrections for
the dynamic conservation of particle number and total energy in the
transport scheme, and describe the extension of {\sc boltztran} to
general relativistic flows \citep{Liebendoerfer_00}. In section \ref{section_physical_model},
we start with the characterization of the equation of state and a
list of the included neutrino-matter interactions, followed by a collection
of the basic equations of general relativistic radiation hydrodynamics
in spherical symmetry. Two exemplary runs from collapse through bounce
and postbounce evolution are also described in this section. In order
to test our code in a broad range of conditions, we launched simulations
from a small \( 13 \) M\( _{\odot } \) progenitor star \citep{Nomoto_Hashimoto_88}
and a very massive \( 40 \) M\( _{\odot } \) progenitor star \citep{Woosley_Weaver_95}.
We investigate the different regions of neutrino emission in the star.
Section \ref{section_computer_model} is entirely devoted to the documentation
of the finite differencing in our computer model. In section \ref{section_code_verification},
we analyze and verify the performance of our code in various example
situations encountered in the evolution of the two simulation runs.
We close the section with the comparison of our results with Bruenn's
multi-group flux-limited diffusion code which implements a recently
developed new flux limiter \citep{Bruenn_02}.

\section{Physics in the Model}\label{section_physical_model}

The physical model can be divided into two parts. On the one hand,
there is the microscopic physics input\---the specification of particle
abundances and reaction cross sections. The microphysics in supernova
models is continuously improving and many uncertainties remain to
be resolved. However, in this methodological paper, we will only shortly
summarize the ingredients that were standard at the time the code
was written and hope that our code {\sc agile-boltztran} will continue
to be useful in future discussions and evaluations of input physics
improvements. On the other hand, there are the radiation hydrodynamics
equations we are solving on the computer. They are the foundation
of our implementation of neutrino transport and receive a detailed
discussion later in this section.

\subsection{Equation of state and weak interactions}

The equation of state describes the thermodynamical state of a fluid
element based on density, \( \rho  \), temperature, \( T \), and
the composition. We use the equation of state of \citet{Lattimer_Swesty_91}.
It assumes nuclear statistical equilibrium and we apply it wherever
the density is larger than \( 10^{7} \) g/cm\( ^{3} \) and the temperature
larger than \( 5\times 10^{9} \)K. This region is described by a
liquid drop model for a representative nucleus with atomic number
\( A \) and charge \( Z \), surrounded by free alpha particles,
protons, and neutrons. The baryons are immersed in an electron and
positron gas in equilibrium with a photon gas. Beyond nuclear density,
where no isolated nuclei are present, the complicated population of
hadrons \citep{Glendenning_85,Pons_et_al_99} is approximated by bulk
nuclear matter comprised of protons, neutrons, and electrons. However,
the central density of the protoneutron star at bounce reaches only
about twice nuclear density and the hadron population may only develop
later, after the very dynamical postbounce phases. At the low temperature
border of nuclear statistical equilibrium, the equation of state is
connected to a Boltzmann gas of silicon atoms. In any of these cases,
once the density and temperature are given, the composition is fully
determined by the specification of the electron fraction \( Y_{e} \). 

Matter is connected to the neutrino radiation field by weak interactions.
We consider neutrinos of all three flavors and assume that they are
massless. The weak interactions enter the collision term in the Boltzmann
equation as energy- and angle-dependent emissivities, opacities, and
scattering kernels. We include the set specified by \citep{Bruenn_85}:
(i) electron-type neutrino absorption on neutrons, (ii) electron antineutrino
absorption on protons, (iii) electron-type neutrino absorption on
nuclei, (iv) neutrino-nucleon scattering, (v) coherent scattering
of neutrinos on nuclei, (vi) neutrino-electron scattering, and (vii)
neutrino production from electron/positron pair annihilation. These
reactions, and their inverses, are implemented in our code as described
by \citet{Mezzacappa_Bruenn_93b,Mezzacappa_Bruenn_93c,Messer_00}.
In the following code description, we will only include emissivities,
\( j \), and opacities, \( \chi  \), because this is sufficient
to describe how the collision term enters the transport and hydrodynamics
equations. In the simulations, the scattering kernels are included
in the collision integral as well.

The particles treated by the equation of state are assumed to react
with each other on very short time scales such that a description
in terms of an instantaneous equilibrium is appropriate. Neutrinos
in high-density regimes can also achieve local thermal and weak equilibrium
with matter if the opacities are sufficiently high. Unlike the equilibrium
with respect to the strong interaction, however, this equilibrium
must be determined within our solution of the transport equation.
For example, in the protoneutron star at densities above \( 10^{12} \)
g cm\( ^{-3} \) and temperatures above \( 5\times 10^{10} \) K the
neutrinos are trapped and are well-described by a Fermi-gas in thermal
equilibrium with the fluid. At lower densities, the thermalization
time scale becomes longer; then the neutrinos can propagate with a
nonequilibrium spectrum throughout these regions, to be absorbed elsewhere
or leave the star. The strong coupling of the neutrinos to the matter
at high densities and the strong coupling between different locations
mediated by neutrino transport complicates the evolution of a numerical
solution. If the problem is separated into independently updated pieces
by operator splitting, the numerical solution will only be stable
if information in the numerical implementation is shared faster between
the independent updates than in the evaluated physical processes.
The fast time scale of neutrino-matter interactions and the propagation
of neutrinos at light speed may severely restrict the time step. The
required coupling can be built directly into the numerical scheme
by an implicit finite differencing of essential parts of the transport
equation. Unfortunately, such a differencing requires knowledge of
the derivatives of the collision term with respect to all independent
state variables\---i.e., density, temperature, electron fraction,
and neutrino distribution functions. Because the emissivities, opacities,
and scattering kernels strongly depend on the neutrino energies and,
in the scattering case, on the neutrino propagation directions, the
numerical evaluation of the collision term and its derivatives becomes
a nonnegligible part of the overall computational effort. \citet{Mezzacappa_Bruenn_93a}
developed a storage scheme that allows the reuse of previously calculated
emissivities, opacities, and scattering kernels by linear interpolation
within a dynamical table in the independent variables of logarithmic
density, \( \log _{10}\left( \rho \right)  \), logarithmic temperature,
\( \log _{10}\left( T\right)  \), and electron fraction, \( Y_{e} \).
If one uses these same independent variables in the implicit formulation
of the Boltzmann equation, the correct partial derivatives of the
reactions directly emerge from the coefficients of the linear interpolation,
without additional computational effort. On the one hand, the reuse
of previously evaluated interactions is straightforward if the transport
equation is solved in the rest frame of the fluid, such that no transformation
of the neutrino energy or angle dependence of the interactions is
required. On the other hand, the transport equations are simpler in
the laboratory frame. In this paper we demonstrate that in the case
of spherical symmetry the complexity of the transport equation is
manageable and proceed with the analysis in a comoving frame (spacetime
coordinates and neutrino four-momentum) to take advantage of the simplifications
in the collision term. On average, we have to evaluate new collision
integrals in about two to three zones per time step (out of a hundred
zones). Although these numbers depend very much on the specific phase
of the simulation, the evaluation of nuclear physics input may still
take about half of the total execution time.

\subsection{Radiation hydrodynamics in spherical symmetry}

Many spherically symmetric simulations of compact objects have been
approached in comoving orthogonal coordinates \citep{Misner_Sharp_64,May_White_66}.
Finite difference schemes of varying complexity were designed in \citep{May_White_67,VanRiper_79,Bruenn_85,Rezzolla_Miller_94,Swesty_95,Liebendoerfer_Rosswog_Thielemann_02},
culminating in an approximate Riemann solver \citep{Yamada_97}. The
left-hand side of the Einstein field equation, the Einstein tensor,
is based on the metric \begin{equation}
\label{eq_comoving_metric}
ds^{2}=-\alpha ^{2}dt^{2}+\left( \frac{r'}{\Gamma }\right) ^{2}da^{2}+r^{2}\left( d\vartheta ^{2}+\sin ^{2}\vartheta d\varphi ^{2}\right) ,
\end{equation}
 where \( r \) is the areal radius and \( a \) is a label corresponding
to an enclosed rest mass (the prime denotes a derivative with respect
to \( a \): \( r'=\partial r/\partial a \)). The proper time lapse
of a comoving observer is related to the coordinate time \( dt \)
by the lapse function \( \alpha  \). We have made the substitution
\( g_{aa}=r'/\Gamma  \), based on a function \( \Gamma \left( t,a\right)  \),
for the space-space component of the metric. The angles \( \vartheta  \)
and \( \varphi  \) describe a two-sphere. We use natural units such
that the velocity of light, \( c \), and the gravitational constant,
\( G \), become \( 1 \).

The right-hand side of the Einstein equations is given by the fluid-
and radiation stress-energy tensor, \( T \). In a comoving orthonormal
basis, it has the components \citep{Lindquist_66}\begin{eqnarray}
T^{tt} & = & \rho \left( 1+e+J\right) \nonumber \\
T^{ta}=T^{at} & = & H\nonumber \\
T^{aa} & = & p+\rho K\nonumber \\
T^{\vartheta \vartheta }=T^{\varphi \varphi } & = & p+\frac{1}{2}\rho \left( J-K\right) .\label{eq_stress_energy_tensor} 
\end{eqnarray}
 The total energy is expressed in terms of the rest mass density,
\( \rho  \), the specific internal fluid energy, \( e \), and the
specific radiation energy, \( J \). The isotropic fluid pressure
is denoted by \( p \), and the radiation stress is composed from
the zeroth (\( J \)) and second (\( K \)) angular moments of the
radiation intensity. Radial net energy transport is accounted for
by the nondiagonal component of the stress-energy tensor, the first
angular moment (\( H \)) of the specific radiation intensity.

We define a velocity \( u \), equivalent to the \( r \) component
of the fluid four-velocity as observed from a frame at constant areal
radius \( r \) \citep{May_White_67}, and identify the total energy
enclosed in a sphere with the gravitational mass, \( m \). In the
special relativistic limit, \( \Gamma =\sqrt{1+u^{2}-2m/r} \) then
becomes the Lorentz factor corresponding to the boost between inertial
and comoving observers. As in nonrelativistic hydrodynamics we can
define a specific volume, \( 1/D \), specific energy, \( \tau  \),
and specific radial momentum, \( S \), by\begin{eqnarray}
\frac{1}{D} & = & \frac{\Gamma }{\rho }\label{eq_specific_volume} \\
\tau  & = & \Gamma \left( e+J\right) +\frac{2}{\Gamma +1}\left( \frac{1}{2}u^{2}-\frac{m}{r}\right) +uH\label{equatino_specific_energy} \\
S & = & u\left( 1+e+J\right) +\Gamma H.\label{eq_specific_momentum} 
\end{eqnarray}
 It has been shown in \citep{Liebendoerfer_Mezzacappa_Thielemann_01}
that these definitions lead to conservation equations (\ref{eq_continuity})-(\ref{eq_momentum})
that are analogous to the continuity equation, the conservation of
total energy, and the conservation of radial momentum%
\footnote{This reference unnecessarily assumes an isotropic radiation stress.
However, we note that the difference between the full stress-energy
tensor (\ref{eq_stress_energy_tensor}) and the isotropic approximation
has exactly the same form as the artificial viscosity tensor introduced
in the same reference to numerically stabilize shock fronts. Hence,
in all derivations in the above reference we may simply use the pressure
\( \widetilde{p}=p+\rho J/3 \) for the isotropic part and set the
viscosity coefficient, \( Q \), to \( \widetilde{Q}=-\rho \left( J/3-K\right)  \),
in order to obtain a description of radiation hydrodynamics that extends
to the case where large radiation energies do not satisfy \( J\not \simeq 3K \).
} :

\begin{eqnarray}
\frac{\partial }{\partial t}\left[ \frac{1}{D}\right]  & = & \frac{\partial }{\partial a}\left[ 4\pi r^{2}\alpha u\right] \label{eq_continuity} \\
\frac{\partial \tau }{\partial t} & = & -\frac{\partial }{\partial a}\left[ 4\pi r^{2}\alpha \left( up+u\rho K+\Gamma \rho H\right) \right] \label{eq_total_energy} \\
\frac{\partial S}{\partial t} & = & -\frac{\partial }{\partial a}\left[ 4\pi r^{2}\alpha \left( \Gamma p+\Gamma \rho K+u\rho H\right) \right] \nonumber \\
 & - & \frac{\alpha }{r}\left[ \left( 1+e+\frac{3p}{\rho }+J+3K\right) \frac{m}{r}-\left( 1-\frac{2m}{r}\right) \left( J-3K\right) \right. \nonumber \\
 & + & \left. 8\pi r^{2}\left( \left( 1+e+J\right) \left( p+\rho K\right) -\rho H^{2}\right) -2\left( \frac{p}{\rho }+K\right) \right] \label{eq_momentum} \\
\frac{\partial V}{\partial a} & = & \frac{1}{D}\label{eq_volume_gradient} \\
\frac{\partial m}{\partial a} & = & 1+\tau \label{eq_mass_gradient} \\
\frac{\partial }{\partial t}\left[ \frac{1}{4\pi r^{2}\rho }H\right]  & = & -\left( 1+e+J\right) \frac{\partial \alpha }{\partial a}-\frac{1}{\rho }\frac{\partial }{\partial a}\left[ \alpha \left( p+\rho K\right) \right] +\frac{\alpha }{3VD}\left( J-3K\right) .\label{eq_lapse_gradient} 
\end{eqnarray}
 The change of the specific volume in Eq. (\ref{eq_continuity}) is
given by the balance in the displacement of the zone boundaries. The
rate of change of total energy in Eq. (\ref{eq_total_energy}) is
determined by the surface luminosity, \( L=4\pi r^{2}\rho H \), and
the work on the surface of the mass shell against the pressure, \( p+\rho K \).
Of leading order in the momentum equation (\ref{eq_momentum}) are
the pressure gradient and the gravitational force, \( m/r^{2} \).
The constraints (\ref{eq_volume_gradient}) and (\ref{eq_mass_gradient})
are most easily understood in the Newtonian limit (the enclosed volume
is defined by \( V=4\pi r^{3}/3 \)), where the first becomes the
definition of the rest mass density and the second the Poisson equation
for the gravitational potential. The time derivative in equation (\ref{eq_lapse_gradient})
is very small; therefore, this equation essentially acts as a constraint
on the lapse function, \( \alpha  \). This equation derives from
the space component of the four-divergence of the stress-energy tensor.
In addition to the evolution of the total energy, we also need an
equation for the evolution of the internal energy that we may derive
from the time component of the four-divergence of the stress-energy
tensor:\begin{equation}
\label{eq_internal_energy}
\frac{\partial }{\partial t}\left[ e+J\right] =-\frac{1}{\alpha }\left[ 4\pi r^{2}\alpha ^{2}\rho H\right] -\left( p+\rho K\right) \frac{\partial }{\partial t}\left( \frac{1}{\rho }\right) -\frac{\alpha u}{r}\left( J-3K\right) .
\end{equation}

Next, we detail the description of the radiation field. We identify
the energy flux \( \rho H \) with a particle flux that is determined
by a Boltzmann transport equation. The transport equation is split
into a left-hand side and a right-hand side. The left-hand side is
the directional derivative of the particle distribution function along
trajectories of free particle propagation. This derivative is equated
to the changes in the distribution function due to collisions, which
are described by the right hand side of the equation. Once a 1+1 decomposition
of space-time \citep{Arnowitt_Deser_Misner_62,Smarr_York_78} and a
basis in the momentum phase space for the particle four-momentum have
been chosen, the directional derivative along the phase flow can be
expressed in terms of partial derivatives of the distribution function
with respect to the space-time coordinates and momenta \citep{Lindquist_66,Mezzacappa_Matzner_89}.
We measure the particle four-momentum in a comoving orthonormal frame,
with components \begin{equation}
\label{eq_neutrino_coordinates}
p^{a}=p\cos \theta ,\; p^{\vartheta }=p\sin \theta \cos \phi ,\; p^{\varphi }=p\sin \theta \sin \phi .
\end{equation}
 In spherical symmetry, the particle energy, \( E \), measured in
a comoving frame, and the cosine of the angle between the particle
momentum and the radial direction, \( \mu =\cos \theta  \), completely
describe the particle phase space. The neutrinos are assumed to have
no mass. In spherical symmetry, the distribution function does not
depend on the three-momentum azimuth angle \( \phi  \). Thus, the
specific particle distribution function depends on four arguments
and describes the number of particles at a given time, \( t \), in
the phase space volume \( E^{2}dEd\mu da \) by \begin{equation}
\label{eq_neutrino_number}
dN=F(t,a,\mu ,E)E^{2}dEd\mu da.
\end{equation}
 With the metric of Eq. (\ref{eq_comoving_metric}), the Boltzmann
equation reads \citep{Yamada_Janka_Suzuki_99,Liebendoerfer_Mezzacappa_Thielemann_01},\begin{equation}
\label{eq_relativistic_boltzmann}
C_{t}+D_{a}+D_{\mu }+D_{E}+O_{\mu }+O_{E}=C_{c},
\end{equation}
with

\begin{eqnarray}
C_{t} & = & \frac{\partial F}{\alpha \partial t}\label{eq_boltzmann_ct} \\
D_{a} & = & \frac{\mu }{\alpha }\frac{\partial }{\partial a}\left[ 4\pi r^{2}\alpha \rho F\right] \label{eq_boltzmann_da} \\
D_{\mu } & = & \Gamma \left( \frac{1}{r}-\frac{1}{\alpha }\frac{\partial \alpha }{\partial r}\right) \frac{\partial }{\partial \mu }\left[ \left( 1-\mu ^{2}\right) F\right] \label{eq_boltzmann_dmu} \\
D_{E} & = & -\mu \Gamma \frac{1}{\alpha }\frac{\partial \alpha }{\partial r}\frac{1}{E^{2}}\frac{\partial }{\partial E}\left[ E^{3}F\right] \label{eq_boltzmann_de} \\
O_{E} & = & \left( \mu ^{2}\left( \frac{\partial \ln \rho }{\alpha \partial t}+\frac{3u}{r}\right) -\frac{u}{r}\right) \frac{1}{E^{2}}\frac{\partial }{\partial E}\left[ E^{3}F\right] \label{eq_boltzmann_oe} \\
O_{\mu } & = & \left( \frac{\partial \ln \rho }{\alpha \partial t}+\frac{3u}{r}\right) \frac{\partial }{\partial \mu }\left[ \mu \left( 1-\mu ^{2}\right) F\right] \label{eq_boltzmann_omu} \\
C_{c} & = & \frac{j}{\rho }-\chi F.\label{eq_boltzmann_cc} 
\end{eqnarray}
 The source on the right-hand side, \( C_{c} \), is the collision
term that describes changes in the particle distribution function
due to local interactions with matter. It is represented here by an
emissivity \( j \) and an opacity \( \chi  \). All other terms stem
from the partial derivatives of the distribution function with respect
to the phase-space coordinates in the directional derivative along
the phase flow. They can all be physically interpreted. The first
term on the left hand side of the equation, \( C_{t} \), is the temporal
change of the particle distribution function. The second term, \( D_{a} \),
counts the particles that are propagating into or out of an infinitesimal
mass shell. The third term, \( D_{\mu } \), accounts for the change
in the neutrino distribution function in an angle interval owing to
the propagation of the neutrinos along geodesics with changing local
angle cosine \( \mu  \). The curved particle trajectories in general
relativity are accounted for by the term proportional to the gradient
of the gravitational potential, \( \Phi  \),\[
\frac{1}{\alpha }\frac{\partial \alpha }{\partial r}=\frac{\partial \Phi }{\partial r}.\]
 The fourth term, \( D_{E} \), expresses the redshift or blueshift
of the particle energy that applies when the particles have a velocity
component in the radial direction (\( \mu \neq 0 \)) and, therefore,
change their position in the gravitational well. The fifth and sixth
term, \( O_{E} \) and \( O_{\mu } \), account for the Doppler shift
and the angular aberration between adjacent comoving observers.

The integration of the Boltzmann equation over momentum space, spanned
by the particle direction cosine and energy, gives the local conservation
laws for particle number and energy. We define \( J^{N} \) and \( H^{N} \)
to represent the zeroth and first \( \mu  \) moments of the distribution
function: \begin{eqnarray}
J^{N} & = & \int ^{1}_{-1}\int ^{\infty }_{0}FE^{2}dEd\mu ,\nonumber \\
H^{N} & = & \int ^{1}_{-1}\int ^{\infty }_{0}FE^{2}dE\mu d\mu .\label{eq_number_moments_definition} 
\end{eqnarray}
 Integration of Eq. (\ref{eq_relativistic_boltzmann}) over \( \mu  \)
and \( E \) with \( E^{2} \) as the measure of integration gives
the following evolution equation for \( J^{N} \): \begin{equation}
\label{eq_neutrino_number_conservation}
\frac{\partial J^{N}}{\partial t}+\frac{\partial }{\partial a}\left[ 4\pi r^{2}\alpha \rho H^{N}\right] -\alpha \int \frac{j}{\rho }E^{2}dEd\mu +\alpha \int \chi FE^{2}dEd\mu =0.
\end{equation}
 The derivatives with respect to the momentum phase space in Eq. (\ref{eq_relativistic_boltzmann})
do not contribute because \( \left( 1-\mu ^{2}\right)  \) vanishes
at \( \mu =\pm 1 \) and \( E^{3}F \) is zero for \( E=0 \) and
\( E=\infty  \). Eq. (\ref{eq_neutrino_number_conservation}) is
a continuity equation analogous to Eq. (\ref{eq_continuity}), extended
by source and sink terms for the radiation particles. One more integration
over the rest mass \( a \) from the center of the star to its surface
gives the evolution equation of the total particle number.

Slightly less straightforward is the derivation of total radiation
energy conservation. We define the energy moments \begin{eqnarray}
J & = & \int FE^{3}dEd\mu \nonumber \\
H & = & \int FE^{3}dE\mu d\mu \nonumber \\
K & = & \int FE^{3}dE\mu ^{2}d\mu \nonumber \\
Q & = & \int FE^{3}dE\mu ^{3}d\mu \label{eq_energy_moment_definition} 
\end{eqnarray}
 and evaluate the evolution of the radiation energy as measured by
an observer at infinity,\begin{equation}
\label{eq_evolution_total_energy}
\frac{\partial }{\partial t}\int \left( \Gamma +u\mu \right) FE^{3}dEd\mu .
\end{equation}
 To this purpose, we integrate Eq. (\ref{eq_relativistic_boltzmann})
again over phase space, but this time with measure of integration
\( \left( \Gamma +u\mu \right) E^{3} \). After performing some integrations
by parts to account for the time and space dependence of \( \Gamma  \)
and \( u \), this leads to the concise result \citep{Liebendoerfer_Mezzacappa_Thielemann_01}\begin{eqnarray}
0 & = & \frac{\partial }{\partial t}\left( \Gamma J+uH\right) +\frac{\partial }{\partial a}\left[ 4\pi r^{2}\alpha \rho \left( uK+\Gamma H\right) \right] +4\pi r\alpha \rho \left( 1+e+\frac{p}{\rho }\right) H\nonumber \\
 & - & \alpha \Gamma \int \left( \frac{j}{\rho }-\chi \right) E^{3}dEd\mu +\alpha u\int \chi FE^{3}dE\mu d\mu .\label{eq_radiation_energy_conservation} 
\end{eqnarray}
 Note that the conserved quantity \( \Gamma J+uH \) is the radiation
energy density in the frame of an observer at infinity. It is expressed
in terms of the momentum moments \( J \) and \( H \) in the comoving
frame. The second term describes the surface work by the radiation
pressure, \( \rho K \), and the energy loss or gain due to the luminosity
\( L=4\pi r^{2}\rho H \) at the boundary. The third term contains
a gravitational term coupling the matter enthalpy with the luminosity
that we neglected in previous work \citep{Liebendoerfer_Mezzacappa_Thielemann_01}.
The source terms in Eq. (\ref{eq_radiation_energy_conservation})
describe the energy exchange with matter by particle emission, absorption,
and radiation stress. The omitted terms from neutrino scattering enter
the equation in a similar form. Beforehand, we found that Eq. (\ref{eq_total_energy})
describes the evolution of the total energy. Then, from the Boltzmann
equation, we derived Eq. (\ref{eq_radiation_energy_conservation})
for the evolution of the radiation energy. Thus, we will find a consistent
equation for the evolution of the hydrodynamics part by the subtraction
of Eq. (\ref{eq_radiation_energy_conservation}) from Eq. (\ref{eq_total_energy}).
The result is Eq. (\ref{eq_hydro_total_energy}) in section \ref{subsection_hydrodynamics}
where we discuss the implementation of the hydrodynamics part. The
same procedure applied to other conserved quantities leads the full
set of consistent hydrodynamics equations. By taking moments of the
Boltzmann equation with the measures of integration \( E^{3}\left( u+\Gamma \mu \right)  \),
\( E^{3}\mu /\left( 4\pi r^{2}\rho \right)  \), and \( E^{3} \),
respectively, we derive equations for the evolution of the radiation
momentum, Eq. (\ref{eq_radiation_momentum}), a radiative contribution
to the lapse function, Eq. (\ref{eq_radiation_lapse}), and the radiation
energy in the comoving frame, Eq. (\ref{eq_radiation_internal_energy}):\begin{eqnarray}
\frac{\partial }{\partial t}\left[ uJ+\Gamma H\right]  & = & -\frac{\partial }{\partial a}\left[ 4\pi r^{2}\alpha \rho \left( \Gamma K+uH\right) \right] \nonumber \\
 & - & \frac{\alpha }{r}\left[ \left( J+3K\right) \frac{m}{r}-\left( 1-\frac{2m}{r}\right) \left( J-3K\right) -2K\right. \nonumber \\
 & + & \left. 4\pi r^{2}\left( J\left( p+\rho K\right) +\left( 1+e+J\right) \rho K-2\rho H^{2}\right) \right] \nonumber \\
 & - & \alpha \Gamma \int \chi FE^{3}dE\mu d\mu +\alpha u\int \left( \frac{j}{\rho }-\chi F\right) E^{3}dEd\mu \label{eq_radiation_momentum} \\
\frac{\partial }{\partial t}\left[ \frac{1}{4\pi r^{2}\rho }H\right]  & = & -J\frac{\partial \alpha }{\partial a}-\frac{1}{\rho }\frac{\partial }{\partial a}\left[ \alpha \rho K\right] +\frac{\alpha }{3VD}\left( J-3K\right) \nonumber \\
 & - & \frac{1}{4\pi r^{2}\rho }\int \chi FE^{3}dE\mu d\mu \label{eq_radiation_lapse} \\
\frac{\partial J}{\partial t} & = & -\frac{1}{\alpha }\left[ 4\pi r^{2}\alpha ^{2}\rho H\right] -\rho K\frac{\partial }{\partial t}\left( \frac{1}{\rho }\right) -\frac{\alpha u}{r}\left( J-3K\right) \nonumber \\
 & + & \alpha \int \left( \frac{j}{\rho }-\chi F\right) E^{3}dEd\mu .\label{eq_radiation_internal_energy} 
\end{eqnarray}
The subtraction of Eq. (\ref{eq_radiation_momentum}) from Eq. (\ref{eq_momentum})
leads to a hydrodynamics equation for the evolution of the momentum,
Eq. (\ref{eq_hydro_momentum}). The subtraction of Eq. (\ref{eq_radiation_lapse})
from Eq. (\ref{eq_lapse_gradient}) leads to a hydrodynamics equation
for the update of the lapse function, Eq. (\ref{eq_hydro_lapse_gradient}).
Finally, the substraction of Eq. (\ref{eq_radiation_internal_energy})
from Eq. (\ref{eq_internal_energy}) leads to a hydrodynamics equation
for the evolution of the internal energy, Eq. (\ref{eq_hydro_internal_energy}).
We will pay attention to preserve this consistency also in our finite
difference representation of the equations of radiation hydrodynamics.
However, before we proceed with the technical details in section \ref{section_computer_model},
we provide in the next subsection an overview of two exemplary simulation
runs to complete the physical context and to illustrate the numerical
challenges we face.

\subsection{Neutrino transport in two representative simulations}\label{section_simulations}

In this section, we provide an overview of the core collapse and postbounce
evolution in our models for the \( 13 \) M\( _{\odot } \) and \( 40 \)
M\( _{\odot } \) stellar progenitors. These provide the physical
context for the code tests in section \ref{section_code_verification}.
A thorough discussion of supernova physics can be found in previous
reviews, e.g. in \citep{Bethe_90,Burrows_Young_00,Mezzacappa_Bruenn_00,Janka_Kifonidis_Rampp_01}.
Earlier runs of the \( 13 \) M\( _{\odot } \) model have been described
in \citep{Mezzacappa_et_al_01,Liebendoerfer_et_al_01} with Newtonian
and general relativistic gravity. Here, we add information on the
formation of the neutrino spectra and report on our first self-consistent
simulation running all the way from core collapse to the onset of
black hole formation in the case of the \( 40 \) M\( _{\odot } \)
model. We have chosen progenitors on the light and massive side with
respect to the range of potential core collapse supernova progenitors.
This demonstrates the spread in the results. The results of our simulations
for intermediate mass progenitors are summarized in \citep{Messer_00,Liebendoerfer_et_al_01a,Liebendoerfer_et_al_02}.
The latest was calculated with the finite differencing described in
this paper.

\begin{figure}
{\centering \resizebox*{0.8\textwidth}{!}{\includegraphics{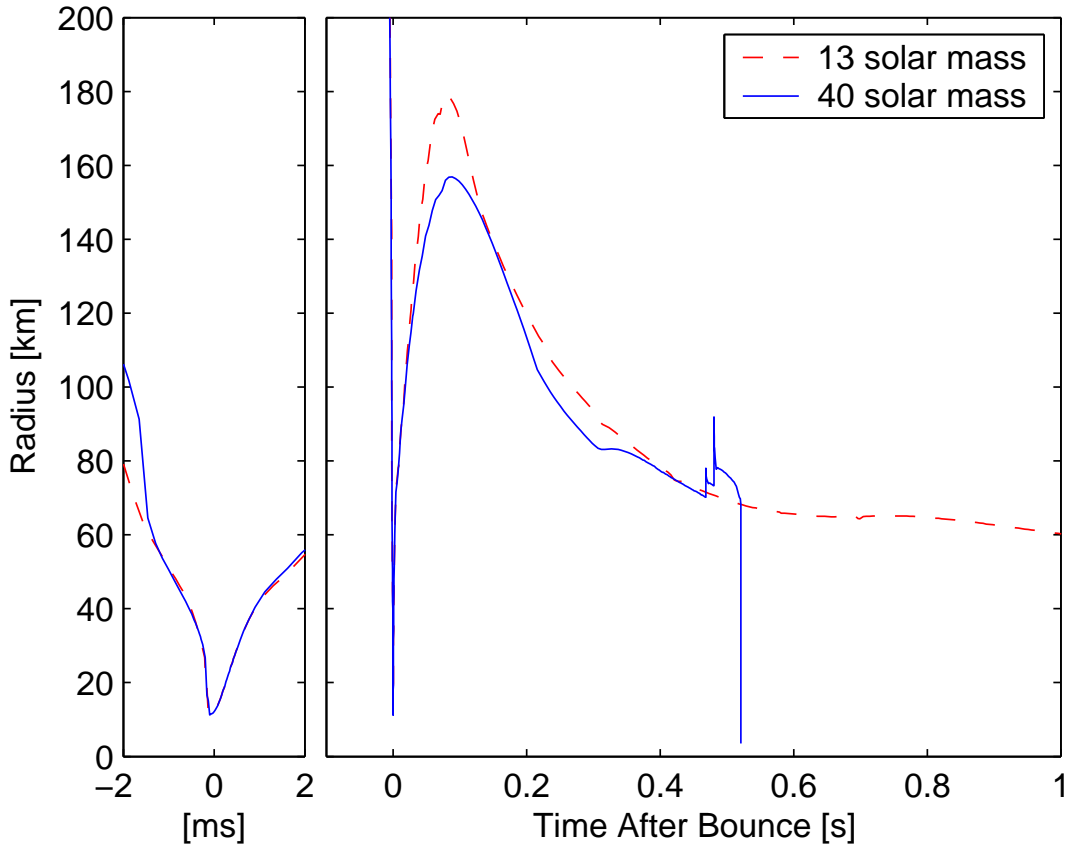}} \par}

\caption{Shown are the shock trajectories of the \protect\( 13\protect \)
M\protect\( _{\odot }\protect \) (dashed line) and \protect\( 40\protect \)
M\protect\( _{\odot }\protect \) (solid line) models. At negative
times, i.e. before bounce, the trajectories indicate the position
of the sonic point instead of the position of the not yet formed shock
wave. The left hand part of the figure zooms in on the time around
bounce to demonstrate that the shock is formed at the sonic point
(no discontinuity in the lines) and that this happens at the same
position in both models. The right hand part shows the shock trajectories
over a longer time scale. After about \protect\( 500\protect \) ms,
the \protect\( 40\protect \) M\protect\( _{\odot }\protect \) star
collapses to a black hole. The perturbation at \protect\( t_{pb}=\sim 0.3\protect \)
s is the consequence of a physical change in the luminosities. The
two perturbations after \protect\( t_{pb}=\sim 0.4\protect \) s are
the result of an artificial increase of the numerical shock width
we had to apply in order to run the simulations to the end.\label{fig_shock.ps}}
\end{figure}

The most prominent characterization of a supernova explosion is the
trajectory of the shock position. We define the shock position as
the location with the maximum infall velocity (i.e. minimum in the
velocity profile). Before the shock is formed after bounce, the location
with maximum infall velocity coincides with the sonic point which
separates the causally connected inner core from the supersonically
infalling outer core. The pressure wave emerging from the center at
bounce turns at this transition point into a shock wave. The trajectory
of maximum infall velocity is therefore continuous across bounce as
shown in the magnifying window on the left hand side of Fig. (\ref{fig_shock.ps}).
 If we compare the position of the sonic point in the \( 13 \) M\( _{\odot } \)
model with its position in the \( 40 \) M\( _{\odot } \) model,
we find that they converge to the same point before bounce. For the
explanation, we recall that the sonic point depends on the Chandrasekhar
mass, which is determined by the electron fraction profile. The electron
fraction profile of the two runs converges due to a strong feedback
of the electron fraction on the free proton abundance \citep{Messer_et_al_03}.
Within the {}``standard'' input physics, it is assumed that electron
capture on nuclei with a full neutron f7/2 shell is Pauli-blocked
\citep{Bruenn_85}. Under such conditions, our simulations allow only
electron capture on free protons. Now, if the electron fraction and/or
temperature in one model would only be slightly higher than in the
other model, this would result in a significantly higher free proton
abundance. It would cause a significantly higher number of electron
captures than in the other model. Hence, the differences between the
models are reduced. It has recently been shown that, due to the finite
temperature in the nuclei and due to correlations, electron capture
on nuclei dominates electron capture on free protons throughout core
collapse \citep{Langanke_et_al_03}. It will be interesting to see
if the described feedback will to the same extent be at work with
these more realistic electron capture rates. We can only expect this
if the electron capture on low abundance nuclei would turn out to
be comparable to, or larger than, the electron capture on the most
abundant nuclei (the quantity to compare would of course always be
the product of the electron capture rate with the abundance of the
target). The right hand side of Fig. (\ref{fig_shock.ps}) shows the
shock trajectories over a longer time interval, up to one second after
bounce. The shocks recede in both models after \( t_{pb}=100 \) ms.
The conditions for a shock revival deteriorate. The intensive neutrino
emission from the cooling region undermines the pressure support below
the heating region and the material is drained from the latter onto
the protoneutron star \citep{Janka_01}. The transient stall in the
receding shock front in the \( 40 \) M\( _{\odot } \) model at \( 0.3 \)
s after bounce is a reaction to an enhanced electron flavor luminosity.
It will be further analyzed below, together with the description of
the evolution of the luminosities. In the evolution of the \( 40 \)
M\( _{\odot } \) progenitor, we encountered a numerical stability
problem after \( t_{pb}=0.4 \) s. The adaptive grid created extremely
small mass zones such that the convergence radius of the Newton-Raphson
algorithm in the implicit hydrodynamics was severely reduced due to
truncation errors. We increased the artificial viscosity in two sequential
steps to widen the shock and continue the run. This numerical shock
widening is responsible for the outward steps in the shock position
\( \sim 0.5 \) s after bounce. Shortly after the collapse of the
protoneutron star to a black hole has set in, our code crashes unavoidably
because of the coordinate singularity in the comoving coordinates
at the Schwarzschild horizon. We will extensively use the hydrodynamic
profiles at \( 0.4 \) s after bounce for testing in later subsections.

\begin{figure}
{\centering \resizebox*{0.7\textwidth}{!}{\includegraphics{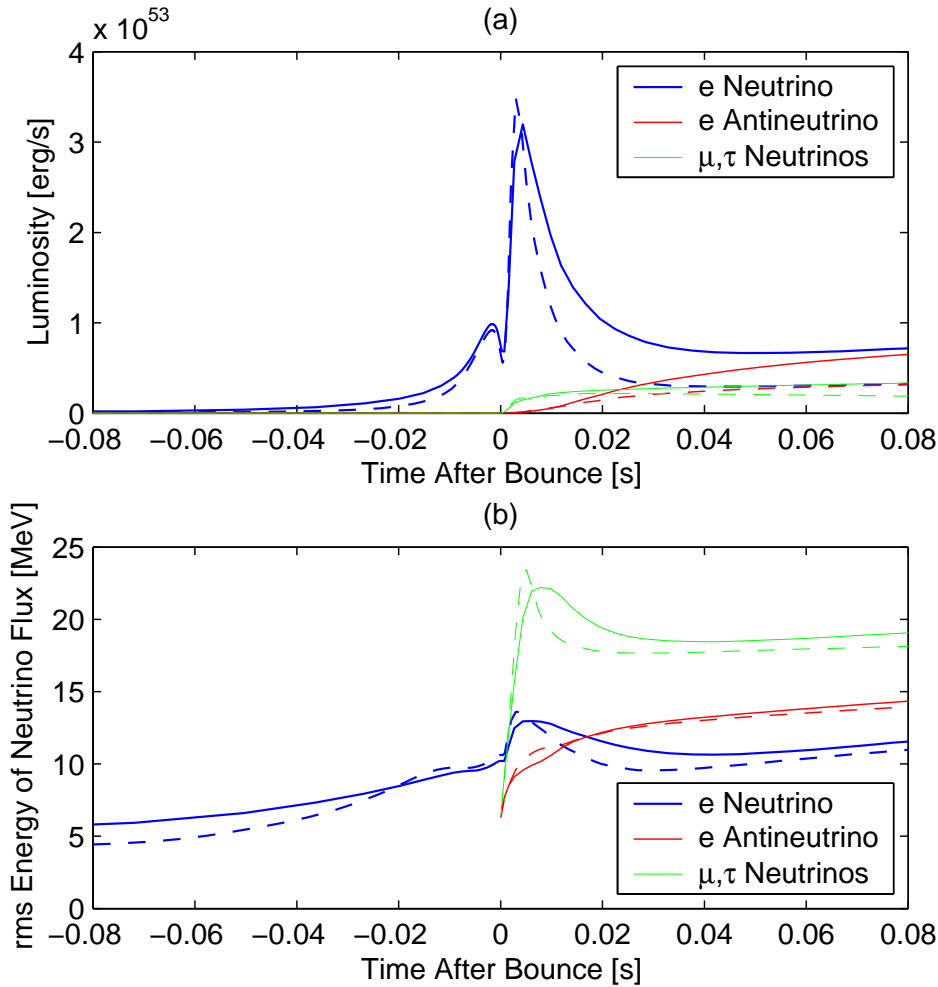}} \par}

\caption{The luminosities and rms energies of the neutrinos are shown as a
function of time. The results of the \protect\( 13\protect \) M\protect\( _{\odot }\protect \)
model are drawn with dashed lines and the results of the \protect\( 40\protect \)
M\protect\( _{\odot }\protect \) model with solid lines. A thick
line belongs to the electron neutrino, a line with medium width to
the electron antineutrino, and a thin line to the \protect\( \mu \protect \)-
and \protect\( \tau \protect \)-neutrinos. We sampled the luminosities
at a radius of \protect\( 500\protect \) km. Also the rms energy
of the neutrino flux (not abundance) was calculated at this location.
We adjusted the time coordinate by \protect\( \Delta t=-500\protect \)
km\protect\( /c\protect \) to account for the (approximate) propagation
time to the sampling radius. The two progenitors show a comparable
neutrino burst with a peak height of \protect\( 3.5\times 10^{53}\protect \)
erg/s. Significant differences appear in later phases. The variations
in the density profiles in the outer layers of the two models determine
the accretion-dominated electron flavor luminosities.\label{fig_lumin1.ps}}
\end{figure}

The neutrino luminosities and rms energies are shown in Fig. (\ref{fig_lumin1.ps}). 
The electron neutrino luminosities rise during core collapse and reach
a level of \( 10^{53} \) erg/s. The collapse is halted at nuclear
densities and a bounce shock propagates outwards through neutrino
opaque material. The neutrino luminosity decays to a \( 30\% \) lower
level during this short period of \( \sim 4 \) ms duration. It has
been assigned to a decrease in the free proton fraction when the shock
is formed \citep{Thompson_Burrows_Pinto_03}. Additionally, while the
shock is running out to the neutrinospheres, it condenses previously
still neutrino emitting material to even more neutrino opaque densities.
When the shock reaches the electron neutrinosphere, an electron neutrino
burst with a peak height of \( 3.5\times 10^{53} \) erg/s is launched
by copious electron capture. As the neutrinos escape quickly, the
freed phase space is refilled with new neutrinos and the matter deleptonizes
rapidly. This phase is very similar in both models because, after
core collapse, the structure of the inner core is so similar. Differences
appear later, when the accretion luminosity dominates over the core
diffusion luminosity with a ratio of about \( 2:1 \) to \( 3:1 \).
The higher densities in the outer layers of the more massive progenitor
produce considerably higher accretion luminosities when they settle
in the gravitational potential on the surface of the protoneutron
star. The rising electron neutrino rms energies before bounce reflect
the conditions in the compactifying material at infall. After the
neutrino burst, the rms energies adjust to the spectrum set by the
shock heated mantle and reflect the conditions at the location of
decoupling. However, we have to keep in mind that the location of
decoupling strongly varies for individual neutrino flavors and neutrino
energies. Thus, the rms energy rather reflects an emission-weighed
sampling of conditions at different locations. Quite generally, the
\( \mu  \) and \( \tau  \) neutrinos decouple deeper because of
the insensitivity of these neutrinos to charged current reactions.
The electron antineutrinos decouple deeper than the electron neutrinos
because of the smaller proton than neutron abundance.

In order to investigate the origin of the neutrino luminosities in
more detail, we introduce in appendix \ref{appendix_interaction_rates}
radius- and energy-dependent attenuation coefficients, \( \xi \left( r,E\right)  \),
that express the probability that a neutrino with energy \( E \)
emitted at radius \( r \) escapes from the computational domain without
a reaction that changes its type or energy. The attenuation coefficients
carry information that is similar to the optical depth, \( \tau \left( r,E\right)  \):
\( \xi \left( r,E\right) \simeq \exp \left( -\tau \left( r,E\right) \right)  \).
However, their evaluation according to Eq. (\ref{eq_attenuation_coefficients})
is fully consistent with the finite differencing of the Boltzmann
equation and takes the variations in the local flux factors into account.
Assume for example, we investigate a reaction \( \ell  \) that produces
at radius \( r \) neutrinos in the energy interval \( dE \) with
an energy emissivity \( E^{3}j^{\ell }(r,E)dE \). With the help of
the attenuation coefficients, we can quantify the contribution of
a given volume element \( 4\pi r^{2}dr \) to the total luminosity,\[
g^{\ell }\left( r,E\right) dEdr=\xi (r,E)j^{\ell }\left( r,E\right) E^{3}dE4\pi r^{2}dr.\]
 The total neutrino luminosity of a given neutrino type is then given
by the integral of \( g^{\ell }\left( r,E\right)  \) over energy
and position for all reactions that contribute, \[
L=\sum _{\ell }\int dr\int g^{\ell }\left( r,E\right) dE.\]
We demonstrate in appendix \ref{appendix_interaction_rates} that
the attenuation coefficients in Eq. (\ref{eq_attenuation_coefficients})
represent the total luminosity in the simulation accurately. In the
following figures, we cumulatively plot the quantity \( g^{\ell }\left( r,E\right) \Delta E \)
in units of erg/s/km. It is natural to choose \( \Delta E \) in accordance
with the width of the \( k_{\rm max}=12 \) energy groups we used
in the simulations. At the bottom of the figure we start with the
electron or positron capture reaction (depending on the neutrino type).
We draw \( g^{\rm capt}\left( r,E_{1}\right) \Delta E_{1} \) for
the lowest energy group and enclose it by a blue line. On top of
it, we add \( g^{\rm capt}\left( r,E_{2}\right) \Delta E_{2} \) for
the next energy group, again enclosed by a blue line. The shading
of the enclosed areas indicates the energy group according to the
legend at the bottom of the figure. Energy groups that do not contribute
collapse to a single line with zero enclosed area. On top of \( g^{\rm capt}\left( r,E_{k_{\rm max}}\right) \Delta E_{k_{\rm max}} \)
we continue with \( g^{\rm pair}\left( r,E_{1}\right) \Delta E_{1} \)
for the pair creation reaction. We enclose this and the following
area elements by a white line to distinguish them from the electron
or positron capture reactions. After the addition of \( g^{\rm pair}\left( r,E_{k_{\rm max}}\right) \Delta E_{k_{\rm max}} \),
we continue with the contributions from neutrino-electron scattering,
i.e. \( g^{\rm scat}\left( r,E_{1}\right) \Delta E_{1} \) to \( g^{\rm scat}\left( r,E_{k_{\rm max}}\right) \Delta E_{k_{\rm max}} \).
We use green lines as separators for the neutrino-electron scattering.
The figures become intuitively accessible as soon as one realizes
that the total shaded area is proportional to the total luminosity
of the star. The total area of a specific energy shading is proportional
to the contribution to the total luminosity of neutrinos from the
corresponding energy group. The total area of a specific reaction
is proportional to the contribution to the total luminosity of neutrinos
with this last inelastic interaction before the escape. A cross section
through the shaded area at a given radius tells about the spectrum
of the neutrinos escaping from that region, and about the probability
of the reaction type they had at that position before the escape.
In order to characterize also the extent of isoenergetic scattering
of neutrinos off nucleons and nuclei, we mark the neutrinospheres
for the energy groups at the top of the figure. The energies are rising
from the left to the right according to the legend at the bottom of
the figure. For the interpretation of the figures it is also useful
to know the thermodynamical conditions at the locations the neutrinos
are emitted. For each density decade we set a marker at the bottom
of the figure. Additionally, we include the electron fraction in the
graph with the electron neutrino analysis, and the entropy in the
graph with the electron antineutrino analysis. The solid line represents
the profile in the simulation, the dashed line represents the equilibrium
value that would be achieved by infinitely long exposure of the stationary
fluid element to the prevailing neutrino abundances. Finally, the
graph with the \( \mu  \)- and \( \tau  \)-neutrino analysis obtains
profiles with the temperature (dashed line) and electron chemical
potential (solid line).

\begin{figure}
{\centering \resizebox*{0.7\textwidth}{!}{\includegraphics{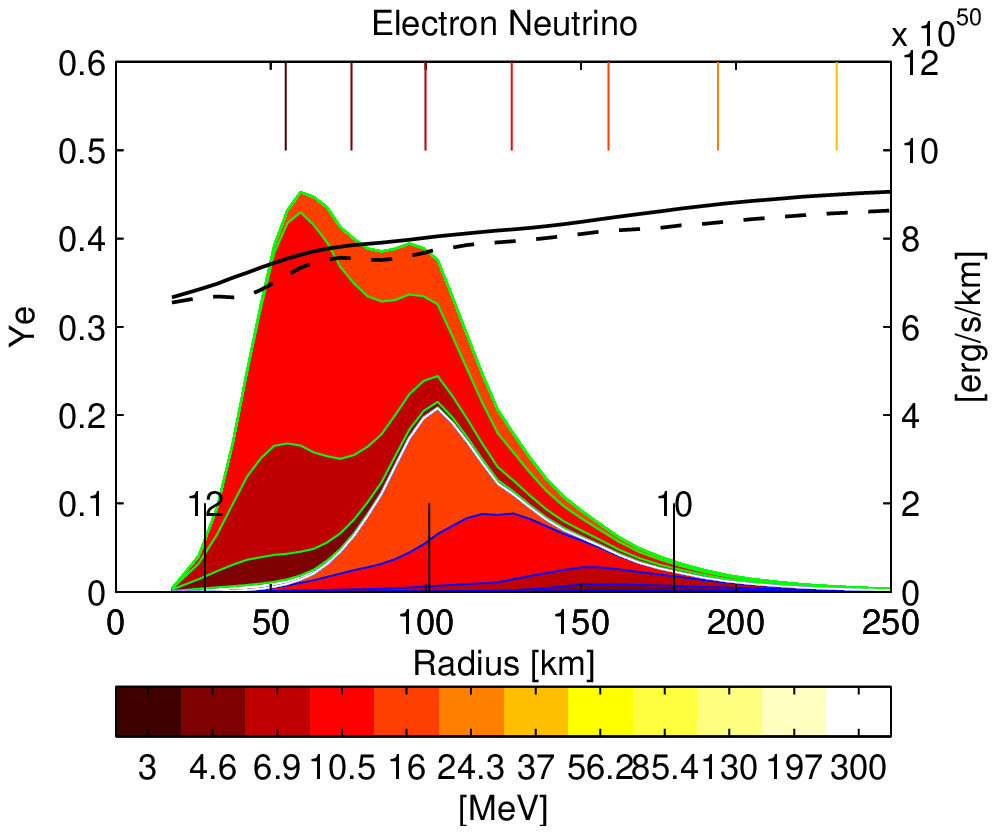}} \par}

\caption{The last inelastic interactions of escaping neutrinos at \protect\( 5\protect \)
ms before bounce in the \protect\( 13\protect \) M\protect\( _{\odot }\protect \)
model. The abscissa of the graph is the radius, ranging from the center
of the star to \protect\( 250\protect \) km radius. The density markers
at the bottom of the graph indicate the position of density decades,
\protect\( \log _{10}\left( \rho \right) \protect \), where \protect\( \rho \protect \)
is given in g/cm\protect\( ^{3}\protect \). The thick solid line
shows the electron fraction profile. The thick dashed line shows the
electron fraction in weak equilibrium under otherwise unchanged conditions
(same density, temperature, neutrino abundances, and spectra). These
quantities belong to the ordinate at the left hand side. The total
shaded area in the graph corresponds to the total electron neutrino
luminosity at \protect\( 5\protect \) ms before bounce in units of
erg/s. The appropriate ordinate on the right hand side then carries
the units erg/s/km. The shaded area is divided into three sections,
according to the type of the last energy-changing reaction of the
escaping neutrinos. Segments separated by blue lines in the lower
part of of the shaded area outline the contribution of electron captures
to the escaping neutrinos. The contribution from pair production is
bordered by white lines. The contribution from neutrino-electron scattering
is shown in the upper part of the shaded area, bordered
by green lines. As there are no contributions from pair production
in the cool and electron-degenerate material at \protect\( 5\protect \)
ms before bounce, the section belonging to the pair production reaction
collapses to a white line between the electron capture section and
the neutrino-electron scattering section. The contribution for each
reaction is further subdivided into contributions from each energy
group in the simulation. The intensity of the shading identifies the
neutrino energy according to the legend at the bottom of the figure.
Each energy group has its own neutrinosphere at optical depth \protect\( \tau =2/3\protect \).
Their locations, marked in the upper part of the figure, statistically
indicate the positions of the last interaction of the neutrinos with
matter, isoenergetic scattering included. The energies rise from the
left to the right according to the legend. The transient ondulations
in the luminosity contributions in this phase are a numerical artefact
caused by the low resolution of the Fermi surface in the energy dimension
of the momentum phase space (see discussion in subsection \ref{subsection_resolution}).\label{fig_lumcomp0025.ps}}
\end{figure}

Fig. (\ref{fig_lumcomp0025.ps}),
for example, shows a snapshot at \( 5 \) ms before bounce in the
collapse of the \( 13 \) M\( _{\odot } \) progenitor. Neutrinos
are escaping from the range between \( 20 \) km and \( 200 \) km
radius, roughly corresponding to a density range of \( 10^{10} \)
g/cm\( ^{3} \) to \( 10^{12} \) g/cm\( ^{3} \). The electron fraction
profile (solid line) reflects the deleptonization that has already
occurred in the more interior regions where \( Y_{e} \) approaches
values around \( 0.3 \). The equilibrium value (dashed line) is still
lower, indicating that the deleptonization is ongoing and that the
deleptonization time scale is slightly slower than the dynamical time
scale. The pair process does not contribute in this collapse phase
of high electron degeneracy. The corresponding area collapses to one
white line in the graph that separates the electron capture contributions
below it from the neutrino-electron scattering contributions above
it. Almost no neutrinos escape directly after an electron capture
at a radius as small as \( 50 \) km. Neutrinos escaping from this
region are thermalized by scattering off electrons. They escape with
quite low energies. Around \( 100 \) km radius we find that about
half of the escaping neutrinos stem directly from electron capture,
while the other half has scattered off an electron. At \( 150 \)
km \( \sim 2/3 \) of the neutrinos escape without further electron-scattering.
Among the neutrinos produced by electron capture, the neutrinos with
higher energies escape from smaller radii than the neutrinos with
lower energies. In this collapse phase, the {}``standard'' input
physics only includes electron captures on free nucleons. The Q value
of the reaction is small and very few low energy neutrinos are directly
produced if the electron chemical potential is large (e.g. \( 17 \)
MeV at \( 100 \) km radius in this time slice). The escaping low
energy neutrinos have scattered off electrons. However, the larger
Q value of more realistic electron capture rates on neutron-rich nuclei
may shift the energy of directly escaping neutrinos to lower values
such that the count of direct escapes is increased \citep{Langanke_et_al_03}.
Finally, we remark that the neutrinos in a given energy group are
produced at a significantly smaller radius than the location of the
corresponding transport sphere. This is the result of the dominance
of the isoenergetic scattering cross section in the collapse phase.
The diffusive propagation of the neutrinos extends to much lower densities
than neutrino-electron scattering or neutrino absorption.

\begin{figure}
{\centering \resizebox*{0.7\textwidth}{!}{\includegraphics{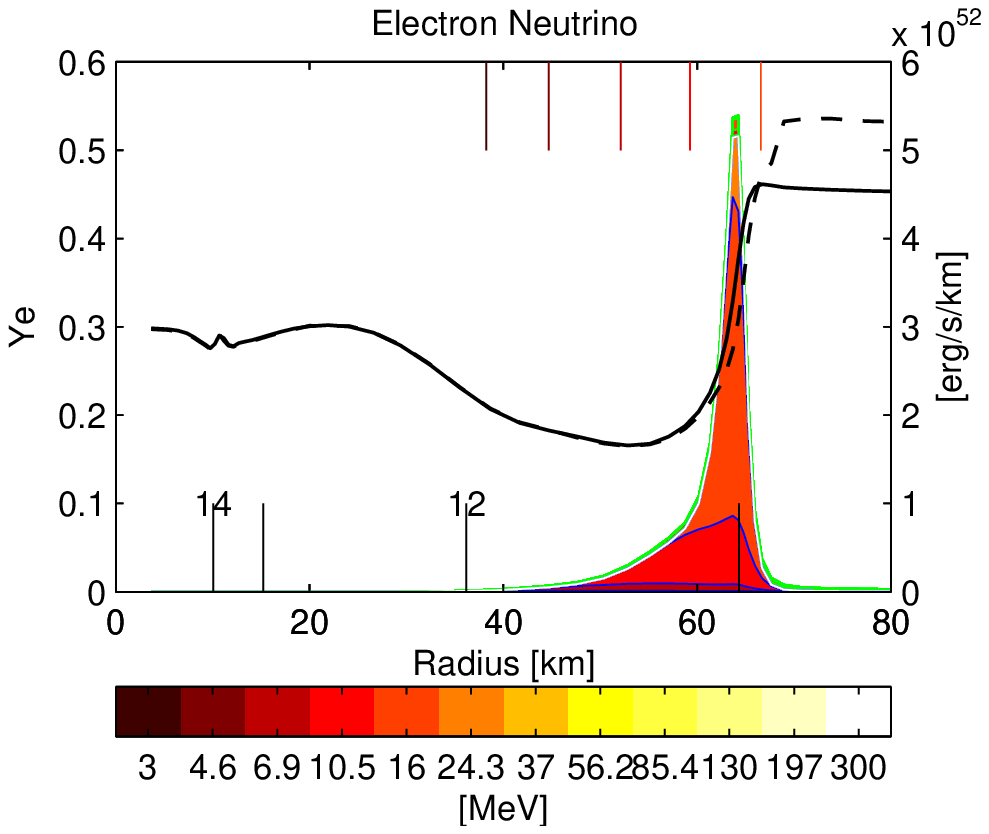}} \par}

\caption{The last inelastic interactions of escaping neutrinos at \protect\( 5\protect \)
ms after bounce in the \protect\( 13\protect \) M\protect\( _{\odot }\protect \)
model. The abscissa of the graph is the radius, ranging from the center
of the star to \protect\( 80\protect \) km radius. The density markers
at the bottom of the graph indicate the position of density decades,
\protect\( \log _{10}\left( \rho \right) \protect \), where \protect\( \rho \protect \)
is given in g/cm\protect\( ^{3}\protect \). The thick solid line
shows the electron fraction profile. The thick dashed line shows the
electron fraction in weak equilibrium under otherwise unchanged conditions
(same density, temperature, neutrino abundances, and spectra). These
quantities belong to the ordinate at the left hand side. The total
shaded area in the graph corresponds to the total electron neutrino
luminosity at \protect\( 5\protect \) ms after bounce in units of
erg/s. The appropriate ordinate on the right hand side then carries
the units erg/s/km. Note that the scale is \protect\( 50\protect \)
times larger than in Fig. (\ref{fig_lumcomp0025.ps}). In this phase,
almost all neutrinos escape directly after their production by electron
capture on free protons (area below the white line). The emission
of the neutrino burst occurs from a very narrow radius interval. This
causes a steep drop in the electron fraction at the same position.
The subdivision of the burst into different energy groups also shows
a narrow energy spectrum of the emitted neutrinos. Inside \protect\( 50\protect \)
km radius, the trapped neutrinos are in weak equilibrium with the
matter. In the burst region, the deleptonization time scale is only
slightly slower than the shock propagation . Outside the shock front,
it is much slower because of the low abundance of free protons.\label{fig_lumcomp00098.ps}}
\end{figure}

We switch to the next interesting phase: the electron neutrino burst
at \( \sim 5 \) ms after bounce (Fig. \ref{fig_lumcomp00098.ps}).
While the region of neutrino emission during core collapse was very
broad, it is extremely narrow (\( \sim 10 \) km) in the burst phase.
In the neutrino burst, the electron neutrinos escape directly from
electron capture. Some neutrino-electron scattering does occur in
front of the shock in an earlier stage and inelastic scattering of
burst neutrinos on nuclei in front of the shock are possible \citep{Bruenn_Haxton_91},
but not included in the standard input physics. Only the energy groups
at \( 10.5 \) MeV and \( 16 \) MeV contribute significantly to the
burst. This is in good agreement with the position of the corresponding
transport sphere in the upper part of the figure (fourth and fifth
from the left, respectively). The density is of order \( 10^{11} \)
g/cm\( ^{3} \). The trapped electron neutrinos inside the region
of main emission are in weak (and thermal) equilibrium with the fluid.
This is evident in the congruence of the electron fraction profile
(solid line) with the equilibrium \( Y_{e} \) (dashed line).

The situation at \( 50 \) ms after bounce is shown in Fig. (\ref{fig_lumcomp0126.ps}).
\begin{figure}
{\centering \resizebox*{0.6\textwidth}{!}{\includegraphics{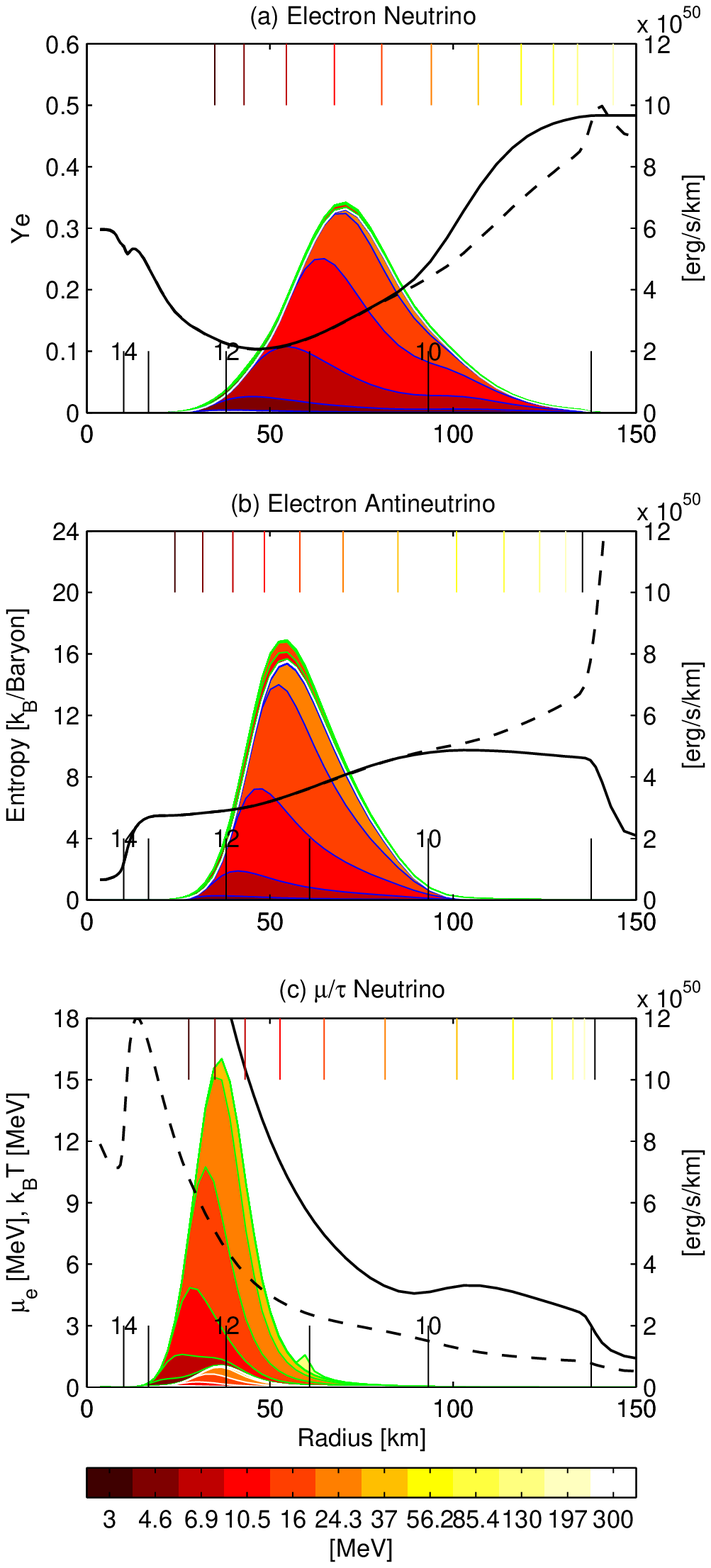}} \par}
\end{figure}
\begin{figure}

\caption{The last inelastic interactions of escaping neutrinos at \protect\( 50\protect \)
ms after bounce in the \protect\( 13\protect \) M\protect\( _{\odot }\protect \)
model. The abscissa of the graph is the radius, ranging from the center
of the star to \protect\( 150\protect \) km radius. The density markers
at the bottom of the graph indicate the position of density decades,
\protect\( \log _{10}\left( \rho \right) \protect \), where \protect\( \rho \protect \)
is given in g/cm\protect\( ^{3}\protect \). The thick solid line
in graph (a) shows the electron fraction profile. In graph (b) it
is the entropy profile, and in graph (c) the electron chemical potential.
The thick dashed line gives the equilibrium \protect\( Y_{e}\protect \)
in graph (a), the equilibrium entropy in graph (b), and the temperature
profile in graph (c). We do not repeat the detailed explanation of
the differently shaded and separated areas indicating the energy-
and reaction-specific contribution of neutrino emissivities to the
total luminosity of the star. Instead, we refer to the caption of
Fig. (\ref{fig_lumcomp0025.ps}) or to the explanation given in the
text. Graph (a) shows the origin of escaping electron neutrinos. Only
electron capture contributes significantly (area below the white line).
By definition, the region of neutrino emission coincides with the
cooling region where the fluid is in weak equilibrium with the neutrino
abundances. In the heating region, between the cooling region and
the shock front at \protect\( 140\protect \) km radius, the reaction
time scales are comparable to, or larger than, the infall time scale.
Graph (b) shows the origin of the electron antineutrino luminosity.
The electron antineutrinos are emitted from a slightly smaller radius.
The contribution of electron-scattered neutrinos (above the white
line) is also slightly larger than for the electron neutrinos. By
following the entropy profile from the right to the left (as an infalling
fluid element would experience it) we find the expected abrupt entropy
increase at the shock position. Behind the shock, the fluid element
would drift more slowly inwards and neutrino absorption indeed leads
to a small increase of the entropy in the region between \protect\( 140\protect \)
km and \protect\( \sim 100\protect \) km radius. However, the equilibrium
entropy is declining towards smaller radii such that cooling becomes
unavoidable once the fluid entropy has joined the equilibrium entropy
in a still infalling state. Graph (c) shows the origin of the \protect\( \mu \protect \)-
and \protect\( \tau \protect \)-luminosities. Almost no neutrinos
escape directly from pair creation (area enclosed by white lines at
the bottom of the figure). Most of the neutrinos have scattered off
electrons before their escape (area enclosed by green lines).\label{fig_lumcomp0126.ps}}
\end{figure}
This is the phase where neutrino heating starts to set in. All luminosities
are fully developed at this time. Graph (a) reveals electron capture
as the almost exclusive source of electron neutrinos. The higher energy
neutrinos emerge from larger radii. The isoenergetic scattering cross
sections are comparable to the neutrino absorption cross sections.
Thus, the regions of emission for the different energy groups are
nicely centered around the corresponding transport sphere. The continued
deleptonization after the neutrino burst caused a rather broad trough
in the electron fraction profile. The region of neutrino emission
coincides with the cooling region where the fluid is in weak equilibrium
with the neutrino abundances. In the heating region, between the cooling
region and the shock front at \( 140 \) km radius, the reaction time
scales are comparable to, or larger than, the infall time scale. Graph
(b) shows the origin of the electron antineutrino luminosity. The
electron antineutrinos are emitted from a slightly smaller radius
at densities exceeding \( 10^{11} \) g/cm\( ^{3} \) (while they
were similar to \( 10^{11} \) g/cm\( ^{3} \) for the electron neutrinos).
The electron antineutrinos decouple at smaller radii because of the
higher neutron than proton abundance. The contribution of electron-scattered
neutrinos (above the white line) is slightly larger than in the case
of the electron neutrinos. The pair production process does not noticeably
contribute to the luminosity at this stage. The equilibrium entropy
(dashed line) is increasing with increasing radius. An infalling fluid
element changes its entropy rather slowly by electron capture until
it hits the accretion shock. At the shock front, most of its kinetic
energy is converted into heat by shock dissipation. The heavy nuclei
are dissociated, mainly into free nucleons which are good neutrino
absorbers. However, before the stage at \( 50 \) ms after bounce,
the entropy of an infalling fluid element has already reached or exceeded
the equilibrium entropy, alone by shock dissipation. Only cooling
is then possible during the continued infall. After \( t_{pb}\sim 50 \)
ms, however, the equilibrium entropy is not reached by the shock dissipation
and the fluid element indeed increases its entropy towards the equilibrium
by neutrino absorption during its flight through the heating region
(solid line in graph (b) \( \sim 120 \) km radius). As the fluid
element also tends to decrease the electron fraction (graph (a)),
electron antineutrinos are preferentially absorbed. Once the equilibrium
is reached, cooling becomes unavoidable if the fluid element is still
infalling because the reaction rates at these densities are faster
than the dynamical infall time. Convection in the heating region is
expected to increase the neutrino heating efficiency (see e.g. the
parameter study of \citet{Janka_Mueller_96}), but not included in
our simulations. We did not find any sign of shock revival in spherical
symmetry with the included input physics. Graph (c) shows the origin
of the \( \mu  \)- and \( \tau  \)-luminosities. The \( \mu  \)-
and \( \tau  \)-neutrinos are produced at an average density slightly
larger than \( 10^{12} \) g/cm\( ^{3} \). Almost no neutrinos escape
directly from pair creation (area enclosed by white lines at the bottom
of the figure). Most of the neutrinos have scattered off electrons
before their escape (area enclosed by green lines). Pair production
is not the dominant source of \( \mu  \)- and \( \tau  \)-neutrinos.
It has been shown, that the production from bremsstrahlung \citep{Thompson_Burrows_Horvath_00}
and from electron flavor neutrino annihilation \citep{Buras_et_al_03a}
exceeds the pair process production rate (both reactions are not included
in our simulations). The latter reference finds that the \( \mu  \)-
and \( \tau  \)-neutrino luminosities show differences of \( 10\%-20\% \)
in the first \( 100 \) ms after bounce and converge to the standard
luminosities afterwards. The spectra are not significantly different.
This finding is also supported by our graph (c): The production site
of the \( \mu  \)- and \( \tau  \)-neutrinos is at a much smaller
radius than their transport sphere. Hence, the neutrino luminosity
is set by the (though energy-dependent) diffusivity between the location
of neutrino production and the transport sphere. Moreover, graph (c)
shows that the majority of escaping neutrinos scattered off electrons
in their last reaction. Differences in the production-spectrum are
likely to be washed out during the thermalization the neutrinos are
experiencing while they are diffusing outwards to the transport sphere.

Finally, we present the situation at \( 500 \) ms after bounce in
Fig. (\ref{fig_f5b.ps}).
\begin{figure}
{\centering \resizebox*{0.6\textwidth}{!}{\includegraphics{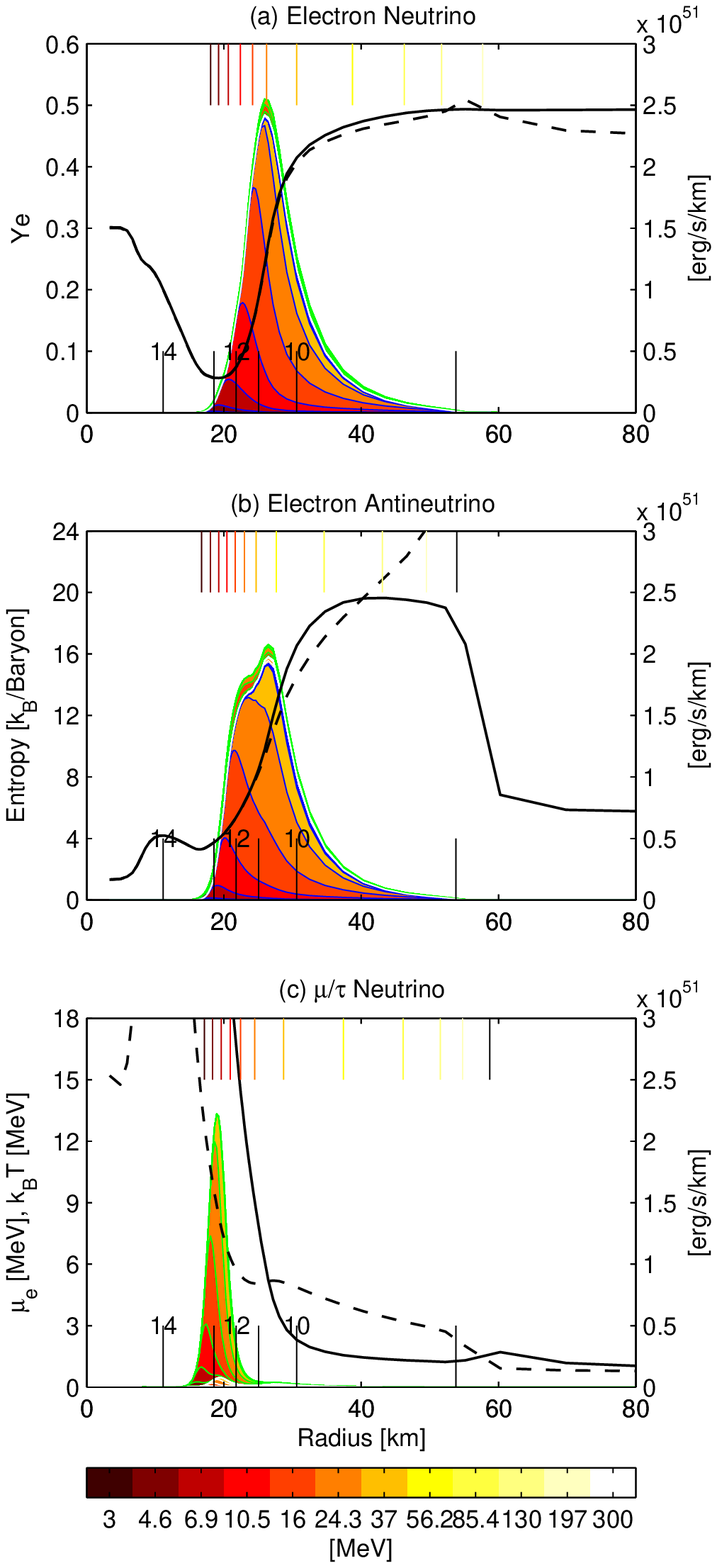}} \par}
\end{figure}
\begin{figure}

\caption{The last inelastic interactions of escaping neutrinos at \protect\( 500\protect \)
ms after bounce in the \protect\( 13\protect \) M\protect\( _{\odot }\protect \)
model. The abscissa of the graph is the radius, ranging from the center
of the star to \protect\( 80\protect \) km radius. The density markers
at the bottom of the graph indicate the position of density decades,
\protect\( \log _{10}\left( \rho \right) \protect \), where \protect\( \rho \protect \)
is given in g/cm\protect\( ^{3}\protect \). The thick solid line
in graph (a) shows the electron fraction profile. In graph (b) it
is the entropy profile, and in graph (c) the electron chemical potential.
The thick dashed line gives the equilibrium \protect\( Y_{e}\protect \)
in graph (a), the equilibrium entropy in graph (b), and the temperature
profile in graph (c). We do not repeat the detailed explanation of
the differently shaded and separated areas indicating the energy-
and reaction-specific contribution of neutrino emissivities to the
total luminosity of the star. Instead, we refer to the caption of
Fig. (\ref{fig_lumcomp0025.ps}) or to the explanation given in the
text. Graph (a) shows the origin of escaping electron neutrinos. Only
electron capture contributes significantly (area below the white line).
Neutrinos with larger energies still escape from larger radii because
of the corresponding staggering of the transport spheres at the top
of the figure. The shock has receded to a radius of \protect\( 57\protect \)
km and all regions are more compact. Graph (b) shows the origin of
the electron antineutrino luminosity. The accreted fluid elements
fall rapidly through the heating region without significant neutrino
heating. Graph (c) shows the origin of the \protect\( \mu \protect \)-
and \protect\( \tau \protect \)-luminosities. Still very few neutrinos
escape directly from pair creation (area enclosed by white lines at
the bottom of the figure). Most of the neutrinos have scattered off
electrons before their escape (area enclosed by green lines). The
continued cooling by \protect\( \mu \protect \)- and \protect\( \tau \protect \)-neutrino
emission becomes now visible in the entropy profile at a radius of
\protect\( 18\protect \) km. At larger radii, however, between \protect\( 25\protect \)
km and the shock position, the temperature is rising and the electrons
are nondegenerate. The overlap of this material with the emission
region of electron antineutrinos in graph (b) enhances the emission
of higher energy antineutrinos.\label{fig_f5b.ps}}
\end{figure}
This is after a long quasi-stationary phase of matter accretion and
shock recession. The volume of neutrino emitting material has considerably
shrunken with respect to the situation at \( 50 \) ms after bounce.
But there are not much qualitative changes. The neutrinos with larger
energies still escape from larger radii because of the corresponding
staggering of the transport spheres in the steep density gradient
at the surface of the protoneutron star. The shock has receded to
a radius of \( 57 \) km. Graph (b) shows the origin of the electron
antineutrino luminosity. The infalling fluid elements are now crossing
the heating region that rapidly (with several thousand km/s) that
there is no time for significant neutrino heating. This can be seen
in the flat top of the entropy curve between \( 30 \) km and \( 50 \)
km radius (solid line). Even the cooling sets in with a slight delay
and thermal balance is only reached at densities larger than \( 10^{11} \)
g/cm\( ^{3} \). Graph (c) shows the origin of the \( \mu  \)- and
\( \tau  \)-luminosities. Still very few neutrinos escape directly
from pair creation (area enclosed by white lines at the bottom of
the figure). Most of the neutrinos have scattered off electrons before
their escape (area enclosed by green lines). The continued cooling
by \( \mu  \)- and \( \tau  \)-neutrino emission becomes now clearly
visible in the entropy profile at a radius of \( 18 \) km, where
an entropy dip develops. At larger radii, however, between \( 25 \)
km and the shock position, the temperature is rising and the electrons
have become nondegenerate. The overlap of this material with the emission
region of electron antineutrinos in graph (b) lets the emission of
higher energy antineutrinos shift to lower densities than before.

\begin{figure}
{\centering \resizebox*{0.7\textwidth}{!}{\includegraphics{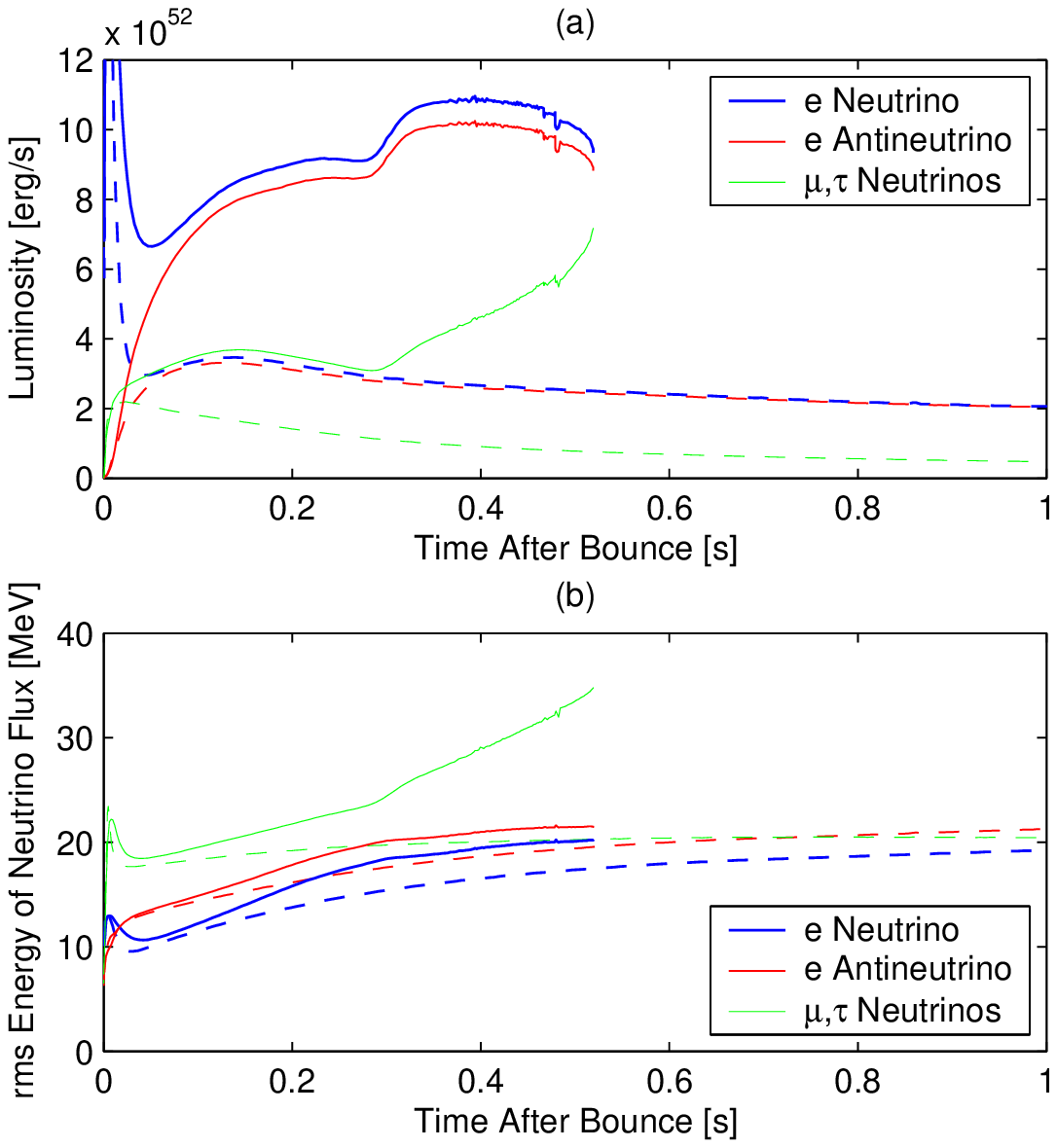}} \par}

\caption{The luminosities and rms energies in the two models are shown on
a longer time scale. In the \protect\( 13\protect \) M\protect\( _{\odot }\protect \)
model (dashed lines), the electron neutrino luminosity (thick line)
and electron antineutrino luminosity (medium width line) converge.
Both are much larger than the \protect\( \mu \protect \)- and \protect\( \tau \protect \)-neutrino
luminosities (thin line) because of the contribution from the accretion
luminosity. The rms energies show the conventional hierarchy at the
beginning. However, at \protect\( 0.7\protect \) s after bounce,
the rms energy of the \protect\( \mu \protect \)- and \protect\( \tau \protect \)-neutrino
luminosity falls below the rms energy of the electron antineutrinos.
This is due to the fact that this low mass protoneutron star is quite
incompressible with respect to the accumulated mass at the given accretion
rate. The continued emission of \protect\( \mu \protect \)- and \protect\( \tau \protect \)-neutrinos
cools the deep layers (around \protect\( r=18\protect \) km in Fig.
(\ref{fig_f5b.ps})) independently from the dynamics of the outer
layers. This is different in the \protect\( 40\protect \) M\protect\( _{odot}\protect \)
mass model (solid lines). The protoneutron star comes closer to its
maximum mass at \protect\( 0.3\protect \) s after bounce such that
it contracts appreciably with continued accretion. An increase in
the accretion rates lets the electron flavor neutrinos step up. The
accelerated transition from a mass accumulating stiff central object
to a contracting compressible core affects the \protect\( \mu \protect \)-
and \protect\( \tau \protect \)-neutrino properties. Shock-heated
material is faster condensed to higher densities and the luminosities
and rms energies of the \protect\( \mu \protect \)- and \protect\( \tau \protect \)-neutrinos
start to rise continuously until the protoneutron star collapses to
a black hole at \protect\( \sim 0.5\protect \) s after bounce.\label{fig_lumin2.ps}}
\end{figure}

\begin{figure}
{\centering \resizebox*{0.7\textwidth}{!}{\includegraphics{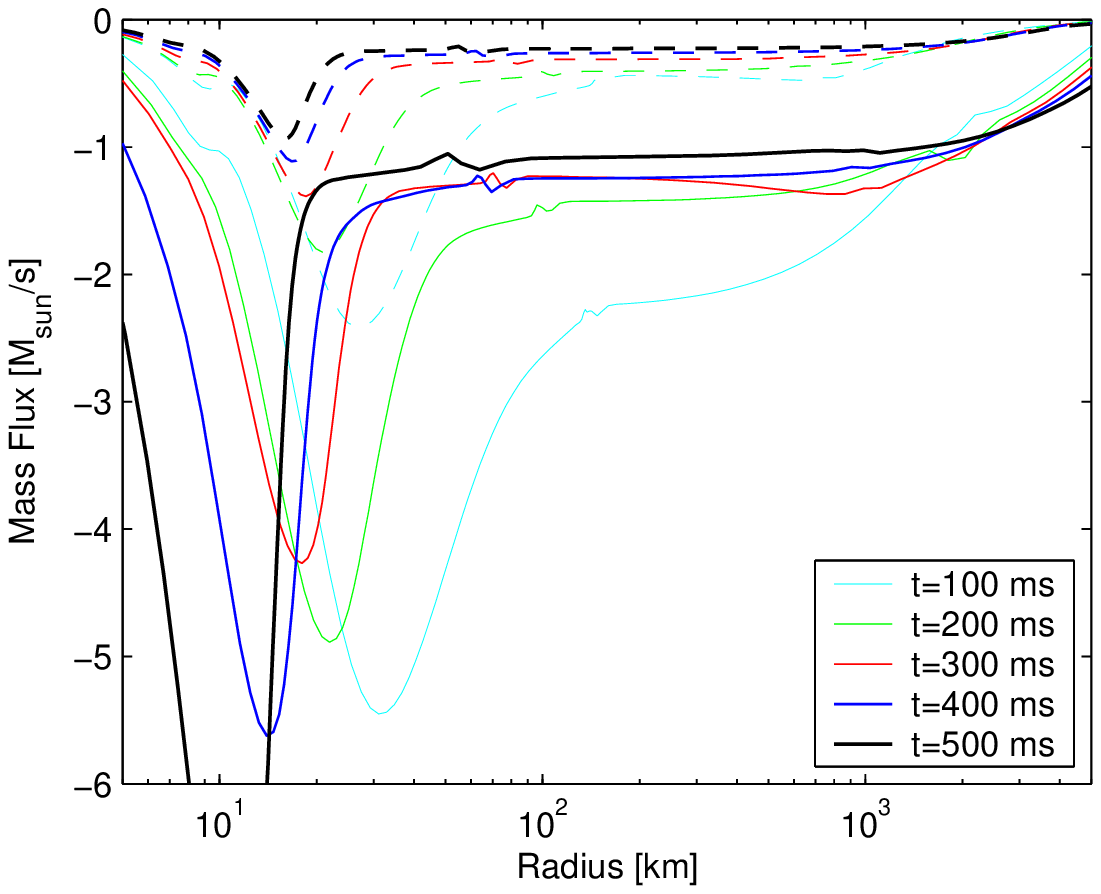}} \par}

\caption{Mass flux through surfaces at constant radii in the \protect\( 13\protect \)
M\protect\( _{\odot }\protect \) model (dashed lines) and \protect\( 40\protect \)
M\protect\( _{\odot }\protect \) model (solid lines). The profiles
are given at \protect\( 0.1\protect \) s, \protect\( 0.2\protect \)
s, \protect\( 0.3\protect \) s, \protect\( 0.4\protect \) s, and
\protect\( 0.5\protect \) s after bounce. While the accreted mass
piles up on the small protoneutron star in the \protect\( 13\protect \)
M\protect\( _{\odot }\protect \) model, it forces the massive protoneutron
star in the \protect\( 40\protect \) M\protect\( _{\odot }\protect \)
model to hydrostatic gravitational contraction after \protect\( t_{pb}=0.3\protect \)
s. An incoming variation in the accretion rate is visible outside
of \protect\( 1000\protect \) km radius.\label{fig_massflux.ps}}
\end{figure}

Figure (\ref{fig_lumin2.ps})
shows the luminosities and rms energies of the neutrino flux after
bounce on a longer time scale. In the \( 13 \) M\( _{\odot } \)
model, the luminosities decrease as a consequence of the declining
accretion rate and continued deleptonization of the core. The electron
flavor luminosities reach very similar values because the lifted electron
degeneracy in a large part of the cooling region (see Fig. (\ref{fig_f5b.ps}bc))
allows the electrons and positrons to be captured from similar chemical
potentials. The luminosities are higher than the luminosities of the
\( \mu  \)- and \( \tau  \)-neutrinos because the latter do not
have an accretion luminosity component. The rms energies show the
usual hierarchy at the beginning, but after \( t_{pb}=0.7 \) s, the
rms energy of the \( \mu  \)- and \( \tau  \)-neutrinos falls below
the rms energy of the electron antineutrino. This is also understood
if one looks again at Fig. (\ref{fig_f5b.ps}bc). While the emission
of high energy electron antineutrinos is aided by shock-heated material
settling at the base of the cooling region with moderate electron
degeneracy, the layers around \( 18 \) km radius, i.e. where the
energy spectra of the \( \mu  \)- and \( \tau  \)-neutrinos are
set, are barely affected by the continued accretion on the still not
very massive protoneutron star. This domain just slowly cools by neutrino
emission (compare the entropy profiles in Fig. (\ref{fig_lumcomp0126.ps}b)
and (\ref{fig_f5b.ps}b)). The result are decreasing luminosities
and rms energies of the \( \mu  \)- and \( \tau  \)-neutrinos. The
more massive \( 40 \) M\( _{\odot } \) model shows qualitatively
different features. In order to understand them, we first discuss
Fig. (\ref{fig_massflux.ps}).
Shown are the profiles of the mass flux through surfaces at constant
radii in the two models. The dashed lines show the mass flux in the
\( 13 \) M\( _{\odot } \) model. Nothing special happens there,
the mass flux generally decreases in accordance with the decreasing
density in the outer layers. Moreover, the contraction of the stiff
core, which is far from its maximum mass, is minimal. A similar evolution
is visible during the first \( 300 \) ms in the \( 40 \) M\( _{\odot } \)
model. However, if we examine the massflux in the \( 300 \) ms time
slice in Fig. (\ref{fig_massflux.ps}) more closely, we find a variation
in the density profile around \( 1000 \) km that falls in (from \( \sim 2000 \)
km at \( 100 \) ms after bounce). The corresponding variation in
the massflux or accretion rate leads to a step in the electron flavor
neutrino luminosities between \( 300 \) ms and \( 350 \) ms after
bounce. The increase is of order \( 20\% \). The slope in their rms
energies flattens slightly. More independent of the details of the
progenitor model, however, might be that the protoneutron star in
the \( 40 \) M\( _{\odot } \) model approaches its maximum mass
much more rapidly because a high accretion rate is maintained when
the outer layers fall in. They have a significantly larger density
in comparison to the \( 13 \) M\( _{\odot } \) model. The fast mass
accumulation in the \( 40 \) M\( _{\odot } \) model becomes evident
in Table (\ref{table_maximum_mass}), which lists the enclosed mass
at \( 100 \) km radius for different time slices in both models.
\begin{table}
\caption{Enclosed mass at \protect\( 100\protect \) km radius.
\label{table_maximum_mass}}
\begin{center}
\begin{tabular}{|c|c|c|}
\hline 
\( t_{pb} \)
\protect\footnote{Shown is the enclosed mass at a radius of \protect\( 100\protect \)
km as a function of time for the \protect\( 13\protect \) M\protect\( _{\odot }\protect \)
and \protect\( 40\protect \) M\protect\( _{\odot }\protect \) model.
In the \protect\( 13\protect \) M\protect\( _{\odot }\protect \)
model, the accretion rate reduces quickly and the mass of the protoneutron
star does not even come close to its maximum mass during the \protect\( \sim 1\protect \)
s time window so far explored with Boltzmann neutrino transport. The
outer layers in the \protect\( 40\protect \) M\protect\( _{\odot }\protect \)
model are much denser. The accretion rate stays high and the simulation
can be performed until the protoneutron star collapses.}&
 \( 13 \) M\( _{\odot } \)&
 \( 40 \) M\( _{\odot } \)\\
\hline
{[}s{]}&
 {[}M\( _{\odot } \){]}&
 {[}M\( _{\odot } \){]}\\
\hline
0.0&
 0.90&
 0.88\\
\hline
0.1&
 1.19&
 1.57\\
\hline
0.2&
 1.25&
 1.77\\
\hline
0.3&
 1.28&
 1.89\\
\hline
0.4&
 1.31&
 2.02\\
\hline
0.5&
 1.33&
 2.20\\
\hline
1.0&
 1.42&
 -- \\
\hline
\end{tabular}
\end{center}
\end{table}
As the accumulated mass in the \( 40 \) M\( _{\odot } \) protoneutron
star gets closer to the maximum mass, the protoneutron star starts
to contract faster by the general relativistic enhancement of the
effective gravitational potential (an effect absent in Newtonian calculations).
We observe in Fig. (\ref{fig_massflux.ps}) that, after an initial
decrease, the mass flux in the inner core is increasing again. This
is by no means a {}``sudden'' change on the short dynamical time
scale of the protoneutron star. The contraction is a hydrostatic adaption
to the accumulated mass in the gravitational potential. The change
is, however, sudden on the time scale of the variations in the neutrino
properties shown in Fig. (\ref{fig_lumin2.ps}). The \( \mu  \)-
and \( \tau  \)-neutrino luminosities and rms energies rise steeply.
This happens because, by the contraction of the protoneutron star,
electron-nondegenerate shock-heated material is condensed to densities
where the main emission of heavy neutrinos occurs. We investigate
the conditions at the locations of the main neutrino emission in more
detail in Fig. (\ref{fig_rmsplot.ps}).
\begin{figure}
{\centering \resizebox*{0.55\textwidth}{!}{\includegraphics{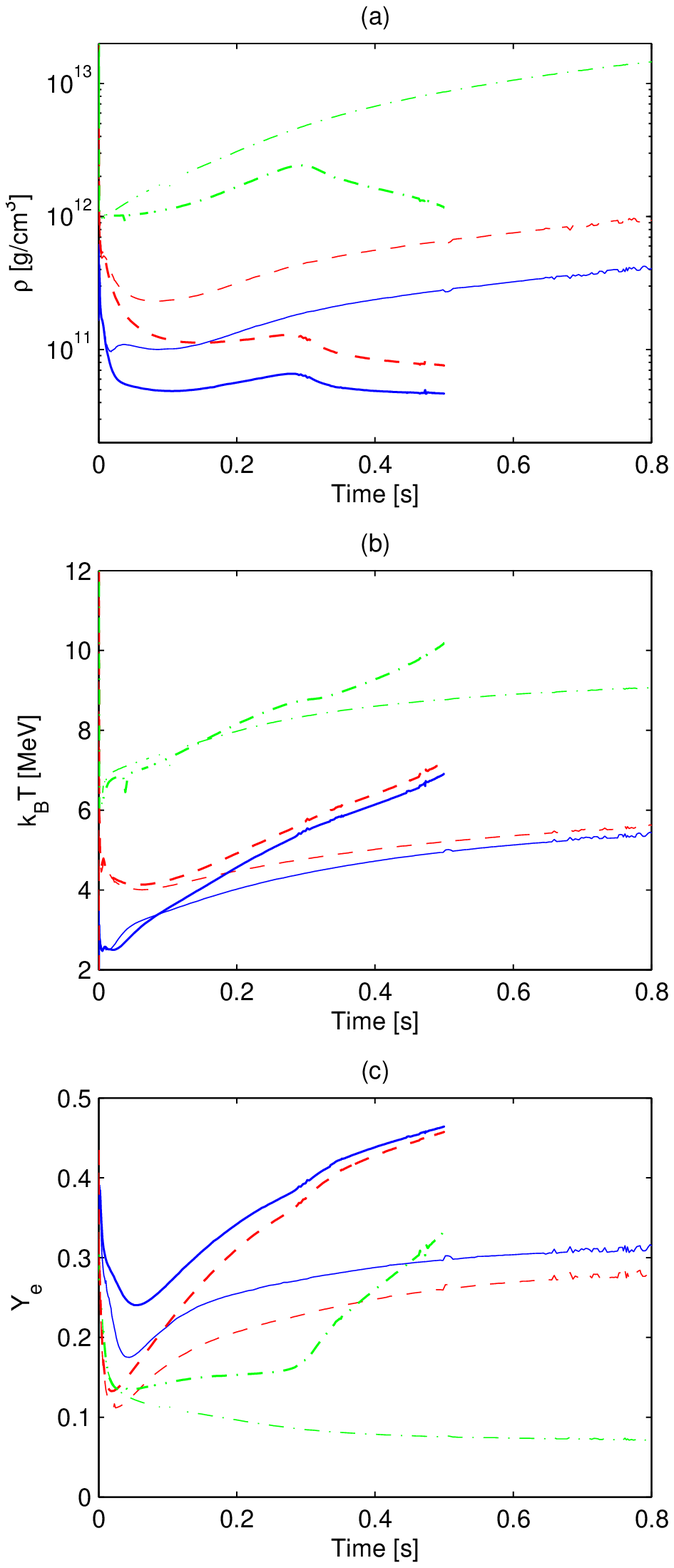}} \par}
\end{figure}
\begin{figure}

\caption{The thermodynamic conditions where escaping neutrinos have their
last inelastic interaction are averaged with the weight of the contribution
of the neutrinos to the total luminosity. Shown is the average neutrino
emission density in graph (a), the average neutrino emission temperature
in graph (b), and the average neutrino emission electron fraction
in graph (c). Inelastic scattering of neutrinos off electrons is also
included as an {}``emission'' reaction. The typical conditions for
electron neutrino emission are represented with a solid line, the
typical conditions for electron antineutrino emission with a dashed
line, and the typical conditions for \protect\( \mu \protect \)-
and \protect\( \tau \protect \)-neutrino emission with a dash-dotted
line. The thin lines belong to the \protect\( 13\protect \) M\protect\( _{\odot }\protect \)
model and the thick lines to the \protect\( 40\protect \) M\protect\( _{\odot }\protect \)
model. The conditions are similar at bounce. They diverge afterwards.
In the \protect\( 13\protect \) M\protect\( _{\odot }\protect \)
model, the main neutrino emission recedes to higher densities during
the evolution of the simulation. In the \protect\( 40\protect \)
M\protect\( _{\odot }\protect \) model, the main neutrino emission
turns back to lower densities after gravitational contraction sets
in at \protect\( 300\protect \) ms after bounce. The emission from
the gravitationally compressed shock-heated matter at moderate densities
competes more successfully with the emission from the cooler core
at higher densities.\label{fig_rmsplot.ps}}
\end{figure}
We calculate the average conditions for the emission of a specific
neutrino type according to Eq. (\ref{eq_rms_radius}). Instead of
the average radius, we calculate here the average density (graph (a)),
temperature (graph (b)), and electron fraction (graph (c)). These
represent the typical conditions where an escaping neutrino makes
its last energy-changing interaction with the matter. The conditions
at the origin of electron neutrino emission are traced with a solid
line, the conditions at the origin of electron antineutrino emission
with a dashed line, and the conditions at the origin of \( \mu  \)-
and \( \tau  \)-neutrino emission with a dash-dotted line. We discuss
first the \( 13 \) M\( _{\odot } \) model (thin lines). After an
initial decrease up to the time of maximum neutrino heating around
\( 100 \) ms after bounce, the average density at the sites of neutrino
production increases steadily. This goes along with a temperature
increase for all neutrino types. At least in the case of the very
temperature-sensitive pair production rates for the \( \mu  \)- and
\( \tau  \)-neutrinos, however, the argumentation should be turned
around: Because the temperature at a given density is slowly decreasing
on the long time scale, the emission from lower density regions decreases
and the average emission conditions shift to increasingly deeper layers,
where the temperatures are higher. Graph (c) shows that the electron
fraction at the place of emission is increasing for the electron flavor
neutrinos. This is due to the rather high electron fraction in the
shock-heated material behind the shock front, where the electron degeneracy
is gradually lifted. The \( \mu  \)- and \( \tau  \)-neutrinos escape
from the floor in the electron fraction profile, which is steadily
decreasing by continued deleptonization. This is reflected in the
declining electron fraction at the location of the production of heavy
neutrinos in graph (c). The contraction of the protoneutron star in
the \( 40 \) M\( _{\odot } \) model causes a qualitatively different
characterization of the regions of main neutrino emission. The temperature
increase by adiabatic compression is no longer fully balanced by neutrino
cooling. The temperature at a given density starts to increase. The
domain of the shock-heated material where the electrons are non-degenerate
reaches down to deeper layers close to the place where the \( \mu  \)-
and \( \tau  \)-neutrinos are produced. The temperature increase
makes these regions to significantly more efficient \( \mu  \)- and
\( \tau  \)-neutrino emitters such that the average emission density
starts to decrease. In spite of this steep density decrease in graph
(a), the temperature in graph (b) still increases at the average emission
condition. The electron fraction in graph (c) rises dramatically as
the mean neutrino emission region moves out of the \( Y_{e} \) trough
in the electron fraction profile. As the production site of heavy
neutrinos moves outwards to lower densities, the fraction of directly
escaping neutrinos (without neutrino-electron scattering) also increases.
Additionally, the forming {}``density cliff'' shortens the escape
path for few high energy \( \mu  \)- and \( \tau  \)-neutrinos that
may leave the star without thermalization. It can be expected that
the not included reactions of \( \mu  \)- and \( \tau  \)-flavor
neutrino production by bremsstrahlung and electron flavor neutrino
annihilation will more significantly influence the \( \mu  \)- and
\( \tau  \)-neutrino spectra in this less opaque density regimes.
Unfortunately, we cannot follow the evolution beyond the collapse
of the protoneutron star because of the coordinate singularity at
the formation of the Schwarzschild horizon. However, the neutrino
luminosities are expected to decay on a short timescale (see e.g.
\citep{Baumgarte_et_al_96}).

\section{Numerical Implementation}\label{section_computer_model}

In this section we motivate and describe our finite difference representation
of spherically symmetric radiation hydrodynamics in numerical computations.
The continuous equations of radiation hydrodynamics are unique. But
the number of possible finite difference representations is large.
We select ours according to the following requirements (with descending
priority):

\begin{enumerate}
\item In the limit of high resolution, the finite difference representation
should converge to the continuous equations.
\item The finite difference representation should lead to a numerically
stable solution.
\item The finite difference representation should not be less accurate than
simpler schemes in regions where simpler schemes are sufficient. 
\item The computed evolution of the particle distribution function should
imply an accurate evolution of expectation values like the particle
density or the average particle energy. 
\end{enumerate}
In our discrete ordinates (S\( _{N} \)) method, we discretize the
full transport equation at once. This takes care of requirement (1).
In the first place, we choose the finite differencing such that the
numerical scheme is stable according to requirement (2). Moreover,
we adjust it such that requirement (3) is met with respect to the
important diffusion limit. In the finite differencing of the free
streaming limit, we focus on accurate particle number luminosities
and accurate particle energy luminosities. The accuracy of the angular
moments of the radiation field is less essential in this domain because
the neutrinos cannot contribute to the supernova mechanism in regions
far outside the neutrinospheres. More accurate angular information
in the free streaming limit with the S\( _{N} \) method would require
a very large number of angular bins, an adaptive grid in angle space
\citep{Yamada_Janka_Suzuki_99}, or a postprocessing step with a ray-tracing
tool. All remaining degrees of freedom in the choice of the finite
difference representation are used to optimize requirement (4). Of
course, point (1) already guarantees that requirement (4) will successfully
be approached if the resolution is sufficiently high. However, as
we solve the transport equation at once, and with an implicit finite
difference representation, resolution is computationally rather expensive.
In the spirit of requirement (3), high resolution may only be required
where the very details of the complete transport equation are relevant
for the physical solution. In order to make low resolution results
reliable, important physical laws must be represented accurately independent
of the resolution setting. This includes the challenge of conserving
lepton number and total energy in the simulations. Indeed, one might
say that the power of a specific implementation of the S\( _{N} \)
method is given by the extent to which requirements (3) and (4) can
be satisfied at low resolution.

A variable Eddington factor method (VEF) is a different numerical
approach than the S\( _{N} \) method. There, requirements (2), (3),
and (4) are satisfied by construction. A series of simpler radiation
moment equations are solved and combined with an Eddington factor
to produce the solution to the equations of radiation hydrodynamics.
The quality of the variable Eddington factor method is determined
by the capability to meet requirement (1). It depends on the interplay
between the different moment equations and the accuracy of the Eddington
factor (which may require the numerical solution of a model Boltzmann
equation \citep{Burrows_et_al_00,Rampp_Janka_02,Thompson_Burrows_Pinto_03}).
With careful implementation, both methods should fulfill requirements
(1)-(4) and therefore lead to similar results \citep{Liebendoerfer_et_al_04}.

\subsection{Conservation Laws and Expectation Value Matching}

In the S\emph{\( _{N} \)} approach, the complete dynamics in the
phase space of the transported particles is determined by \emph{one}
kinetic equation (\ref{eq_relativistic_boltzmann}). Its integration
over the momentum phase space is equivalent to a continuity equation
(\ref{eq_neutrino_number_conservation}), its integration weighed
with particle energies, \( \varepsilon  \), is equivalent to an energy
equation (\ref{eq_radiation_energy_conservation}), and its integration
weighed with the particle direction cosine and energy would lead to
a momentum equation. These derived equations certainly fulfill the
macroscopic conservation laws as given in Eq. (\ref{eq_total_energy}).
But it is a challenge to obtain the same level of consistency in a
finite difference representation of the Boltzmann equation. We illustrate
this with the following little example:

For simplicity we assume that a distribution function, \( f(t,x) \),
of particles propagating with speed \( c \) only depends on time,
\( t \), and on one dimension in the phase space, \( x \). The simplest
form of a transport equation would then relate the time derivative
of the distribution function to an advection term,\begin{equation}
\label{eq_toy_cont_transport}
\frac{\partial f}{\partial t}+c\frac{\partial f}{\partial x}=0.
\end{equation}
If we now ask for the time evolution of the expectation value of the
distribution function with respect to an operator, \( g(t,x) \),
we apply integrations by parts to find\begin{equation}
\label{eq_toy_cont_expectation}
\frac{\partial }{c\partial t}\int gfdx=\int \left( \frac{\partial g}{c\partial t}+\frac{\partial g}{\partial x}\right) fdx-\left[ gf\right] _{\partial x}.
\end{equation}
Here, the time evolution of the expectation value is described by
other expectation values and the value of \( gf \) at the boundary
of the domain of integration. This equation has an immediate physical
interpretation: The change of the expectation value along the characteristics
of the particle flow is simply given by the change of the weight along
the characteristics of the particle flow because the phase space volume
stays constant, \begin{equation}
\label{eq_expectation_along_characteristic}
\left( \frac{\partial }{c\partial t}+\frac{\partial }{\partial x}\right) \int gfdx=\int \left( \frac{\partial g}{c\partial t}+\frac{\partial g}{\partial x}\right) fdx.
\end{equation}
The second term on the left hand side is equivalent to the boundary
term \( -\left[ gf\right] _{\partial x} \) in Eq. (\ref{eq_toy_cont_expectation}).
Let us now assume that numerical stability requires upwind differencing
of the advection term in Eq. (\ref{eq_toy_cont_transport}), and,
for simplicity, that the wind is unidirectional. A finite difference
representation would then read like\begin{equation}
\label{eq_toy_fd_transport}
\frac{\partial f_{i'}}{\partial t}+c\frac{f_{i'}-f_{i'-1}}{dx_{i'}}=0.
\end{equation}
We use the convention that a prime at the index points to the zone
center value \( i+1/2 \) while an integer index \( i \) points to
the zone edge. We simplify the example once more and assume that the
operator \( g \) is not time-dependent. With this, we track above
integration by parts in the finite difference representation and calculate
the evolution of the expectation value of operator \( g(x) \),\begin{eqnarray}
\frac{\partial }{c\partial t}\sum ^{n}_{i=1}g_{i'}f_{i'}dx_{i'} & = & \sum ^{n}_{i=1}g_{i'}\frac{\partial f_{i'}}{c\partial t}dx_{i'}=-\sum ^{n}_{i=1}g_{i'}\left( f_{i'}-f_{i'-1}\right) \nonumber \\
 & = & -\sum ^{n}_{i=1}g_{i'}f_{i'}+\sum ^{n-1}_{i=0}g_{i'+1}f_{i'}\nonumber \\
 & = & \sum ^{n}_{i=1}\frac{g_{i'+1}-g_{i'}}{dx_{i'}}f_{i'}dx_{i'}-\left[ g_{n'+1}f_{n'}-g_{0'+1'}f_{0'}\right] .\label{eq_toy_fd_expectation} 
\end{eqnarray}
Firstly, we remark that there is a discrete analogue to the integration
by parts and that Eq. (\ref{eq_toy_fd_expectation}) has exactly the
same structure as Eq. (\ref{eq_toy_cont_expectation}). Secondly,
we note that the choice of finite differencing for the advection term
in Eq. (\ref{eq_toy_fd_transport}) determines the finite differencing
of the weight function \( \left( g_{i'+1}-g_{i'}\right) /dx_{i'} \)
in the expectation value on the right hand side of Eq. (\ref{eq_toy_fd_expectation}).
In the following, we will frequently use this type of investigation
to optimize for requirements (3) and (4): First we choose a stable
finite difference representation of a term in the transport equation.
Then we evaluate relevant expectation values by performing discrete
{}``integrations by parts'', analogously to this simple example.
The emerging finite difference representations of the expectation
values along one phase space dimension are then used as prefactors
for the finite differencing of terms related to other phase space
dimensions in order to build a consistent finite difference representation
of the full transport equation.

In order to construct a conservative finite difference representation
of Eq. (\ref{eq_relativistic_boltzmann}), we analyze the structure
of the conservation equations in the continuum world, where insight
is not buried in discretization indices. The emergence of lepton number
conservation in Eq. (\ref{eq_neutrino_number_conservation}) is straightforward,
because the observer corrections are already written in a form that
allows an immediate integration over energy or the angle cosine. Number
conservation will also emerge naturally in the finite difference representation
when we finite difference the same basic structure of the equation.
The energy conservation equation in the frame of a distant observer,
however, introduces the weight \( g(t,a,\mu ,E)=E\left( \Gamma +u\mu \right)  \)
in Eq. (\ref{eq_evolution_total_energy}). It depends on all four
phase space dimensions and leads to the contribution of several expectation
values of the distribution function in the evaluation of the energy
conservation equation, \begin{equation}
\label{eq_conservation_outline}
\int \left( \Gamma +u\mu \right) E\left[ C_{t}+D_{a}+D_{\mu }+D_{E}+O_{E}+O_{\mu }-C_{c}\right] d\mu E^{2}dE=0.
\end{equation}
 All terms in the Boltzmann equation (\ref{eq_relativistic_boltzmann}),
\( C_{t} \), \( D_{a} \), \( D_{\mu } \), \( D_{E} \), \( O_{E} \),
and \( O_{\mu } \), are written as the derivative of an expression
with respect to time, rest mass, angle cosine, or particle energy.
We now perform an integration by parts with respect to these integration
variables \( x \) along the lines of Eq. (\ref{eq_toy_cont_expectation}).
A multitude of correction terms proportional to \( \partial \left( \Gamma +u\mu \right) E/\partial x \)
arise. As in Eq. (\ref{eq_expectation_along_characteristic}), they
describe the evolution of the weight \( g(t,x)=\left( \Gamma +u\mu \right) E \)
along the characteristic of the particle flow. However, we showed
in \citep{Liebendoerfer_Mezzacappa_Thielemann_01} that \( \left( \Gamma +u\mu \right) E \)
is nearly a constant of motion along the characteristics. Therefore,
it is natural that most of the partial terms \( \partial \left( \Gamma +u\mu \right) E/\partial x \)
actually cancel in the total evolution of the radiation energy in
the frame of a distant observer. If, like in Eq. (\ref{eq_toy_fd_expectation}),
the integrations by parts are also carefully followed in the finite
difference representation, mutual cancellations of important expectation
values can be forced to be exact---independently of the resolution.
Because these canceling terms individually reach large values around
and after bounce, this expectation value matching is an essential
step for the conservation of energy in the finite difference representation.
We perform the integrations by parts in Eq. (\ref{eq_conservation_outline})
and start with the identification of canceling contributions in the
following overview of contributions from \( C_{t} \), \( D_{a} \),
\( D_{\mu } \), \( D_{E} \), \( O_{E} \), \( O_{\mu } \), \( C_{c} \)
in the transport equation (\ref{eq_relativistic_boltzmann}):

\begin{eqnarray}
C_{t}: &  & \frac{\partial }{\alpha \partial t}\left( \Gamma J\right) -\frac{\partial \Gamma }{\alpha \partial t}J+\frac{\partial }{\alpha \partial t}\left( uH\right) -\frac{\partial u}{\alpha \partial t}H\nonumber \\
D_{a}: &  & \frac{\partial }{\alpha \partial a}\left[ 4\pi r^{2}\alpha \rho \Gamma H\right] -4\pi r^{2}\rho \frac{\partial \Gamma }{\partial a}H\nonumber \\
 &  & +\frac{\partial }{\alpha \partial a}\left[ 4\pi r^{2}\alpha \rho uK\right] -4\pi r^{2}\rho \frac{\partial u}{\partial a}K\nonumber \\
D_{\mu }: &  & -\Gamma \frac{u}{r}\left( J-K\right) +\Gamma u\frac{\partial \Phi }{\partial r}\left( J-K\right) \nonumber \\
D_{E}: &  & \Gamma \frac{\partial \Phi }{\partial r}\left( \Gamma H+uK\right) \nonumber \\
O_{E}: &  & -\left[ \frac{\partial \ln \rho }{\alpha \partial t}+\frac{2u}{r}\right] \left( \Gamma K+uQ\right) +\Gamma \frac{u}{r}\left( J-K\right) +\frac{u^{2}}{r}\left( H-Q\right) \nonumber \\
O_{\mu }: &  & -\left[ \frac{\partial \ln \rho }{\alpha \partial t}+\frac{2u}{r}\right] u\left( H-Q\right) -\frac{u^{2}}{r}\left( H-Q\right) \nonumber \\
C_{c}: &  & \Gamma \int \left( \frac{j}{\rho }-\chi F\right) E^{3}dEd\mu -u\int \chi FE^{3}dE\mu d\mu .\label{eq_cancellation_table} 
\end{eqnarray}
 We will neglect terms that are nonlinear in the radiation field,
i.e. terms that describe a gravitational interaction of the radiation
field with itself. From the hydrodynamics equations in \citep{Misner_Sharp_64,May_White_66}
we then derive the following useful relationships for the coefficients
in Eq. (\ref{eq_cancellation_table}).\begin{eqnarray}
\frac{1}{\Gamma }\frac{\partial \Gamma }{\alpha \partial t} & = & u\frac{\partial \Phi }{\partial r}\label{eq_misner_dGdt} \\
-\left( \frac{\partial \ln \rho }{\alpha \partial t}+\frac{2u}{r}\right)  & = & \frac{4\pi r^{2}\rho }{\Gamma }\frac{\partial u}{\partial a}\label{eq_misner_drhodt} \\
\frac{\partial u}{\alpha \partial t} & = & \Gamma ^{2}\frac{\partial \Phi }{\partial r}-\frac{m}{r^{2}}-4\pi rp.\label{eq_misner_dudt} 
\end{eqnarray}
We may now identify cancellations in Eq. (\ref{eq_cancellation_table})
and label them for later reference: The fourth term of \( D_{a} \)
cancels with the first term of \( O_{E} \) (\( D^{4}_{a}O^{1}_{E} \))
by Eq. (\ref{eq_misner_drhodt}). The first and second term of \( D_{\mu } \)
cancel with the third and fourth term in \( O_{E} \) (\( D^{12}_{\mu }O^{34}_{E} \)).
These are the only {\cal O}\( (v/c) \) cancellations and therefore
the most important ones. In an analogous notation, we find the following
higher order cancellations: (\( C_{t}^{2}D^{3}_{\mu } \)) by Eq.
(\ref{eq_misner_dGdt}), (\( D^{4}_{\mu }D^{2}_{E} \)), (\( O_{E}^{2}O^{2}_{\mu } \)),
(\( O_{E}^{56}O^{34}_{\mu } \)). The remaining terms, \( (C_{t}^{4}D_{a}^{2}D_{E}^{1}O_{\mu }^{1}) \),
reduce by Eq. (\ref{eq_misner_dudt}) and the definition of \( \Gamma  \)
to the general relativistic term, \begin{equation}
\label{eq_hydro_cancellation}
\left( -\frac{\partial u}{\alpha \partial t}-\Gamma \frac{\partial \Gamma }{\partial r}+\Gamma ^{2}\frac{\partial \Phi }{\partial r}+u\frac{\partial u}{\partial r}\right) H=4\pi r\rho \left( 1+e+p/\rho \right) H.
\end{equation}
 They enter in this form the radiation energy conservation equation
(\ref{eq_radiation_energy_conservation}).

We will adopt the following strategy in the finite differencing of
Eq. (\ref{eq_relativistic_boltzmann}) (the reader is invited to draw
lines in Eq. (\ref{eq_cancellation_table}) to visualize the relationship
between the different terms): (i) The finite differencing of \( C_{t} \)
and \( C_{c} \) are straightforward. (ii) Appreciable experience
has been gathered in previous work (see e.g. \citep{Lewis_Miller_84})
with the finite differencing of the {\cal O}\( (v/c) \) terms in
\( D_{a} \) and \( D_{\mu } \). They are assumed to be well-chosen
in \citep{Mezzacappa_Bruenn_93a} and therefore not subject to changes.
(iii) Based on this, the cancellation (\( D^{4}_{a}O^{1}_{E} \))
dictates the finite difference representation of \( \left( \partial \ln \rho /\alpha \partial t+2u/r\right)  \)
in \( O^{1}_{E} \) and \( O^{2}_{E} \). The cancellation (\( O_{E}^{2}O^{2}_{\mu } \))
sets the finite difference representation of \( \left( \partial \ln \rho /\alpha \partial t+2u/r\right)  \)
in \( O^{2}_{\mu } \) and \( O_{\mu }^{1} \). (iv) The cancellation
(\( D^{12}_{\mu }O^{34}_{E} \)) dictates the finite difference representation
of \( u/r \) in \( O_{E}^{34} \) and \( O_{E}^{56} \). The cancellation
(\( O_{E}^{56}O^{34}_{\mu } \)) propagates the finite difference
representation to \( u/r \) in \( O_{\mu }^{34} \). (v) The evaluation
of \( (C_{t}^{4}D_{a}^{2}D_{E}^{1}O_{\mu }^{1}) \) according to Eq.
(\ref{eq_hydro_cancellation}) constrains the finite difference representation
of \( \Gamma \partial \Phi /\partial r \) in \( D_{E}^{1} \) and
therewith defines \( D_{E}^{2} \). (vi) And finally we choose a different
finite difference representation for \( \Gamma u\partial \Phi /\partial r \)
in \( D^{3}_{\mu } \) and \( D^{4}_{\mu } \), as suggested by the
cancellation (\( C_{t}^{2}D^{34}_{\mu }D_{E}^{2} \)). By these six
chains, all terms in Eq. (\ref{eq_relativistic_boltzmann}) become
constrained.

\subsection{Adaptive Grid}\label{subsection_adaptive_grid}

{\sc agile-boltztran} has the capability to dynamically adapt the
computational grid to the evolution of the stellar profile. Each radial
zone contributes to the computationally expensive implicit solution
of the Boltzmann equation. The adaptive grid allows the concentration
of these zones to domains where radiation transport is actually active
and important. It also allows a proper shock capturing and propagation.
The fundamental ideas of the dynamically adaptive grid have been developed
in \citep{Winkler_Norman_Mihalas_84,Dorfi_Drury_87}. The first reference
describes how the continuous equations of radiation hydrodynamics
are formulated on an adaptive grid and the second reference provides
a numerically stable prescription how to adapt the local grid point
concentration to steep gradients in the solution vector. The details
of our implementation of the adaptive grid are described in \citep{Liebendoerfer_Rosswog_Thielemann_02}.
Here we resume the key points and describe the extension of the approach
from the hydrodynamics code {\sc agile} to the radiation quantities.

We concentrate on observers at rest in their slice. They reside in
local orthogonal reference frames which we assume to be collinear
with global coordinates \( A \). Consequently, coordinates \( A \)
have vanishing shift vectors everywhere. We introduce another system
of global coordinates, \( B \), having arbitrary shift vectors on
the same time slices. Let us select fixed grid labels in coordinates
\( B \): \( \{q_{i}\in \textrm{R},i=1\ldots n\} \). The fixed labels
in coordinates \( B \) define a moving grid in coordinate system
\( A \): \( \{a(t,q_{i})\} \). The grid labels in coordinate system
\( B \) cut the continuum into zones \( \Delta q \) on each time
slice. These zones contain a time dependent selection of observers
in coordinate system \( A \): \( \Delta a=\{a(t,q),q\in \Delta q\} \).
We define a zone integral of the observations \begin{equation}
\langle y(t,a)\rangle =\int _{\Delta a}y(t,a)da
\end{equation}
that is based on the observations \( y \) of single observers \( a \)
in the zone. In coordinate system \( A \), the borders of the zones
change with time according to the shift vectors in coordinate system
\( B \). In Newtonian parlance, one can identify the moving zones
in coordinate system \( A \) with the motion of an adaptive grid.
The important point is that the adaptive grid only regroups the observers
into new zones. As stated already by \citep{Winkler_Norman_Mihalas_84},
the adaptive grid never changes the reference frame of the observations.
It solely determines the width and location of a zone on the time
slice for the zone-integration of the observations \( y \) in the
comoving frame \( A \).

In order to formulate time evolution equations, we are interested
in the time change of the zone-integrated quantity \( \langle y(t,a)\rangle  \)
in a specific zone \( \Delta q \). Taking into account that the boundaries
of the integration over \( \Delta a \) are time-dependent, this leads
to \begin{equation}
\label{eq_reynolds_theorem}
\frac{\partial }{\partial t}\int _{a(t,q_{i})}^{a(t,q_{i+1})}yda=\int _{a(t,q_{i})}^{a(t,q_{i+1})}\frac{dy(t,a)}{dt}da+\left[ y\frac{\partial a(t,q)}{\partial t}\right] _{a(t,q_{i})}^{a(t,q_{i+1})}
\end{equation}
The notation \( \partial /\partial t \) is used for time derivatives
at constant grid label \( q \) while Lagrangian derivatives, \( d/dt \),
are evaluated at constant baryon mass \( a \). The left hand term
is the time change of the zone-integrated variable \( y \) between
fixed grid labels in the numerically accessible coordinate system
\( B \). The first term on the right hand side is the time derivative
in orthogonal coordinates \( A \) as used in the physical equations,
e.g. on the left hand side of Eqs. (\ref{eq_continuity})-(\ref{eq_momentum}).
The second term on the right hand side of Eq. (\ref{eq_reynolds_theorem})
is due to the motion of the grid. It is evaluated at the zone boundaries
and accounts for the observers that fall into or drop out of the zone
during an infinitesimal time step. The rate of coordinate change at
a fixed grid label, \( u^{\rm rel}=-\partial a(t,q)/\partial t \),
defines a {}``grid velocity'' with respect to coordinates \( A \).
Note that the adaptive grid corrections in Eq. (\ref{eq_reynolds_theorem})
enter in a conservative form of fluxes from zone to zone. If the time
change of the variable \( y \) is integrated over the whole domain,
all grid corrections at intermediate zone boundaries cancel. The adaptive
grid does not affect conservation properties of the variable \( y \).
This important feature is maintained in the finite difference representation
of Eq. (\ref{eq_reynolds_theorem}),\begin{equation}
\label{eq_reynolds_theorem_fd}
\frac{y_{i'}\Delta a_{i'}-\bar{y}_{i'}\overline{\Delta a}_{i'}}{\Delta t}=\left( \frac{dy}{dt}\right) _{i'}\Delta a_{i'}+y_{i+1}\frac{a_{i+1}-\bar{a}_{i+1}}{\Delta t}-y_{i}\frac{a_{i}-\bar{a}_{i}}{\Delta t}.
\end{equation}
The overbars in this equation mark the quantities that are evaluated
at the old time, before the time step \( \Delta t \).

The application of Eq. (\ref{eq_reynolds_theorem_fd}) for the evaluation
of Lagrangian time derivatives on the adaptive grid is usually straightforward.
However, a side-remark concerning our ad hoc {}``burning'' of silicon
to nuclear statistical equilibrium (NSE) at the edge of the iron core
should be made here. In a Lagrangian scheme, we would simply convert
zone \( i+1 \) to NSE during infall as soon as the burning time scale
becomes considerably smaller than the evolution time scale. The conversion
from a given composition to NSE is handled such that energy remains
conserved with the respective equations of state. If the adaptive
grid is applied to the energy equation, it automatically performs
this conversion also to the continuous mass advection between zone
\( i \) and \( i+1 \). This is perfect if the grid moves outwards
such that advected silicon is continuously converted to NSE. It is
less perfect if the grid moves inwards, because this leads to the
advection of material labelled with {}``NSE'' from zone \( i \)
to {}``silicon'' in zone \( i+1 \) until the whole zone \( i+1 \)
carries the label {}``NSE''. At this point the whole zone is converted
back to NSE. Without additional measures, the advected NSE material
would therefore transiently be treated as {}``silicon''. By energy
conservation, the temperature of zone \( i+1 \) would drop because
energy would be invested to unphysically {}``unburn'' the advected
NSE material. We eliminate this effect by externally feeding the zone
with the difference in the nuclear binding energy between advected
and internal material such that no sudden changes in the thermodynamical
state will take place. As soon as the zone is converted back to NSE,
we extract the same amount of energy from the burning energy to restore
the balance.

If the adaptive grid technique is applied to radiation quantities
that depend on momentum phase space variables like the particle energy,
\( E \), or the angle cosine, \( \mu  \), further considerations
are required. A particle, that is observed in the comoving frame of
zone one with an energy \( E \) and a propagation angle \( \mu  \),
is observed in the comoving frame of an adjacent zone two with energy
\( E'\neq E \) and \( \mu '\neq \mu  \) because of frequency shift
and angular aberration. If the moving zone boundary sweeps over the
particle such that it is assigned to the adjacent zone two, the momentum
phase space state of the particle in zone two should be changed to
\( E' \) and \( \mu ' \) in order to describe the same physical
state the particle had when it was a member of zone one. The motion
of the zone boundary has a second effect: The location of the center
of zone two moves towards zone one and fluid momentum is advected
from zone one to zone two. This means that the particles that do not
change the zone are as well subject to frequency shift and angular
aberration corrections in order to maintain their exact original state
after the redefinition of the zones. In the more general derivation
of the adaptive grid equations in Winkler and Norman \citep{Winkler_Norman_Mihalas_84},
these two corrections do not appear. In order to convince ourselves
that they cancel, we calculate them explicitly.

We assume that we have a global description of the particle state
in the momentum phase space that does not depend on the location of
the particle or the velocity of the fluid element. We call these global
parameters {}``impact parameter'', \( b \), and {}``energy at
infinity'', \( \varepsilon  \), because the quantities\begin{eqnarray}
b & = & r\frac{\sqrt{1-\mu ^{2}}}{\Gamma +u\mu }\nonumber \\
\varepsilon  & = & \left( \Gamma +u\mu \right) E.\label{eq_inertial_frame_transformation} 
\end{eqnarray}
are in good approximation constants of motion along the general relativistic
particle trajectory \citep{Liebendoerfer_Mezzacappa_Thielemann_01}.
Once the description of the particle momentum phase space is invariant,
it is safe to reassign neutrinos between zones according to the adaptive
grid equation (\ref{eq_reynolds_theorem}). We substitute the general
variable \( y(t,a) \) by the distribution function \( \hat{F}(t,a,b,\varepsilon ) \),
\begin{equation}
\label{eq_fbar_reynolds}
\frac{\partial }{\partial t}\int ^{a(q_{2})}_{a(q_{1})}\hat{F}da+\left[ -\hat{F}\frac{\partial a}{\partial t}\right] ^{q_{2}}_{q_{1}}=\int ^{a(q_{2})}_{a(q_{1})}\frac{d\hat{F}}{dt}da.
\end{equation}
 Evaluating the first term by the chain rule gives \begin{eqnarray}
\frac{\partial }{\partial t}\int ^{a(q_{2})}_{a(q_{1})}\hat{F}da & = & \frac{\partial }{\partial t}\int ^{a(q_{2})}_{a(q_{1})}Fda\nonumber \\
+\int ^{a(q_{2})}_{a(q_{1})}\left[ \frac{\partial F}{\partial \mu }\frac{\partial \mu }{\partial t}+\frac{\partial F}{\partial E}\frac{\partial E}{\partial t}\right] da & + & \int ^{q_{2}}_{q_{1}}\left( \hat{F}-F\right) \frac{\partial }{\partial t}\left( \frac{\partial a}{\partial q}\right) dq.\label{eq_firstterm} 
\end{eqnarray}
 The second term can be written as \begin{eqnarray}
\left[ \hat{F}\frac{\partial a}{\partial t}\right] ^{q_{2}}_{q_{1}} & = & \left[ F\frac{\partial a}{\partial t}\right] ^{q_{2}}_{q_{1}}\nonumber \\
+\int ^{a(q_{2})}_{a(q_{1})}\left[ \frac{\partial F}{\partial \mu }\frac{\partial \mu }{\partial a}+\frac{\partial F}{\partial E}\frac{\partial E}{\partial a}\right] \frac{\partial a}{\partial t}da & + & \int ^{a(q_{2})}_{a(q_{1})}\left( \hat{F}-F\right) \frac{\partial }{\partial a}\left( \frac{\partial a}{\partial t}\right) da.\label{eq_secondterm} 
\end{eqnarray}
 Finally, the term on the right hand side evaluates to \begin{equation}
\label{eq_thirdterm}
\int ^{a(q_{2})}_{a(q_{1})}\frac{d\hat{F}}{dt}da=\int ^{a(q_{2})}_{a(q_{1})}\frac{dF}{dt}da+\int ^{a(q_{2})}_{a(q_{1})}\left[ \frac{\partial F}{\partial \mu }\frac{d\mu }{dt}+\frac{\partial F}{\partial E}\frac{dE}{dt}\right] da.
\end{equation}
 Substitution of expressions (\ref{eq_firstterm})-(\ref{eq_thirdterm})
into Eq. (\ref{eq_fbar_reynolds}) shows that the correction terms
indeed cancel. Hence, we can as well directly apply Eq. (\ref{eq_reynolds_theorem})
to the neutrino distribution functions, \( F\left( t,a,\mu ,E\right)  \),
defined in the comoving frame \begin{equation}
\label{eq_f_reynolds}
\frac{\partial }{\partial t}\int ^{a(q_{2})}_{a(q_{1})}Fda=\int ^{a(q_{2})}_{a(q_{1})}\frac{dF}{dt}da+\left[ F\frac{\partial a}{\partial t}\right] ^{q_{2}}_{q_{1}}.
\end{equation}
 We note however, that energy conservation in the adaptive grid corrections
will become resolution dependent if we finite difference Eq. (\ref{eq_f_reynolds})
instead of the much more complicated Eq. (\ref{eq_fbar_reynolds}).
Particle number conservation is still guaranteed at machine precision.
We quantify this effect in our energy conservation test in section
\ref{subsection_energy_conservation}. 

The implementation of an adaptive grid consists of two parts: We discussed
above how the physical equations are evolved on a dynamically moving
grid. Here we summarize how the motion of the grid itself is controlled.
On the one hand, it has to increase the local grid point concentration
where steep gradients in the solution vector need to be resolved.
On the other hand, it has to capture self-similar flows and to propagate
them through the computational domain. If there are traveling features
in the flow, it is advantageous if the adaptive grid can just change
the zone locations instead of significantly changing all the physical
variables describing the state in the zones. A useful grid equation
has been found in \citep{Dorfi_Drury_87}. We transferred the approach
from Eulerian coordinates in Newtonian space to comoving coordinates
in general relativistic time slices such that we retrieve a Lagrangian
scheme if the adaptive grid is switched off. The grid equation of
Dorfi and Drury is based on a resolution function, \begin{equation}
\label{eq_grid_resolution_function}
R(t,a)=\left( 1+\sum \left( \frac{a^{\rm scl}}{y^{\rm scl}}\frac{\partial y(t,a)}{\partial a}\right) ^{2}\right) ^{\frac{1}{2}},
\end{equation}
 where the sum includes a selection of relevant variables \( y \).
Currently, our grid position is guided by gradients in the velocity
and logarithmic density profiles. The velocity contribution focuses
on shocks and the density contribution draws grid points to steep
density gradients where the resolution is important for the radiation
transport. We set the global scale in this equation according to the
total variation, \( y^{\rm scl}=\sum _{i}\left| y_{i+1}-y_{i}\right|  \).
If we define a grid point concentration, \( n=a^{\rm scl}(\partial a(t,q)/\partial q)^{-1} \),
the grid equation requires that the the grid point concentration is
proportional to the requested resolution (\ref{eq_grid_resolution_function}),
\begin{equation}
\label{eq_adaptive_grid}
\frac{\partial }{\partial q}\left( \frac{n}{R}\right) =0.
\end{equation}
 After the substitution of \begin{equation}
\frac{n}{R}=\left( \left( \frac{\partial a(t,q)}{a^{{\textrm{scl}}}\partial q}\right) ^{2}+\sum \left( \frac{\partial y(t,q)}{y^{{\textrm{scl}}}\partial q}\right) ^{2}\right) ^{-\frac{1}{2}},
\end{equation}
 we note that this prescription enforces a constant generalized arc
length per grid point interval \( \Delta q \). In the same way as
\citet{Dorfi_Drury_87}, we apply operators for space smoothing and
time retardation to the grid point concentration before solving Eq.
(\ref{eq_adaptive_grid}) implicitly with the hydrodynamics equations.
As the adaptive grid requires corrections in the evaluation of time
derivatives, we have payed attention to reduce the occurrence of time
derivatives to a minimum in the implementation of the Boltzmann equation
on the adaptive grid. Therefore, we can relax the notation in the
following: We will use the notation \( \partial /\partial t \) for
the time derivatives at constant enclosed mass. This makes the discussion
of the radiation transport equation more consistent because the enclosed
mass is just one of four dimensions in the phase space, and the derivatives
will always be partial. The few time differences at constant zone
labels will explicitly be written out in the finite difference expressions.

\subsection{Finite differencing of the transport equation}

We split the description of the finite differencing of the Boltzmann
equation (\ref{eq_relativistic_boltzmann}) into separate terms according
to the list in Eq. (\ref{eq_boltzmann_ct})-(\ref{eq_boltzmann_cc}).
Each term is prepared in its own subsection. In section \ref{subsection_code_flow}
we describe how all the individual terms are gathered to form one
implicit solution of the complete transport equation and an operator
split implicit solution of the hydrodynamics equations.

The knowledge of a complete set of primitive variables at a certain
time allows the derivation of all other quantities of interest. Our
choice of primitive variables is listed in table (\ref{tab_primitive_variables}).
The discretization in space is indicated by the index \( i \). We
adopt the convention that integer values of \( i \) point to zone
edges while \( i+1/2 \) refers to the mass center of the zone between
edge \( i \) and \( i+1 \). We abbreviate half-valued indices with
a prime, \( i'\equiv i+1/2 \), in order to not further challenge
the readability of the finite difference equations. Of order \( 100 \)
spatial zones are used on a dynamically adaptive grid. The particle
distribution functions, \( F_{i',j',k'} \), additionally depend on
the particle momentum phase space, i.e. the angle cosine of the particle
propagation direction, \( \mu  \), and the particle energy, \( E \),
measured in the comoving frame. Both momentum phase space dimensions
are discretized with zone center values, \( \mu _{j'} \) and \( E_{k'} \),
in compliance with the above-mentioned conventions. We use Gaussian
quadrature in the angles \( \mu _{j'} \) with weights \( w_{j'} \).
They are derived from Legendre polynomials and normalized such that
\( \int _{-1}^{1}d\mu =\sum w_{j'}=2 \). The range from inwards to
outwards propagation is currently resolved with \( 6 \) different
propagation angles, but we are free to choose a larger number of directions
for higher resolution (see section \ref{subsection_resolution}).
The energy grid is set by geometrically increasing zone edge values
\( E_{k} \). Following \citet{Bruenn_02}, we define the zone center
energies by\[
E_{k'}=\sqrt{\frac{E_{k+1}^{2}+E_{k+1}E_{k}+E_{k}^{2}}{3}}.\]
 This choice implements the phase space volume \( E_{k'}^{2}dE_{k'}=\left( E_{k+1}^{3}-E_{k}^{3}\right) /3 \)
in an exact manner. The zone center energies also form a geometric
series. We currently use \( 12 \) zone center energies ranging from
\( 3 \) MeV to \( 300 \) MeV, but this number can also be increased
for higher resolution (see section \ref{subsection_resolution}).
Thermodynamical and compositional quantities are obtained from the
primitive variables \( \rho  \), \( T \), and \( Y_{e} \) by the
equation of state. Properties of the particle radiation field are
obtained by the evaluation of the expectation value of the corresponding
particle property with the primitive particle distribution function
\( F \).
\begin{table}
\caption{Primitive variables.}
\label{tab_primitive_variables}
\begin{center}
\begin{tabular}{|c|l|c|l|}
\hline 
&
 primitive variable
\protect\footnote{The primitive variables are listed with an index \protect\( i\protect \)
if they live on zone edges and an index \protect\( i'\protect \)
if they live on zone centers. The momentum phase space for the neutrino
distribution function is labelled with an index \protect\( j'\protect \)
for the angle cosine of the propagation direction and an index \protect\( k'\protect \)
for the neutrino energy.}&
primitive variable \\
\hline
\( a_{i} \)&
 enclosed rest mass&
 \( \rho _{i'} \)&
 rest mass density\\
\hline
\( r_{i} \)&
 radius&
 \( T_{i'} \)&
 temperature\\
\hline
\( u_{i} \)&
 velocity&
 \( Y_{e,i'} \)&
 electron fraction\\
\hline
\( m_{i} \)&
 enclosed gravitational mass&
 \( \alpha _{i'} \)&
 lapse function\\
\hline
&
&
 \( F_{i',j',k'} \)&
 neutrino distribution function \\
\hline
\end{tabular}
\end{center}
\end{table}

\subsubsection{Time derivative of the particle distribution function}

The time derivative of the particle distribution function, Eq. (\ref{eq_boltzmann_ct}),
is implemented in the following way:\begin{equation}
\label{eq_ct_fd}
C_{t}=\frac{1}{\alpha _{i'}da_{i'}}\left\{ \frac{F_{i',j',k'}da_{i'}-\bar{F}_{i',j',k'}\overline{da}_{i'}}{dt}+\left[ u_{i+1}^{{\rm rel}}F_{i+1,j',k'}^{*}-u_{i}^{{\rm rel}}F_{i,j',k'}^{*}\right] \right\} .
\end{equation}
 The quantities with an overbar refer to the previous value of the
variables. All other quantities refer to the updated values after
the time step \( dt=t-\bar{t} \). On a Lagrangian grid, the mass
in a spherical shell, \( da_{i'}=a_{i+1}-a_{i} \), is a constant
and the relative velocity (in mass) between the matter and the grid
\( u_{i}^{\rm rel}=-\left( a_{i}-\bar{a}_{i}\right) /dt \) vanishes.
In this case the expression would simply reduce to \( C_{t}=\alpha _{i}^{-1}\left( F_{i',j',k'}-\bar{F}_{i',j',k'}\right) /dt \).
On the adaptive grid, however, we apply Eq. (\ref{eq_reynolds_theorem_fd})
and obtain Eq. (\ref{eq_ct_fd}). The additional particle fluxes across
moving zone edges, \( u_{i}^{\{\rm rel\}}F_{i,j',k'}^{*} \), are
evaluated by upwind differencing in order to guarantee numerical stability:\begin{equation}
\label{eq_adaptive_F_advect}
F^{*}_{i,j',k'}=\left\{ \begin{array}{ll}
F_{i'-1,j',k'} & \mbox {if}\quad u_{i}^{\rm rel}\geq 0\\
F_{i',j',k'} & \mbox {otherwise}.
\end{array}\right. 
\end{equation}
Note that this particle advection is determined by the grid velocity
alone, and not by any physical particle speed. The grid velocity can
in principle become arbitrarily large. Unlike the advection due to
particle propagation, it does not depend on the particle mean free
path.

\subsubsection{Particle propagation in space}

The term \( D_{a} \) in Eq. (\ref{eq_boltzmann_da}) accounts for
the particles that are propagating into and out of a spherical mass
shell. From first considerations it is evident that this term has
to be proportional to the dot product between the particle propagation
direction and the shell boundary, i.e. the angle cosine \( \mu  \).
By this property, the contribution becomes proportional to the gradient
of the first angular moment of the particle distribution function
when the Boltzmann equation (\ref{eq_relativistic_boltzmann}) is
integrated over the propagation angles. In the free streaming limit,
the advected particle number in a time step (always in the perspective
of a comoving observer) is large with respect to the particle number
in the zone. Upwind differencing of the advection terms is appropriate
to limit destabilizing errors in the advection fluxes. For discrete
angle cosines, \( \mu _{j'} \), the direction of the particle {}``wind''
is simply given by the sign of \( \mu _{j'} \) (A particle with \( \mu =0 \)
propagates tangential to the mass shell). In diffusive conditions,
an asymmetric differencing can lead to an overestimate of the first
angular moment because of improper cancellations among the contributions
of the isotropic component of the particle field. Based on transport
coefficients \( \beta _{i,k'} \), we therefore interpolate between
upwind differencing in free streaming regimes (\( \beta _{i,k'}=1 \))
and centered differencing (\( \beta _{i,k'}=1/2 \)) in diffusive
regimes \citep{Mezzacappa_Bruenn_93a}. The chosen transport coefficients,
however, have the disadvantage, that they are not perfectly one half
in the diffusive regime inside the neutrino sphere. This can cause
a spurious numerical contribution to the diffusion flux in the hydrodynamic
limit where the particle density, multiplied by the propagation speed,
exceeds the particle flux by orders of magnitude. In order to satisfy
the equilibrium constraint (73) in \citep{Mezzacappa_Bruenn_93a} everywhere
inside the neutrinosphere, we use transport coefficients according
to\begin{equation}
\label{eq_transport_coefficients_fd}
\beta _{i,k'}=\left\{ \begin{array}{cc}
1/2 & {\rm if}\quad 2dr_{i}>\lambda _{i},\\
\left( 2dr_{i}/\lambda _{i,k'}+1\right) ^{-1} & {\rm otherwise}.
\end{array}\right. 
\end{equation}
That is, the transport coefficients at zone edges depend on the ratio
between the distance \( dr_{i} \) between zone centers and the neutrino
energy dependent mean free path \( \lambda _{i,k'} \). We finite
difference the propagation term as in \citep{Mezzacappa_Bruenn_93a}:
\begin{equation}
\label{eq_fd_da}
D_{a}=\frac{\mu _{j'}}{\alpha _{i'}da_{i'}}\left[ 4\pi r^{2}_{i+1}\alpha _{i+1}\rho _{i+1}F_{i+1,j',k'}-4\pi r^{2}_{i}\alpha _{i}\rho _{i}F_{i,j',k'}\right] 
\end{equation}
 with\[
\alpha _{i}\rho _{i}F_{i,j',k'}=\beta _{i,j',k'}\alpha _{i'-1}\rho _{i'-1}F_{i'-1,j',k'}+\left( 1-\beta _{i,j',k'}\right) \alpha _{i'}\rho _{i'}F_{i',j',k'}\]
for outwards propagating particles \( \left( \mu _{j'}>0\right)  \),
and\begin{equation}
\label{eq_Fi_interpolation}
\alpha _{i}\rho _{i}F_{i,j',k'}=\left( 1-\beta _{i,j',k'}\right) \alpha _{i'-1}\rho _{i'-1}F_{i'-1,j',k'}+\beta _{i,j',k'}\alpha _{i'}\rho _{i'}F_{i',j',k'}
\end{equation}
for inwards propagating particles \( \left( \mu _{j'}<0\right)  \).

Based on this finite difference representation, we calculate now the
expectation value \( D^{4}_{a} \) in Eq. (\ref{eq_cancellation_table})
for a later match with the observer corrections described in subsections
\ref{subsection_frequency_shift} and \ref{subsection_angular_aberration}.
For \( D_{a}^{4} \) we identify in Eq. (\ref{eq_cancellation_table})
the weight \( u_{i+1}\mu _{j'}E_{k'} \) in front of the space derivative
in Eq. (\ref{eq_fd_da}). As in Eq. (\ref{eq_toy_fd_expectation}),
we perform an integration by parts over \( da \) in the finite difference
representation,\begin{eqnarray*}
 &  & \frac{\mu ^{2}_{j'}E_{k'}}{\alpha _{i'}da_{i'}}\left[ 4\pi r^{2}_{i+1}\alpha _{i+1}\rho _{i+1}u_{i+1}F_{i+1,j',k'}-4\pi r^{2}_{i}\alpha _{i}\rho _{i}u_{i+1}F_{i,j',k'}\right] \\
 & = & \frac{\mu _{j'}^{2}E_{k'}}{\alpha _{i'}da_{i'}}\left[ 4\pi r^{2}_{i+1}\alpha _{i+1}\rho _{i+1}u_{i+1}F_{i+1,j',k'}-4\pi r^{2}_{i}\alpha _{i}\rho _{i}u_{i}F_{i,j',k'}\right] \\
 & - & \frac{4\pi r_{i}^{2}}{\alpha _{i'}}\frac{u_{i+1}-u_{i}}{da_{i'}}\mu _{j'}^{2}E_{k'}\alpha _{i}\rho _{i}F_{i,j',k'}.
\end{eqnarray*}
 If we would additionally integrate over the momentum phase space,
with \( E_{k'}^{2} \) as the usual measure of integration, the last
term would correspond to \( D^{4}_{a}=-4\pi r^{2}\rho \frac{\partial u}{\partial a}K \).
We insert in this term the interpolation prescription (\ref{eq_Fi_interpolation})
for \( \alpha _{i}\rho _{i}F_{i,j',k'} \) on zone edges and collect
the two contributions with the zone center index \( i' \) in the
distribution function to obtain a finite difference representation
for \( A=-4\pi r^{2}\rho /\Gamma \cdot \partial u/\partial a \):
\[
A_{i',k'}=\frac{4\pi \rho _{i'}}{\Gamma _{i+1}da_{i'}}\left( r^{2}_{i+1}\left( u_{i+2}-u_{i+1}\right) \left( 1-\beta _{i+1,k'}\right) +r^{2}_{i}\left( u_{i+1}-u_{i}\right) \beta _{i,k'}\right) \]
 if \( \mu _{j'}\leq 0 \), and \begin{equation}
\label{eq_A_fd}
A_{i',k'}=\frac{4\pi \rho _{i'}}{\Gamma _{i+1}da_{i'}}\left( r^{2}_{i+1}\left( u_{i+2}-u_{i+1}\right) \beta _{i+1,k'}+r^{2}_{i}\left( u_{i+1}-u_{i}\right) \left( 1-\beta _{i,k'}\right) \right) 
\end{equation}
 if \( \mu _{j'}>0 \). We use exactly the same analysis to evaluate
the finite difference representation of \( D_{a}^{2}=-4\pi r^{2}\rho \frac{\partial \Gamma }{\partial a}H \)
in Eq. (\ref{eq_cancellation_table}). The weight for this term in
front of Eq. (\ref{eq_fd_da}) is \( \Gamma _{i+1}E_{k'} \). The
result is a finite difference representation for \( \Lambda =-4\pi r^{2}\rho /\Gamma \cdot \partial \Gamma /\partial a \):
\[
\Lambda _{i',k'}=\frac{4\pi \rho _{i'}}{\Gamma _{i+1}da_{i'}}\left( r^{2}_{i+1}\left( \Gamma _{i+2}-\Gamma _{i+1}\right) \left( 1-\beta _{i+1,k'}\right) +r^{2}_{i}\left( \Gamma _{i+1}-\Gamma _{i}\right) \beta _{i,k'}\right) \]
 if \( \mu _{j'}\leq 0 \), and \begin{equation}
\label{eq_Lambda_fd}
\Lambda _{i',k'}=\frac{4\pi \rho _{i'}}{\Gamma _{i+1}da_{i'}}\left( r^{2}_{i+1}\left( \Gamma _{i+2}-\Gamma _{i+1}\right) \beta _{i+1,k'}+r^{2}_{i}\left( \Gamma _{i+1}-\Gamma _{i}\right) \left( 1-\beta _{i,k'}\right) \right) 
\end{equation}
otherwise. These definitions allow to write the contribution of space
propagation to the total energy conservation, \( \sum _{j,k}\left( \Gamma _{i+1}+u_{i+1}\mu _{j'}\right) D_{a,i'}w_{j'}E_{k'}^{3}dE_{k'} \),
in the form of Eq. (\ref{eq_cancellation_table}),\begin{eqnarray}
D^{1-4}_{a}: &  & \frac{1}{\alpha _{i'}da_{i'}}\left[ 4\pi r_{i+1}^{2}\alpha _{i+1}\rho _{i+1}\Gamma _{i+1}H_{i+1}-4\pi r_{i}^{2}\alpha _{i}\rho _{i}\Gamma _{i}H_{i}\right] \nonumber \\
 & + & \sum _{j,k}\Gamma _{i+1}\Lambda _{i',k'}F_{i',j',k'}\mu _{j'}w_{j'}E_{k'}^{3}dE_{k'}\nonumber \\
 & + & \frac{1}{\alpha _{i'}da_{i'}}\left[ 4\pi r_{i+1}^{2}\alpha _{i+1}\rho _{i+1}u_{i+1}K_{i+1}-4\pi r_{i}^{2}\alpha _{i}\rho _{i}u_{i}K_{i}\right] \nonumber \\
 & + & \sum _{j,k}\Gamma _{i+1}A_{i',k'}F_{i',j',k'}\mu _{j'}^{2}w_{j'}E_{k'}^{3}dE_{k'}.\label{eq_cancellation_table_da_fd} 
\end{eqnarray}

\subsubsection{Angular advection from spatial propagation}

If a particle propagates in outwards direction, its direction cosine
increases towards unity. Hence, particle propagation causes angular
advection relative to a fixed grid of particle direction cosines \( \mu _{j'} \).
This contribution to the time evolution of the particle distribution
function is described in the term \( D_{\mu } \) in Eq. (\ref{eq_boltzmann_dmu}).
The finite difference representation of this term is chosen such that
an equilibrium between an isotropic radiation field and stationary
matter remains undisturbed \citep{Lewis_Miller_84,Mezzacappa_Bruenn_93a}.
Following this approach, we set\begin{equation}
\label{eq_dmu_fd}
D_{\mu }=\left( \Gamma _{i+1}\frac{3\left[ r^{2}_{i+1}-r^{2}_{i}\right] }{2\left[ r^{3}_{i+1}-r^{3}_{i}\right] }-G_{i+1}^{\mu }\right) \frac{1}{w_{j'}}\left( \zeta _{j+1}F_{i',j+1,k'}-\zeta _{j}F_{i',j,k'}\right) .
\end{equation}
 Characteristic for the general relativistic equation is the gravitational
term \( G^{\mu }=\Gamma \partial \Phi /\partial r \) in Eq. (\ref{eq_dmu_fd}).
It implements gravitational bending. We outline its finite difference
representation below in Eq. (\ref{eq_Gmu_fd}). The differencing of
the coefficients, \( \zeta =1-\mu ^{2} \), is defined by\begin{equation}
\label{eq_def_angular_diff_coff}
\zeta _{j+1}-\zeta _{j}=-2\mu _{j'}w_{j'}.
\end{equation}
The angular integration of the term \( D_{\mu } \) produces the zeroth
and second angular moments of the particle distribution function.
Its finite difference representation is therefore not as sensitive
to cancellations in the diffusive limit as in the case of the spatial
advection term \( D_{a} \). Upwind differencing can be justified.
The angular \char`\"{}wind\char`\"{} always points towards \( \mu =1 \).
However, for reasons of completeness and consistency, we also use
centered differencing in the diffusive regime. With angular transport
coefficients \( \gamma _{i',k'}:=\beta _{i',k'} \), we interpolate
the values of the neutrino distribution function on angular zone edges
by\begin{equation}
\label{eq_Fj_interpolation}
F_{i',j,k'}=\gamma _{i',k'}F_{i',j'-1,k'}+\left( 1-\gamma _{i',k'}\right) F_{i',j',k'}.
\end{equation}

Again, the terms \( D^{1-4}_{\mu } \) in Eq. (\ref{eq_cancellation_table})
arise from an integration by parts of the Boltzmann equation (\ref{eq_relativistic_boltzmann})
with weight \( \left( \Gamma _{i+1}+u_{i+1}\mu _{j'}\right) E_{k'} \).
But this time the integration is over the angle cosines, \( \mu  \).
The part with the weight \( \Gamma _{i+1}E_{k'} \) does not depend
on the angle cosine, its contribution vanishes. For the calculation
of the other part we introduce the abbreviation\[
\Upsilon _{i+1}=\Gamma _{i+1}\frac{3\left[ r^{2}_{i+1}-r^{2}_{i}\right] }{2\left[ r^{3}_{i+1}-r^{3}_{i}\right] }-G^{\mu }_{i+1}\]
for the prefactor in Eq. (\ref{eq_dmu_fd}) and perform the integration
by parts again in its finite difference representation:\begin{eqnarray}
 &  & \sum _{j,k}u_{i+1}\Upsilon _{i+1}\left( \zeta _{j+1}\mu _{j'}F_{i',j+1,k'}-\zeta _{j}\mu _{j'}F_{i',j,k'}\right) E_{k'}^{3}dE_{k'}\nonumber \\
 & = & \sum _{j,k}u_{i+1}\Upsilon _{i+1}\left( \zeta _{j+1}\mu _{j'}F_{i',j+1,k'}-\zeta _{j}\mu _{j'-1}F_{i',j,k'}\right) E_{k'}^{3}dE_{k'}\nonumber \\
 & - & \sum _{j,k}u_{i+1}\Upsilon _{i+1}\zeta _{j}F_{i',j,k'}\left( \mu _{j'}-\mu _{j'-1}\right) E_{k'}^{3}dE_{k'}\nonumber \\
 & = & -\sum _{j,k}u_{i+1}\Upsilon _{i+1}F_{i',j',k'}E_{k'}^{3}dE_{k'}\nonumber \\
 & \times  & \left[ \gamma _{i',k'}\zeta _{j+1}\left( \mu _{j'+1}-\mu _{j'}\right) +\left( 1-\gamma _{i',k'}\right) \zeta _{j}\left( \mu _{j'}-\mu _{j'-1}\right) \right] .\label{eq_dmu_moments} 
\end{eqnarray}
The expression in the second line vanishes because of mutual cancellations
in the sum and \( \zeta _{1}=\zeta _{j_{{\rm max}}+1}=0 \). With
\( \zeta =1-\mu ^{2} \), we can identify the result as a special
finite difference representation of the terms \( D_{\mu }^{12} \)
and \( D_{\mu }^{34} \) in Eq. (\ref{eq_cancellation_table}). Following
the strategy of the previous paragraph, we extract from Eq. (\ref{eq_dmu_moments})
a finite difference expression for the expression \( B=\left( 1-\mu ^{2}\right) u/r \),\begin{equation}
\label{eq_B_fd}
B_{i',j',k'}=\frac{3}{2}\frac{r^{2}_{i+1}-r^{2}_{i}}{r^{3}_{i+1}-r^{3}_{i}}\frac{u_{i+1}}{w_{j'}}\left[ \gamma _{i',k'}\zeta _{j+1}\left( \mu _{j'+1}-\mu _{j'}\right) +\left( 1-\gamma _{i',k'}\right) \zeta _{j}\left( \mu _{j'}-\mu _{j'-1}\right) \right] .
\end{equation}
With this definition, we write the contribution \( \sum _{j,k}\left( \Gamma _{i+1}+u_{i+1}\mu _{j'}\right) D_{\mu ,i'}w_{j'}E_{k'}^{3}dE_{k'} \)
to the total energy evolution in the form of its continuous analogue
in Eq. (\ref{eq_cancellation_table}),\begin{eqnarray}
D_{\mu }: & - & \Gamma _{i+1}\sum _{j,k}B_{i',j',k'}F_{i',j',k'}w_{j'}E_{k'}^{3}dE_{k'}+u_{i+1}G_{i+1}^{\mu }\sum _{j,k}F_{i',j',k'}E_{k'}^{3}dE_{k'}\nonumber \\
 & \times  & \left[ \gamma _{i',k'}\zeta _{j+1}\left( \mu _{j'+1}-\mu _{j'}\right) +\left( 1-\gamma _{i',k'}\right) \zeta _{j}\left( \mu _{j'}-\mu _{j'-1}\right) \right] .\label{eq_cancellation_table_dmu_fd} 
\end{eqnarray}

The finite difference representation of \( G_{i+1}^{\mu } \) for
the gravitational bending still needs to be defined. We make its definition
depend on the finite differencing of \( G^{E}=\Gamma \partial \Phi /\partial r \)
in Eq. (\ref{eq_Ge_fd}), which will account for the gravitational
frequency shift of the particle energies. In the term \( D_{E}^{2} \)
in Eq. (\ref{eq_cancellation_table}) it contributes with \( u_{i+1}G_{i+1}^{E}K_{i'} \)
to total energy conservation. If we define\begin{equation}
\label{eq_Gmu_fd}
G_{i+1}^{\mu }=\frac{3}{2}\frac{\Gamma _{i+1}}{\alpha _{i'}}\frac{\alpha _{i'+1}-\alpha _{i'}}{r_{i'+1}-r_{i'}}-\frac{1}{2}G_{i+1}^{E},
\end{equation}
 this will support the desired cancellation \( C_{t}^{2}+D^{34}_{\mu }+D_{E}^{2}=0 \)
under two conditions: (i) The hydrodynamics equations should accurately
implement Eq. (\ref{eq_misner_dGdt}), i.e.\begin{equation}
\label{eq_misner_dGdt_fd}
C_{t}^{2}\equiv -\frac{\partial \Gamma _{i+1}}{\alpha _{i'}\partial t}J_{i'}=-u_{i+1}\frac{\Gamma _{i+1}}{\alpha _{i'}}\frac{\alpha _{i'+1}-\alpha _{i'}}{r_{i'+1}-r_{i'}}J_{i'}.
\end{equation}
If this is not the case, an alternative discretization based on the
time derivative of \( \Gamma  \) as a function of the velocity and
the gravitational potential has also provided satisfactory results
(see Eq. (\ref{eq_adaptive_velocity}) for the evaluation of \( \partial u/\partial t \)
on the adaptive grid),\[
C_{t}^{2}\equiv -\frac{\partial \Gamma _{i+1}}{\alpha _{i'}\partial t}J_{i'}=\frac{u_{i+1}}{\Gamma _{i+1}}\left( \frac{\partial u}{\alpha _{i'}\partial t}+4\pi r_{i+1}p_{i'+1}+\frac{m_{i+1}}{r_{i+1}\bar{r}_{i+1}}\right) J_{i'}.\]
(ii) The particle distribution function should be close to isotropic.
In the isotropic limit we can use the relation \( J=3K \) and obtain
from Eq. (\ref{eq_dmu_moments}) a simplified contribution to energy
conservation \begin{eqnarray*}
D_{\mu }^{34}: &  & -u_{i+1}G_{i+1}^{\mu }\sum _{k}F_{i',k'}E_{k'}^{3}dE_{k'}\sum _{j}\mu _{j'}\left( \zeta _{j+1}-\zeta _{j}\right) \\
 & = & u_{i+1}G_{i+1}^{\mu }\sum _{k}F_{i',k'}E_{k'}^{3}dE_{k'}\sum _{j}\mu _{j'}^{2}w_{j'}\\
 & = & \frac{2}{3}u_{i+1}G_{i+1}^{\mu }J_{i'}.
\end{eqnarray*}
Note that the Gaussian quadrature in the angle discretization provides
the exact integral for \( \int \mu ^{2}d\mu =2/3 \). Both conditions
are fulfilled in the high density domain of the protoneutron star
where discretization errors most dramatically affect energy conservation
because of the large radiation energy densities. Of course, Eq. (\ref{eq_Gmu_fd})
provides also a valid finite differencing of gravitational bending
outside of the protoneutron star, where the effect is physically important.
But the finite differencing is not specially tuned to enforce the
cancellation \( C_{t}^{2}+D^{34}_{\mu }+D_{E}^{2}=0 \) outside of
the diffusive regime.

\subsubsection{Frequency shift from observer motion}\label{subsection_frequency_shift}

We continue our description with the observer correction \( O_{E} \)
in Eq. (\ref{eq_boltzmann_oe}). We start with a review of the finite
differencing procedure presented in \citep{Mezzacappa_Bruenn_93a}
in order to extend it from local to global energy conservation. Global
energy conservation requires the cancellation of observer correction
terms in Eq. (\ref{eq_cancellation_table}) with terms from space
propagation and angular advection. Therefore, it is convenient to
express the observer corrections in terms of the same physical quantities
\( A \) and \( B \) we have used before, \begin{eqnarray}
A & = & \left( \frac{\partial \ln \rho }{\alpha \partial t}+\frac{2u}{r}\right) \nonumber \\
B & = & \left( 1-\mu ^{2}\right) \frac{u}{r}.\nonumber \\
O_{E} & = & \left( \mu ^{2}A-B\right) \frac{1}{E^{2}}\frac{\partial }{\partial E}\left[ E^{3}F\right] .\label{eq_AB_definition} 
\end{eqnarray}
 In the following, we solve the equation for the frequency shift correction,
\( C_{t}+O_{E}=0 \), by the method of characteristics. We convert
the time derivative at constant energy to a time derivative along
the characteristic in the energy dimension of the momentum phase space.
This eliminates the partial derivative with respect to energy and
facilitates a conservative finite differencing of \( O_{E} \). Note
that the characteristic used in this section implements the frequency
shift correction alone, it differs from the free propagation characteristic
prescribed by the full Boltzmann equation. As in \citep{Bruenn_85,Mezzacappa_Bruenn_93a}
we write the prefactor of the correction as a time derivative of a
quantity \( R_{f}=r^{\left( 3\mu ^{2}-1\right) }\rho ^{\left( \mu ^{2}\right) } \),\[
\frac{\partial \ln R_{f}}{\alpha \partial t}=\mu ^{2}A-B,\]
such that \( C_{t}+O_{E}=0 \) becomes\begin{equation}
\label{eq_lagrangian_energy_derivative}
0=E^{3}\left( \frac{\partial F}{\partial t}\right) _{E}+\frac{\partial \ln R_{f}}{\partial t}E\frac{\partial }{\partial E}\left[ E^{3}F\right] .
\end{equation}
It is now possible to transform from the {}``Eulerian'' variable,
\( x\equiv E \), to a {}``Lagrangian'' variable, \( y\equiv E/R_{f} \),
along the characteristic by the chain rule \( \left( \partial /\partial t\right) _{x}=\left( \partial /\partial t\right) _{y}+\left( \partial y/\partial t\right) _{x}\left( \partial /\partial y\right)  \)
(the subscript of the bracket denotes the variable that is kept constant
for the differentiation). Eq. (\ref{eq_lagrangian_energy_derivative})
simplifies to\begin{eqnarray*}
0 & = & \left( \frac{\partial }{\partial t}\left[ E^{3}F\right] \right) _{E}+\frac{\partial R_{f}}{R^{2}_{f}\partial t}E\times R_{f}\frac{\partial }{\partial E}\left[ E^{3}F\right] \\
 & = & \left( \frac{\partial }{\partial t}\left[ E^{3}F\right] \right) _{E}-\left( \frac{\partial \left[ E/R_{f}\right] }{\partial t}\right) _{E}\frac{\partial \left[ E^{3}F\right] }{\partial \left[ E/R_{f}\right] }=\left( \frac{\partial }{\partial t}\left[ E^{3}F\right] \right) _{E/R_{f}}.
\end{eqnarray*}
 For a small section of the energy phase space \( E^{2}\Delta E=\left( E^{3}_{2}-E^{3}_{1}\right) /3 \),
this relationship leads to \begin{equation}
\label{eq_bunch_enumber_evolution}
\left( \frac{\partial }{\partial t}\left[ E^{2}F\Delta E\right] \right) _{E/R_{f}}=0.
\end{equation}
The validity of Eq. (\ref{eq_bunch_enumber_evolution}) for arbitrary
distribution functions \( F \) leads to the following interpretation:
The observer correction shifts the particles that initially reside
in the energy interval \( E^{2}\Delta E \) along the characteristic
with constant \( E/R_{f} \) in the energy phase space. This allows
us to determine the evolution of any other particle property in analogy
to Eq. (\ref{eq_expectation_along_characteristic}). For example,
the specific energy of the particles in this phase space interval,
\( d\epsilon =E^{3}F\Delta E \), evolves according to \begin{equation}
\label{eq_specific_energy_change}
\left( \frac{\partial }{\partial t}\left[ E^{3}F\Delta E\right] \right) _{E/R_{f}}=E^{2}F\Delta E\left( \frac{\partial E}{\partial t}\right) _{E/R_{f}}=\frac{\partial \ln R_{f}}{\partial t}d\epsilon .
\end{equation}
 A finite difference representation of Eqs. (\ref{eq_bunch_enumber_evolution})
and (\ref{eq_specific_energy_change}) has been given in \citep{Mezzacappa_Bruenn_93a}.
Consider a particle energy group \( k' \), with a neighbor group
\( k'+dk \), \( dk=\pm 1 \). Eq. (\ref{eq_bunch_enumber_evolution})
tells us that the number of particles before the correction, \( F_{i',j',k'}E^{2}_{k'}dE_{k'} \),
is equal to the number of particles after the correction. The distribution
after the correction is represented by a diminished number of particles
\( F_{i',j',k'}E_{k'}^{2}dE_{k'}-n_{i',j',k'}^{-} \) in group \( k' \)
and an additional number of particles \( n_{i',j',k'+dk}^{+} \) in
the neighbor group \( k'+dk, \)\begin{equation}
\label{eq_bunch_enumber_fd}
F_{i',j',k'}E_{k'}^{2}dE_{k'}-\left[ \left( F_{i',j',k'}E_{k'}^{2}dE_{k'}-n^{-}_{i',j',k'}\right) +n^{+}_{i',j',k'+dk}\right] =0.
\end{equation}
 Eq. (\ref{eq_specific_energy_change}) now defines a similar relationship
for the particle energies\begin{eqnarray}
F_{i',j',k'}E_{k'}^{3}dE_{k'} & - & \left[ \left( F_{i',j',k'}E_{k'}^{3}dE_{k'}-E_{k'}n^{-}_{i',j',k'}\right) +E_{k'+dk}n^{+}_{i',j',k'+dk}\right] \nonumber \\
 & = & -\left( \mu ^{2}_{j'}A_{i',k'}-B_{i',j',k'}\right) F_{i',j',k'}E^{3}_{k'}dE_{k'}\alpha _{i'}dt,\label{eq_specific_energy_change_fd} 
\end{eqnarray}
 where \( A_{i',k'} \) and \( B_{i',j',k'} \) stand for a finite
difference representation of Eq. (\ref{eq_AB_definition}). Eqs. (\ref{eq_bunch_enumber_fd})
and (\ref{eq_specific_energy_change_fd}) uniquely define the solution
\begin{eqnarray}
n^{-}_{i',j',k'} & = & \left( \mu _{j'}^{2}A_{i',k'}-B_{i',j',k'}\right) \frac{dE_{k'}}{E_{k'+dk}-E_{k'}}E^{3}_{k'}F_{i',j',k'}\alpha _{i'}dt\nonumber \\
n^{+}_{i',j',k'} & = & n_{i',j',k'-dk}^{-},\label{eq_oe_deltaplusminus} 
\end{eqnarray}
which leads, by the update \( F_{i',j',k'}=\bar{F}_{i',j',k'}+\left( n^{+}_{i',j',k'}-n^{-}_{i',j',k'}\right) /\left( E_{k'}^{2}dE_{k'}\right)  \),
to the following finite difference representation of the frequency
shift term in the Boltzmann equation (\ref{eq_relativistic_boltzmann}):
\begin{eqnarray}
O_{E} & = & \frac{1}{E^{2}_{k'}dE_{k'}}\nonumber \\
 & \times  & \left[ \left( \mu _{j'}^{2}A_{i',k'-dk}-B_{i',j',k'-dk}\right) \frac{dE_{k'-dk}}{E_{k'}-E_{k'-dk}}E_{k'-dk}^{3}F_{i',j',k'-dk}\right. \nonumber \\
 & - & \left. \left( \mu _{j'}^{2}A_{i',k'}-B_{i',j',k'}\right) \frac{dE_{k'}}{E_{k'+dk}-E_{k'}}E_{k'}^{3}F_{i',j',k'}\right] .\label{eq_oe_fd} 
\end{eqnarray}
Finally, we calculate the contribution of the frequency shift to energy
conservation, \[
\sum _{j,k}\left( \Gamma _{i+1}+u_{i+1}\mu _{j'}\right) O_{E,i'}w_{j'}E_{k'}^{3}dE_{k'}.\]
 The summation over \( k \) in Eq. (\ref{eq_oe_fd}) with weight
\( E_{k'} \) and measure of integration \( E_{k'}^{2} \) can be
simplified with a discrete {}``integration by parts'' as in Eq.
(\ref{eq_toy_fd_expectation}). If we neglect for the time being the
boundary terms because of a small \( E_{1'}^{3} \) in the lowest
energy group and a small \( F_{i',j',k'_{\rm max}} \) in the highest
energy group (a correction for these terms will be made later in subsection
\ref{subsection_boundary_corrections}), the energy contribution from
\( O_{E} \) becomes\begin{equation}
\label{eq_cancellation_table_oe_fd}
O^{1-4}_{E}:\qquad -\sum _{j,k}\left( \Gamma _{i+1}+u_{i+1}\mu _{j'}\right) \left( \mu _{j'}^{2}A_{i',k'}-B_{i',j',k'}\right) F_{i',j',k'}w_{j'}E_{k'}^{3}dE_{k'}.
\end{equation}
If we use the finite difference representations (\ref{eq_A_fd}) and
(\ref{eq_B_fd}) for \( A_{i',k'} \) and \( B_{i',j',k'} \) respectively,
we find in the comparison of Eq. (\ref{eq_cancellation_table_oe_fd})
with Eqs. (\ref{eq_cancellation_table_da_fd}) and (\ref{eq_cancellation_table_dmu_fd})
that we have indeed matched the terms \( (D_{a}^{4}O_{E}^{1}) \)
and \( (D_{\mu }^{12}O_{E}^{34}) \) to machine precision \citep{Liebendoerfer_00}.

\subsubsection{Angular aberration from observer motion}\label{subsection_angular_aberration}

Motivated by this success, we try the method of characteristics also
on the angular aberration corrections. In this case, the characteristic
is described by \( C_{t}+O_{\mu }=0 \). We write the prefactor of
the angular aberration correction in Eq. (\ref{eq_boltzmann_omu})
as a function of the quantities \( A \) and \( B \) defined in Eq.
(\ref{eq_AB_definition}) to obtain with \( \zeta =1-\mu ^{2} \),
\[
O_{\mu }=\left( A+B/\zeta \right) \frac{\partial }{\partial \mu }\left[ \zeta \mu F\right] .\]
As before, we try to eliminate the partial derivative with respect
to the angle cosine with a time derivative along the characteristic
in the angle part of the momentum phase space \citep{Liebendoerfer_00}.
For the quantity \( R_{a}=r^{3}\rho  \), we find the relation\[
\frac{\partial \ln R_{a}}{\alpha \partial t}=A+B/\zeta .\]
With this, we transform from the {}``Eulerian'' variable \( x\equiv \mu  \)
to the {}``Lagrangian'' variable \( y\equiv \zeta ^{-1/2}\mu /R_{a} \).
After a multiplication with \( \zeta \mu  \), the angular aberration
correction \( 0=C_{t}+O_{\mu } \) becomes:\begin{eqnarray}
0 & = & \zeta \mu \left[ \left( \frac{\partial F}{\partial t}\right) _{\mu }+\frac{\partial \ln R_{a}}{\partial t}\frac{\partial }{\partial \mu }\left[ \zeta \mu F\right] \right] \nonumber \\
 & = & \left( \frac{\partial }{\partial t}\left[ \zeta \mu F\right] \right) _{\mu }+\zeta ^{-1/2}\mu \frac{\partial R_{a}}{R^{2}_{a}\partial t}\times \zeta ^{3/2}R_{a}\frac{\partial }{\partial \mu }\left[ \zeta \mu F\right] \nonumber \\
 & = & \left( \frac{\partial }{\partial t}\left[ \zeta \mu F\right] \right) _{\mu }-\left( \frac{\partial \left[ \zeta ^{-1/2}\mu /R_{a}\right] }{\partial t}\right) _{\mu }\frac{\partial \left[ \zeta \mu F\right] }{\partial \left[ \zeta ^{-1/2}\mu /R_{a}\right] }=\left( \frac{\partial }{\partial t}\left[ \zeta \mu F\right] \right) _{\zeta ^{-1/2}\mu /R_{a}}.\nonumber 
\end{eqnarray}
 The particles initially residing in the angular interval \( \left( 1-3\mu ^{2}\right) \Delta \mu =\zeta _{2}\mu _{2}-\zeta _{1}\mu _{1} \)
are shifted by angular aberration along characteristics with constant
\( \mu /\left( \sqrt{\zeta }R_{a}\right)  \):\begin{equation}
\label{eq_bunch_lnumber_evolution}
\left( \frac{\partial }{\partial t}\left[ \left( 1-3\mu ^{2}\right) F\Delta \mu \right] \right) _{\zeta ^{-1/2}\mu /R_{a}}=0.
\end{equation}
We have to keep in mind that any correction to the particle propagation
direction also affects the energy conservation in the frame of a distant
observer. Therefore, it is desirable to construct a numerical implementation
of angular aberration that prescribes the changes in the specific
luminosity \( \sum _{j}\mu _{j'}F_{i',j',k'}w_{j'} \) to machine
precision. Hence, we evaluate the change of the specific luminosity,
\( d\ell =\left( 1-3\mu ^{2}\right) \mu F\Delta \mu  \), along the
characteristic in analogy to Eq. (\ref{eq_expectation_along_characteristic}),
\begin{equation}
\label{eq_specific_luminosity_change}
\left( \frac{\partial }{\partial t}\left[ \left( 1-3\mu ^{2}\right) \mu F\Delta \mu \right] \right) _{\zeta ^{-1/2}\mu /R_{a}}=\left( 1-3\mu ^{2}\right) F\Delta \mu \left( \frac{\partial \mu }{\partial t}\right) _{\zeta ^{-1/2}/R_{a}}=\zeta \frac{\partial \ln R_{a}}{\partial t}d\ell .
\end{equation}
 We identify the bin size \( \left( 1-3\mu _{j'}^{2}\right) \Delta \mu _{j'}=w_{j'} \)
with our Gaussian quadrature weights. In the finite difference representation,
Eq. (\ref{eq_bunch_lnumber_evolution}) leads to the condition for
number conservation\[
F_{i',j',k'}w_{j'}-\left[ \left( F_{i',j',k'}w_{j'}-n^{-}_{i',j',k'}\right) +n^{+}_{i',j'+dj,k'}\right] =0\]
and Eq. (\ref{eq_specific_luminosity_change}) to a prescription for
the numerical evolution of the specific luminosity\begin{eqnarray*}
F_{i',j',k'}\mu _{j'}w_{j'} & - & \left[ \left( F_{i',j',k'}\mu _{j'}w_{j'}-\mu _{j'}n^{-}_{i',j',k'}\right) +\mu _{j'+dj}n^{+}_{i',j'+dj,k'}\right] \\
 & = & -\left( \zeta _{j'}A_{i',k'}+B_{i',j',k'}\right) F_{i',j',k'}\mu _{j'}w_{j'}\alpha _{i'}dt.
\end{eqnarray*}
 The direction of the differencing can be chosen by \( dj=\pm 1 \).
The change in the neutrino distribution function from angular aberration
is then: \[
F_{i',j',k'}=\bar{F}_{i',j',k'}+\left( n_{i',j',k'}^{+}-n_{i',j',k'}^{-}\right) /w_{j'}\]
 with \begin{eqnarray*}
n^{-}_{i',j',k'} & = & \left( A_{i',k'}+B_{i',j',k'}/\zeta _{j'}\right) \frac{w_{j'}}{\mu _{j'+dj}-\mu _{j'}}\zeta _{j'}\mu _{j'}F_{i',j',k'}\alpha _{i'}dt\\
n^{+}_{i',j',k'} & = & n^{-}_{i',j'-dj,k'}.
\end{eqnarray*}
Combined into one update, this leads to the following finite difference
representation of the angular aberration term in the Boltzmann Eq.
(\ref{eq_relativistic_boltzmann}):\begin{eqnarray}
O_{\mu } & = & \frac{1}{w_{j'}}\left[ \left( A_{i',k'}+B_{i',j'-dj,k'}/\zeta _{j'-dj}\right) \frac{w_{j'-dj}}{\mu _{j'}-\mu _{j'-dj}}\zeta _{j'-dj}\mu _{j'-dj}F_{i',j'-dj,k'}\right. \nonumber \\
 & - & \left. \left( A_{i',k'}+B_{i',j',k'}/\zeta _{j'}\right) \frac{w_{j'}}{\mu _{j'+dj}-\mu _{j'}}\zeta _{j'}\mu _{j'}F_{i',j',k'}\right] .\label{eq_omu_fd} 
\end{eqnarray}
 We apply the aberration corrections with \( dj=+1 \) for \( \mu \leq 0 \)
and \( dj=-1 \) for \( \mu >0 \). This is not upwind differencing
and, therefore, runs the risk of producing negative particle distribution
functions. However, there are two good reasons to accept this shortcoming:
(i) The angular aberration correction is generally small, with the
exception of aberration in the vicinity of strong shocks with large
velocity gradients. (ii) The chosen direction of \( dj \) guarantees
that no particles are shifted off the grid. This is a prerequisite
for number and energy conservation. If we calculate the contribution
of \( \textrm{O}_{\mu } \) to the energy evolution, \( \sum _{j,k}\left( \Gamma _{i+1}+u_{i+1}\mu _{j'}\right) O_{\mu ,i'}w_{j'}E_{k'}^{3}dE_{k'} \),
along the lines of Eq. (\ref{eq_toy_fd_expectation}), this choice
of \( dj \) causes the perfect vanishing of boundary terms in the
discrete {}``integration by parts'' with respect to the angle cosine.
We obtain\begin{equation}
\label{eq_cancellation_table_omu_fd}
O^{1-4}_{\mu }:\qquad \sum _{j,k}u_{i+1}\mu _{j'}\left( -A_{i',k'}+\mu _{j'}^{2}A_{i',k'}-B_{i',j',k'}\right) F_{i',j',k'}w_{j'}E_{k'}^{3}dE_{k'}.
\end{equation}
The second and third terms in the parenthesis cancel exactly with
the yet unmatched terms from \( O^{1-4}_{E} \) in Eq. (\ref{eq_cancellation_table_oe_fd}).
Thus, also the cancellations \( (O_{E}^{2}O_{\mu }^{2}) \) and \( (O_{E}^{56}O_{\mu }^{34}) \)
are guaranteed in Eq. (\ref{eq_cancellation_table}). However, in
failed supernova simulations, the shock recedes at late time to very
small radii and becomes very strong because of high infall velocities.
We have encountered numerical fluctuations in the neutrino distribution
function at the shock position at this late time that are due to this
fixed choice of differencing in the angular advection in the aberration
terms. They disturb the progress of the simulation with large time
steps. In future long term simulations, we will use strict upwind
differencing also in the angular aberration terms and allow a deviation
from perfect energy conservation in these terms (we still conserve
particle number). In section \ref{subsection_energy_conservation},
we demonstrate that the expected energy violations from angular aberration
are unlikely to dominate the violations of energy conservation by
the adaptive grid or a mismatch in \( \left( C_{t}^{2}D_{\mu }^{34}D_{E}^{2}\right)  \).

\subsubsection{Frequency shift in the gravitational potential}\label{subsection_gravitational_potential}

The action of the gravitational potential on the propagation of the
particles, \( C_{t}+D_{E}=0 \), in Eq. (\ref{eq_boltzmann_de}),
is finite differenced in full analogy with Eq. (\ref{eq_oe_fd}),
\begin{equation}
\label{eq_de_fd}
D_{E}=-\frac{\mu _{j'}G^{E}_{i+1}}{E^{2}_{k'}dE_{k'}}\left[ \frac{dE_{k'-dk}}{E_{k'}-E_{k'-dk}}E_{k'-dk}^{3}F_{i',j',k'-dk}-\frac{dE_{k'}}{E_{k'+dk}-E_{k'}}E_{k'}^{3}F_{i',j',k'}\right] .
\end{equation}
This is a valid finite difference representation of \( D_{E}=-\mu G^{E}/E^{2}\cdot \partial \left( E^{3}F\right) /\partial E \).
We apply the corrections from Eqs. (\ref{eq_oe_fd}) and (\ref{eq_de_fd})
with \( dk=+1 \) (blueshift) if the two following conditions are
met: (i) \( \sum _{j,k}\left( \mu _{j'}^{2}A_{i',k'}-B_{i',j',k'}\right) w_{j'}E_{k'}^{2}dE_{k'} \)
is positive, and (ii), its absolute value is larger than \( \left| \sum _{j,k}\mu _{j'}G_{i+1}^{E}w_{j'}E_{k'}^{2}dE_{k'}\right|  \).
Otherwise, we choose \( dk=-1 \) for a frequency redshift. With this,
the chosen sign of \( dk \) in Eqs. (\ref{eq_oe_fd}) and (\ref{eq_de_fd})
implements upwind differencing for the angle-integrated distribution
function. Note, that the direction of the finite differencing in energy
space must not depend on the individual angle bins. An angle dependence
would destroy the important cancellation of redshift and blueshift
in the limit of an isotropic distribution function. 

In Eq. (\ref{eq_Gmu_fd}), we have introduced the placeholder \( G^{E}=\Gamma \frac{\partial \Phi }{\partial r} \),
whose finite differencing we will now define. Eq. (\ref{eq_hydro_cancellation})
describes how the terms \( (C_{t}^{4}D_{a}^{2}D_{E}^{1}O_{\mu }^{1}) \)
should combine to the expression \( 4\pi r\rho \left( 1+e+p/\rho \right) H \).
The latter is used in the equation for the total energy evolution.
A matching of the energy changes in \( C_{t}^{4} \), \( D_{a}^{2} \),
\( O_{\mu }^{1} \), and \( D_{E}^{1} \) to this expression is the
last constraint of the list in Eq. (\ref{eq_cancellation_table})
that we have not yet implemented. The term \( C_{t}^{4} \) is determined
by \( \alpha ^{-1}\partial u/\partial t \) in the hydrodynamics equations.
The finite differencing of the terms \( D_{a}^{2} \) and \( O_{\mu }^{1} \)
has also been determined in the second line of Eq. (\ref{eq_cancellation_table_da_fd})
and in the first term of Eq. (\ref{eq_cancellation_table_omu_fd})
respectively. We simply resolve for \( G^{E} \) and obtain\begin{eqnarray}
G_{i+1}^{E} & = & \frac{1}{\Gamma _{i+1}}\left( \frac{\partial u}{\alpha _{i'}\partial t}+4\pi r_{i+1}\rho _{i'}\left( 1+e_{i'}+p_{i'+1}/\rho _{i'}\right) \right) \nonumber \\
 & + & \frac{1}{H_{i'}}\sum _{j,k}\left( A_{i',k'}-\Lambda _{i',k'}\right) F_{i',j',k'}\mu _{j'}w_{j'}E_{k'}^{3}dE_{k'}.\label{eq_Ge_fd} 
\end{eqnarray}
The evaluation of the Lagrangian time derivative \( \partial u/\partial t \)
on the adaptive grid requires the appropriate corrections. The application
of Eq. (\ref{eq_reynolds_theorem_fd}) to the velocity leads to\begin{equation}
\label{eq_adaptive_velocity}
\frac{\partial u}{\partial t}=\frac{1}{da_{i}}\left\{ \frac{u_{i}da_{i}-\bar{u}_{i}\overline{da}_{i}}{dt}+\left[ u_{i'}^{\rm rel}u_{i'}^{*}-u_{i'-1}^{\rm rel}u_{i'-1}^{*}\right] \right\} ,
\end{equation}
where we have once more used the grid velocity with respect to mass,
\( u^{\rm rel}_{i'}=-\left( a_{i'}-\bar{a}_{i'}\right) /dt \). The
advected velocity is evaluated by upwind differencing:\[
u^{*}_{i'}=\left\{ \begin{array}{ll}
u_{i} & \mbox {if}\quad u_{i'}^{\rm rel}\geq 0\\
u_{i+1} & \mbox {otherwise}.
\end{array}\right. \]

\subsubsection{Boundary corrections and the enforcement of Fermi statistics}\label{subsection_boundary_corrections}

\( k_{{\rm max}} \) energy groups are subject to a frequency shift,
but there are only \( k_{\rm max}-1 \) enclosed group edges to shift
particles across. In the energy phase space, upwind differencing is
necessary for the stability. Therefore, the straightforward application
of Eqs. (\ref{eq_oe_fd}) and (\ref{eq_de_fd}) leads to particles
that fall aside the grid. The lost particle number is proportional
to the particle population in the energy groups at the boundaries
of the grid. At the lower boundary it is small because the phase space
volume \( E^{2}dE \) is small. At the upper boundary it is small
because the neutrino distribution function is small. The particle
loss at both boundaries would approach zero when the coverage of the
energy grid would be extended. Nevertheless, we want to correct for
this effect and avoid that particles and energy unaccountedly disappear
from the radiation field. On the one hand, we would like to support
requirement (iii) in the introduction of this section with an accurate
evolution of the radiation moments, even and especially at low energy
resolution. On the other hand, conservation to machine precision in
wellknown terms is an existential practical asset in the hunt of unknown
energy leaks in a complex code. To resolve this issue, we simply do
not apply the energy shift to the boundary group if this would advect
particles off the grid. This exception fixes particle number conservation.
The induced error in the energy balance of the radiation field is
then restored by the application of minimal energy shifts to the population
of all other energy groups.

Before we describe the details of this correction, we mention another
exceptional correction to the frequency shift. It is applied when
the Fermi statistics of the radiation particles is violated by numerical
effects. The collision term \( C_{c} \) includes all necessary blocking
factors to avoid the overpopulation of states in the fermionic particle
gas by emission or inscattering. If inelastic interactions occur on
a very short time scale, they will efficiently restore numerical deviations
from perfect Fermi statistics in the radiation field. However, an
overpopulation of energy groups can still occur in the numerical implementation.
Consider for example a hydrostatic situation with a collisionless
degenerate particle gas. The Boltzmann equation (\ref{eq_relativistic_boltzmann})
reduces in this case to \( D_{a}+D_{\mu }+D_{E}=0 \). It is straightforward
to demonstrate that \( \rho F=1 \) is a solution of this equation.
In the numerical implementation, however, it is unlikely that the
redshift towards low energy groups, \( D_{E} \), is to machine precision
compensated by the space propagation term, \( D_{a} \), and angular
advection term, \( D_{\mu } \). Even if the collision term is large,
overpopulated energy groups may occur as a consequence of operator
splitting. Our previous code version \citep{Mezzacappa_Bruenn_93a}
implemented the observer corrections in Eq. (\ref{eq_oe_fd}), (\ref{eq_omu_fd}),
and (\ref{eq_de_fd}) in an operator split explicit update to the
implicit solution of the transport equation \( C_{t}+D_{a}+D_{\mu }=C_{c} \).
In this case, overpopulation is more likely to occur. The scheme described
below has been developed for this context. For simulations with massive
progenitors in general relativity, however, we have expanded the implicit
solver to include all terms of the transport equation \citep{Liebendoerfer_et_al_02}.
For cosmetic reasons we still strictly enforce Fermi statistics when
necessary. We performed a control run with both corrections switched
off. At any time before and including bounce, deviations in the luminosity
are smaller than \( 2\% \). Differences in the rms energies are smaller
than \( 0.5\% \), and in any hydrodynamic quantity smaller than \( 0.2\% \).
When the shock propagates through the neutrino-degenerate core, the
\( 2\% \) differences in the luminosity cause transient differences
of the same size in parts of the velocity and the density profile.
Later, the differences are smaller than \( 1\% \) in any quantity
and at any time. 

As in \citep{Mezzacappa_Bruenn_93a}, we recursively shift overpopulated
particles into the next higher energy group. This procedure is number
conservative, but it introduces an artificial increase of the internal
energy of the radiation field. The details of the correction scheme
for low resolution boundary effects and overpopulation are the following:
First, we quantify the energy correction: \[
R_{i',j'}=\left( \mu ^{2}_{j'}A_{i',k'}-B_{i',j',k'}-\mu _{j'}G_{i+1}^{E}\right) \bar{F}_{i',j',k'}E^{3}_{k'}dE_{k'},\]
with \( k=k_{\rm max} \), if \( dk=1 \) (blueshift); and \( k=1 \),
if \( dk=-1 \) (redshift). The corrections stem in most cases from
an energy redshift. Thus, we set \( dk=-1 \) and initialize the accumulation
of corrections in \( \Delta _{i',j',k'} \) with \( \Delta =0 \).
Then, we loop over \( k \) in ascending order and carry out the following
procedure: (i) Distribute the correction \( R_{i',j'} \) over all
groups \( \ell ' \) with \( k+1\leq \ell \leq k_{\rm max} \) according
to \[
\Delta _{i',j',\ell '}=\Delta _{i',j',\ell '}+\frac{R_{i',j'}}{\sum _{s=k+1,k_{\rm max}}\bar{F}_{i',j',s'}E_{s'}^{3}dE_{s'}}.\]
(ii) Apply the correction \( \Delta _{i',j',k'+1} \) according to
the conservative finite differencing used in Eq. (\ref{eq_oe_deltaplusminus})\[
n^{+}_{i',j',k'}=-\Delta _{i',j',k'+1}\frac{dE_{k'+1}}{E_{k'+1}-E_{k'}}E_{k'+1}^{3}\bar{F}_{i',j',k'+1}\]
and update the particle distribution function\begin{equation}
\label{eq_correction_fd}
\frac{F_{i',j',k'}-\bar{F}_{i',j',k'}}{\alpha _{i'}dt}=\frac{n^{+}_{i',j',k'}-n^{+}_{i',j',k'-1}}{E_{k'}^{2}dE_{k'}}.
\end{equation}
(iii) If the distribution function \( F_{i',j',k'} \) is larger than
\( 1/\rho _{i'} \), correct the particle flux into this group by
the amount of overpopulation, \( n^{+}_{i',j',k'}=n^{+}_{i',j',k'}+\left( 1/\rho _{i'}-F_{i',j',k'}\right) E^{2}_{k'}dE_{k'} \),
and evaluate the energy change owing to this flux correction, \[
R_{i',j'}=\left( 1/\rho _{i'}-F_{i',j',k'}\right) \left( E_{k'+1}-E_{k'}\right) E^{2}_{k'}dE_{k'}.\]
Set \( F_{i',j',k'} \) to the maximum value, \( 1/\rho _{i'} \)
and apply the energy correction \( R_{i',j'} \) with the next iteration
of the loop over \( k \).

\subsubsection{Collision term}

The finite difference representation of the collision term, Eq. (\ref{eq_boltzmann_cc}),
is described in detail in \citep{Mezzacappa_Bruenn_93b,Mezzacappa_Bruenn_93c,Mezzacappa_Messer_98,Messer_00}.
We represent it here with a simple placeholder, namely emissivity
\( j_{i',k'} \) and opacity \( \chi _{i',k'} \) as a function of
the thermodynamical state of the fluid,\begin{equation}
\label{eq_cc_fd}
C_{c}=\frac{j_{i',k'}(\rho _{i'},T_{i'}^{*},Y_{e,i'}^{*})}{\rho _{i'}}-\chi _{i',k'}(\rho _{i'},T_{i'}^{*},Y_{e,i'}^{*})F_{i',j',k'}.
\end{equation}
The quantities with a star refer to quantities that have consistently
been updated with the particle distribution functions before the actual
hydrodynamics update. We outline the time sequence of the evaluation
of equations in section \ref{subsection_code_flow}.

\subsection{Coupling between the radiation field and the fluid}\label{subsection_hydrodynamics}

The neutrino radiation field in the supernova is tightly coupled to
the high density fluid. In order to obtain a well defined link between
the evolution of radiation quantities and the evolution of hydrodynamical
quantities, we recall that the solution of the microscopic Boltzmann
equation updates the radiation moments according to Eqs. (\ref{eq_radiation_momentum})-(\ref{eq_radiation_internal_energy}).
We may now subtract these updates from the global evolution equations
(\ref{eq_continuity})-(\ref{eq_internal_energy}) to find the hydrodynamical
part of the evolution equations

\begin{eqnarray}
\frac{\partial }{\partial t}\left[ \frac{\Gamma }{\rho }\right]  & = & \frac{\partial }{\partial a}\left[ 4\pi r^{2}\alpha u\right] \label{eq_hydro_contiunity} \\
\frac{\partial }{\partial t}\left[ \Gamma \left( 1+e\right) \right]  & = & -\frac{\partial }{\partial a}\left[ 4\pi r^{2}\alpha up\right] \nonumber \\
 & - & \alpha \Gamma \int \left( \frac{j}{\rho }-\chi F\right) E^{3}dEd\mu +\alpha u\int \chi FE^{3}dE\mu d\mu \label{eq_hydro_total_energy} \\
\frac{\partial }{\partial t}\left[ u\left( 1+e\right) \right]  & = & -4\pi r^{2}\frac{\partial }{\partial a}\left[ \alpha \Gamma p\right] \nonumber \\
 & - & \frac{\alpha }{r}\left[ \left( 1+e\right) \left( 1+\frac{6Vp}{m}\right) \frac{m}{r}+4\pi r^{2}\left( \left( 1+e\right) \rho K+pJ\right) \right. \nonumber \\
 & + & \left. \left( 2u^{2}-\frac{m}{r}\right) \frac{2p}{\rho }\right] \nonumber \\
 & + & \alpha \Gamma \int \chi FE^{3}dE\mu d\mu -\alpha u\int \left( \frac{j}{\rho }-\chi F\right) E^{3}dEd\mu \label{eq_hydro_momentum} \\
\frac{\partial }{\partial t}Y_{e} & = & \mp \alpha \int \left( \frac{j}{\rho }-\chi F\right) E^{2}dEd\mu \label{eq_hydro_Ye} \\
\frac{\partial V}{\partial a} & = & \frac{\Gamma }{\rho }\label{eq_hydro_volume_gradient} \\
\frac{\partial m}{\partial a} & = & \Gamma \left( 1+e+J\right) +uH\label{eq_hydro_mass_gradient} \\
\rho \left( 1+e\right) \frac{\partial \alpha }{\partial a} & = & -\frac{\partial }{\partial a}\left[ \alpha p\right] +\frac{\alpha }{4\pi r^{2}}\int \chi FE^{3}dE\mu d\mu \label{eq_hydro_lapse_gradient} \\
\frac{\partial e}{\partial t} & = & -p\frac{\partial }{\partial t}\left( \frac{1}{\rho }\right) -\alpha \int \left( \frac{j}{\rho }-\chi F\right) E^{3}dEd\mu .\label{eq_hydro_internal_energy} 
\end{eqnarray}
 The equations are written such that the presence of the radiation
field only enters in terms of energy and momentum exchange. In the
limit of a decoupled radiation and matter flow we therefore solve
for ideal hydrodynamics and free streaming in an independent and numerically
stable manner, no matter what the size of the radiation field is.
The only remaining interactions between the radiation field and the
fluid are of a gravitational nature, for example in the contribution
of the radiation field to the common gravitational mass in Eq. (\ref{eq_hydro_mass_gradient}),
or in the term \( 4\pi r^{2}\left( \left( 1+e\right) \rho K+pJ\right)  \)
in the momentum equation (\ref{eq_hydro_momentum}). The detailed
discretization of Eqs. (\ref{eq_hydro_contiunity})-(\ref{eq_hydro_internal_energy})
has been described in \citep{Liebendoerfer_Rosswog_Thielemann_02}.
The interaction terms between the radiation field and the fluid are\begin{eqnarray}
e_{i'}^{\{\rm ext\}} & = & \sum _{j',k'}\left( \frac{j_{k'}(\rho _{i'},T_{i'}^{*},Y_{e,i'}^{*})}{\rho _{i'}}-\chi _{k'}(\rho _{i'},T_{i'}^{*},Y_{e,i'}^{*})F_{i',j',k'}\right) E_{k'}^{3}dE_{k'}w_{j'}\nonumber \\
S_{i+1}^{\{\rm ext\}} & = & -\sum _{j',k'}\chi _{k'}(\rho _{i'},T_{i'}^{*},Y_{e,i'}^{*})F_{i',j',k'}E_{k'}^{3}dE_{k'}\mu _{j'}w_{j'}\nonumber \\
Y_{e,i'}^{\{\rm ext\}} & = & \mp \sum _{j',k'}\left( \frac{j_{k'}(\rho _{i'},T_{i'}^{*},Y_{e,i'}^{*})}{\rho _{i'}}-\chi _{k'}(\rho _{i'},T_{i'}^{*},Y_{e,i'}^{*})F_{i',j',k'}\right) E_{k'}^{2}dE_{k'}w_{j'},\label{eq_interaction_fd} 
\end{eqnarray}
where the minus sign in the \( Y_{e} \)-term is used for electron
neutrinos and the plus sign for electron antineutrinos. Emission and
absorption of \( \mu - \) and \( \tau  \)-neutrinos do not change
the electron fraction. The star superscript for the temperature and
electron fraction variables indicates again an evaluation of the cross
sections for emission and absorption with a consistently updated thermodynamical
state as described in the next section.

\subsection{Implicit Solution}\label{subsection_code_flow}

We have now a finite difference representation for every term of the
transport equation. In this section we join them to one implicitly
determined solution. Each time step cycles through an implicitly finite
differenced update of the neutrino distribution functions and the
thermodynamic state of the matter, an implicitly finite differenced
update of the hydrodynamics variables and the adaptive grid, and an
explicitly finite differenced cosmetic correction. The intermediately
updated values do not belong to uniquely defined physical times. We
denote them by fractional values of the cycle counter \( n \). In
detail, the update of the primitive variables for one cycle proceeds
as follows:

\begin{enumerate}
\item We start with the old neutrino distribution functions \( \bar{F}:=F(n-1/2) \)
for the electron flavor neutrinos. They are given on a Lagrangian
mass grid \( \bar{a} \). We assume that we know the old hydrodynamic
variables, \( \{\bar{a},\bar{r},\bar{u},\bar{m},\bar{\rho },\bar{T},\bar{Y}_{e},\bar{\alpha }\}:=H(n-1) \),
and that the new hydrodynamic variables, \( \{a,r,u,m,\rho ,T,Y_{e},\alpha \}:=H(n) \),
have been calculated in step (3) of the previous cycle. In one implicitly
finite differenced update, we solve for \( F(n):=F \) as prescribed
by the transport equation \( C_{t}+D_{a}+D_{\mu }+D_{E}+O_{E}+O_{\mu }-C_{c}=0 \).
The finite differencing is chosen according to Eqs. (\ref{eq_ct_fd}),
(\ref{eq_fd_da}), (\ref{eq_dmu_fd}), (\ref{eq_oe_fd}), (\ref{eq_omu_fd}),
(\ref{eq_de_fd}), and (\ref{eq_cc_fd}). This implicit update additionally
includes equations for the evolution of the temperature and electron
fraction due to particle-fluid interactions. These quantities are
strongly coupled to the radiation field. We solve for \begin{eqnarray}
\frac{e_{i'}^{*}-e_{i'}}{dt} & = & \alpha _{i'}e_{i'}^{\rm ext}\label{eq_eupdate_fd} \\
\frac{Y_{e,i'}^{*}-Y_{e,i'}}{dt} & = & \alpha _{i'}Y_{e,i'}^{\rm ext},\label{eq_yeupdate_fd} 
\end{eqnarray}
where the source terms are given in Eq. (\ref{eq_interaction_fd}).
Note that Eq. (\ref{eq_eupdate_fd}) also determines the temperature
because the internal energy and temperature in the implicit update
are always consistent with the equation of state, \( e^{*}(\rho ,T^{*},Y_{e}^{*}) \)
and \( e(\rho ,T,Y_{e}) \). We repeat step (1) for the muon flavor
neutrinos. The tauon flavor neutrinos are included by counting the
appropriate muon quantities twice. But we always distinguish neutrinos
from antineutrinos.
\item In this minor second update, we correct for discretization errors
caused by the advection scheme in the energy phase space and for numerical
violations of Fermi statistics. This update is not related to any
physical terms and only relevant in low resolution simulations. We
set \( \bar{F}:=F(n) \), and update \( F(n+1/2):=F \) as described
by Eq. (\ref{eq_correction_fd}).
\item In the update of the hydrodynamics variables, the primitives \( \{\bar{a},\bar{r},\bar{u},\bar{m},\bar{\rho },\bar{T},\bar{Y}_{e},\bar{\alpha }\}:=H(n) \)
are evolved to \( H(n+1):=\{a,r,u,m,\rho ,T,Y_{e},\alpha \} \) in
a fully implicit step according to Eqs. (\ref{eq_hydro_contiunity})-(\ref{eq_hydro_internal_energy}).
Note that we use the neutrino distribution function \( F(n) \) rather
than \( F(n+1/2) \) to evaluate the moments \( J \), \( H \), and
\( K \) for the gravitational coupling. The former are the direct
result from the implicit solution of the transport equation and represent
more accurately the quasi-stationary radiation quantities. The hydrodynamics
update is based on the transfer of internal energy, \( e^{\rm ext} \),
electron fraction, \( Y^{\rm ext}_{e} \), and the neutrino stress,
\( S^{\rm ext} \), as evaluated in the first step of this cycle.
The detailed finite difference representation in {\sc agile} has been
outlined in \citep{Liebendoerfer_Rosswog_Thielemann_02}.
\end{enumerate}
Steps \( 1 \), \( 2 \), and \( 3 \) form literally a cycle; there
is no position where \emph{all} data are consistently updated. After
step \( 1 \), the neutrino distribution functions \( F(n) \) are
in accurate equilibrium with the matter. After step \( 2 \), the
energy conserving corrections from energy advection are included and
the distribution functions \( F(n+1/2) \) obey Fermi statistics exactly.
After step \( 3 \), the overall energy conservation is balanced,
but the new hydrodynamic variables \( H(n+1) \) already live on a
different spatial grid than the neutrino distribution function. We
chose to set the integer value of \( n \) at the instance we dump
the result into files. This is between step \( 1 \) and step \( 2 \),
where all quantities live on the same grid positions and where all
rates and neutrino distribution functions accurately reflect equilibrium
conditions if the mean free paths are small.

Above description is complete with respect to what we compute. It
does not address the question of how we compute. We apply a Newton-Raphson
scheme to solve the implicitly finite differenced nonlinear equations
in step (1) and step (3). The solution vectors for the two updates
are\begin{equation}
\label{eq_solution_vector}
y^{\rm rad}_{i'}=\left( \begin{array}{c}
F^{\nu }_{i',j=1'-j'_{\rm max},1'}\\
.\\
.\\
F^{\nu }_{i',1'-j'_{\rm max},k'_{\rm max}}\\
F^{\bar{\nu }}_{i',j=1'-j'_{\rm max},1'}\\
.\\
.\\
F^{\bar{\nu }}_{i',1'-j'_{\rm max},k'_{\rm max}}\\
T_{i'}\\
Y_{e,i'}
\end{array}\right) ,\qquad y_{i'}^{\rm hyd}=\left( \begin{array}{c}
a_{i}\\
r_{i}\\
u_{i}\\
m_{i}\\
\rho _{i'}\\
T_{i'}\\
Y_{e,i'}\\
\alpha _{i'}
\end{array}\right) .
\end{equation}
At the present resolution with \( 103 \) zones, \( 6 \) angular
bins, and \( 12 \) energy groups, the radiation solution vector in
step (1), \( y^{\rm rad} \), has the length \( i_{\rm max}\times \left( 2\times j_{\rm max}\times k_{\rm max}+2\right) =15038 \)
and the hydrodynamics solution vector in step (3), \( y^{\rm hyd} \),
has the length \( \left( i_{\rm max}+1\right) \times 8=832 \). The
radiation system is much larger than the hydrodynamics system and
dominates the computational effort. The hydrodynamics system with
the adaptive grid, however, shows a more complicated coupling between
the variables. Common to both cases is that, in one time step, we
have to determine the vector \( y^{n+1} \) at time \( t^{n+1}=t^{n}+dt \)
which solves a nonlinear system of equations, \( R(y^{n},y^{n+1};dt)=0 \).
The Newton-Raphson scheme is based on the linearization of the system
of equations around a guessed solution vector, \( \tilde{y}^{n+1} \)
,

\begin{equation}
\label{eq_taylor_expansion}
R(y^{n},\tilde{y}^{n+1}+\Delta y;dt)=R(y^{n},\tilde{y}^{n+1};dt)+\frac{\partial R(y^{n},\tilde{y}^{n+1};dt)}{\partial y^{n+1}}\Delta y+O((\Delta y)^{2}).
\end{equation}
 The residual on the right hand side of Eq. (\ref{eq_taylor_expansion})
can be neutralized by a specific choice of corrections, \begin{equation}
\label{eq_newton_raphson_correction}
\widetilde{\Delta y}=-\left( \frac{\partial R(y^{n},\tilde{y}^{n+1};dt)}{\partial y^{n+1}}\right) ^{-1}R(y^{n},\tilde{y}^{n+1};dt).
\end{equation}
 \( \tilde{y}^{n+1}+\widetilde{\Delta y} \) is then a better guess
because it fulfills \[
R(y^{n},\tilde{y}^{n+1}+\Delta y;dt)=0+O((\Delta y)^{2}).\]
 This procedure is iterated until a norm of the correction vector
\( |\Delta y|/y^{n,\rm scl} \) becomes sufficiently small. In the
solution of the radiation system we stop the iterations when the corrections
\( \Delta \left( \rho F_{i',j',k'}\right)  \), \( \Delta e_{i'}/e_{i'} \),
and \( \Delta Y_{e,i'}/Y_{e,i'} \) are smaller than \( 10^{-10} \).
If the radiation is rather stationary, (e.g. in early phases of the
gravitational collapse or some time after bounce during neutrino heating),
the solution of the transport equation requires about three iterations.
Around bounce, the iteration count is typically \( 4 \) and may only
occasionally reach \( 6 \). Convergence problems appear only at very
late time (around a second after bounce, depending on the progenitor
model) when the adaptive grid demands tiny mass zones (smaller than
\( \sim 10^{-4} \)M\( _{\odot } \)) to resolve the steep density
gradient developing at the surface of the protoneutron star. Of order
half of the computation time is spent on the construction of the Jacobian
\( \partial R(y^{n},\tilde{y}^{n+1};dt)/\partial y^{n+1} \), while
the other half is required to solve the linear system in Eq. (\ref{eq_newton_raphson_correction}).
We solve the linear system by direct Gauss elimination. The zone-wise
arrangement of the solution vector in Eq. (\ref{eq_solution_vector})
leads to a block-tridiagonal Jacobi matrix because every equation
depends on input from at most three adjacent zones. A straightforward
solver for band-diagonal matrices (e.g. from \citet{Press_et_al_92}
or from the {\sc lapack} linear algebra package) works perfectly if
the variables and equations are scaled such that the maximum element
in each row of the Jacobian is of order unit. A solver for band matrices
is not only simpler than a solver for block-tridiagonal matrices,
but it adapts even better to the boundary of the off-diagonal blocks
because the latter have nonzero coefficients only on their diagonal.
Iterative solvers require careful preconditioning to solve the described
equations satisfactorily \citep{Azevedo_et_al_02}. In order to calculate
the exact Jacobi matrix, we need derivatives of the equations with
respect to all primitive variables. Different methods are used to
build the Jacobi matrix for the derivatives with respect to the particle
distribution functions, the derivatives with respect to temperature
and electron fraction, and the derivatives of the hydrodynamics equations:

The left hand side of the Boltzmann equation (\ref{eq_relativistic_boltzmann})
is a linear operator on the particle distribution function. Thus,
one can inquire Eqs. (\ref{eq_ct_fd}), (\ref{eq_fd_da}), (\ref{eq_dmu_fd}),
(\ref{eq_oe_fd}), (\ref{eq_omu_fd}), and (\ref{eq_de_fd}) for the
occurrences of \( F_{i'\pm 1,j',k'} \), \( F_{i',j'\pm 1,k'} \),
and \( F_{i',j',k'\pm 1} \). The corresponding prefactors, separately
collected, form the contribution to the Jacobian for this part of
the transport equation. There are no other dependencies on the left
hand side of the transport equation. It requires more tedious efforts
to gather the derivatives of the collision term. On the one hand,
the interactions can in principle couple all angle bins and energy
groups within one zone. On the other hand, blocking factors lead to
nonlinear dependencies in some reactions. This applies as well to
the derivatives of Eqs. (\ref{eq_eupdate_fd}) and (\ref{eq_yeupdate_fd})
with respect to particle distribution functions. The coefficients
for the Jacobian have to be read off from extensive listings of summations
over angle bins and energy groups, as outlined for example in Eq.
(67) and (94) in \citep{Mezzacappa_Messer_98}.

Too complicated to be handled manually are the derivatives of the
collision term in Eq. (\ref{eq_cc_fd}), Eq. (\ref{eq_eupdate_fd}),
and (\ref{eq_yeupdate_fd}) with respect to temperature and electron
fraction. The dynamic table developed in \citep{Mezzacappa_Bruenn_93a}
helps in this case. We illustrate its mechanism with the most simple
example, the derivative of the left hand side of Eq. (\ref{eq_eupdate_fd})
with respect to temperature,\begin{equation}
\label{eq_jacobian_energy_derivative}
\frac{\partial }{\partial T_{i'}}\left( \frac{e_{i'}^{*}\left( \rho _{i'},T_{i'},Y_{e,i'}\right) -\bar{e}_{i'}}{dt}\right) .
\end{equation}
 We discretize the three-dimensional space with dimensions \( \rho  \),
\( T \), \( Y_{e} \) into cubes around the thermodynamical states
of the fluid zones \( \left( \rho _{i'},T_{i'},Y_{e,i'}\right)  \).
The eight cube corners \[
\left\{ \left( \rho _{l},T_{m},Y_{e,n}\right) ,\quad l,m,n=1,2\right\} \]
 satisfy\begin{eqnarray*}
\log _{10}\left( \rho _{2}/\rho _{1}\right)  & = & N_{\rho }^{-1}\\
\log _{10}\left( T_{2}/T_{1}\right)  & = & N_{T}^{-1}\\
Y_{2}-Y_{1} & = & N_{Y}^{-1}.
\end{eqnarray*}
The numbers \( N_{\rho } \), \( N_{T} \), \( N_{Y} \) define the
resolution of the cube. In our applications, we use \( N_{\rho }=10 \),
\( N_{T}=40 \), \( N_{Y}=50 \). If a quantity has to be evaluated
as a function of \( \rho _{i'} \), \( T_{i'} \), and \( Y_{e,i'} \),
we fetch the cube for zone \( i' \) and evaluate the quantity on
all eight cube corners if this has not already been done in an earlier
cycle. For the evaluation of the internal energy, for example, eight
calls to the interactive Lattimer-Swesty equation of state \citep{Lattimer_Swesty_91}
would be required to store the internal energy on the cube corners.
The internal energy of the zone is then found by tri-linear interpolation
in the logarithmic internal energy,\begin{eqnarray}
\log _{10}e_{i'} & = & \left( 1-C_{\rho }\right) \left( 1-C_{T}\right) \left( 1-C_{Y}\right) \log _{10}e\left( \rho _{1},T_{1},Y_{e,1}\right) \nonumber \\
 & + & \left( 1-C_{\rho }\right) \left( 1-C_{T}\right) C_{Y}\log _{10}e\left( \rho _{1},T_{1},Y_{e,2}\right) \nonumber \\
 & + & \left( 1-C_{\rho }\right) C_{T}\left( 1-C_{Y}\right) \log _{10}e\left( \rho _{1},T_{2},Y_{e,1}\right) \nonumber \\
 & + & \left( 1-C_{\rho }\right) C_{T}C_{Y}\log _{10}e\left( \rho _{1},T_{2},Y_{e,2}\right) \nonumber \\
 & + & C_{\rho }\left( 1-C_{T}\right) \left( 1-C_{Y}\right) \log _{10}e\left( \rho _{2},T_{1},Y_{e,1}\right) \nonumber \\
 & + & C_{\rho }\left( 1-C_{T}\right) C_{Y}\log _{10}e\left( \rho _{2},T_{1},Y_{e,2}\right) \nonumber \\
 & + & C_{\rho }C_{T}\left( 1-C_{Y}\right) \log _{10}e\left( \rho _{2},T_{2},Y_{e,1}\right) \label{eq_state_variable_interpolation} \\
 & + & C_{\rho }C_{T}C_{Y}\log _{10}e\left( \rho _{2},T_{2},Y_{e,2}\right) \nonumber 
\end{eqnarray}

where\begin{eqnarray*}
C_{\rho } & = & \frac{\log _{10}\left( \rho _{i'}/\rho _{1}\right) }{\log _{10}\left( \rho _{2}/\rho _{1}\right) }\\
C_{T} & = & \frac{\log _{10}\left( T_{i'}/T_{1}\right) }{\log _{10}\left( T_{2}/T_{1}\right) }\\
C_{Y} & = & \frac{Y_{e,i'}-Y_{e,1}}{Y_{e,2}-Y_{e,1}}.
\end{eqnarray*}
The linearity of Eq. (\ref{eq_state_variable_interpolation}) makes
the determination of the derivatives straightforward. For example,
in Eq. (\ref{eq_jacobian_energy_derivative}) we need\begin{equation}
\label{eq_state_derivative_example}
\frac{\partial e_{i'}}{\partial T_{i'}}=\frac{e_{i'}}{T_{i'}}\frac{\partial \log _{10}e_{i'}}{\partial \log _{10}T_{i'}}=N_{T}\frac{e_{i'}}{T_{i'}}\frac{\partial \log _{10}e_{i'}}{\partial C_{T}}
\end{equation}
where \( \partial \log _{10}e_{i'}/\partial C_{T} \) can simply be
read off from Eq. (\ref{eq_state_variable_interpolation}). This scheme
has important advantages. First, the relation between the values of
the interpolated variable and its derivatives inside the cube is exact,
both being derived from the same interpolation formula. This helps
convergence in the multidimensional Newton-Raphson scheme. Second,
the subroutine generating the variable on the cube corners need not
be used at each time step, but only if the point \( \left( \rho _{i'},T_{i'},Y_{e,i'}\right)  \)
moves outside the cube, in which case a new, adjacent, cube is generated.
As the complexity of the scheme does not depend on the complexity
of the function that generates the variable on the cube corner, we
apply the scheme to all reaction rates and scattering kernels, whose
derivatives with respect to \( \rho _{i'} \), \( T_{i'} \), and
\( Y_{e,i'} \) are then straightforward to determine as in the example
of Eq. (\ref{eq_state_derivative_example}). Given that only a few
percent of the cubes have to be regenerated during one time step,
these dynamic tables enhance the computational efficiency of our method
by a factor of \( \sim 50 \). Third, we avoid discontinuous changes
in the variables and any of its derivatives, which can be traumatic
for a Newton-Raphson scheme. Note, that we occasionally allow extrapolation
if the thermodynamic state leaves the cube during the iterations for
the solution of the radiation equation in step (1). The cube boundaries
are only updated afterwards in the hydrodynamics step (3). In the
solution of the hydrodynamics equations, however, we update the cube
boundaries for every iteration of the Newton-Raphson scheme. As the
hydrodynamics equations only rely on the pressure and internal energy
from the equation of state, this is not a computationally expensive
measure, although it occasionally deteriorates the convergence behavior
of the implicit solution. It is necessary because an extrapolated
energy in the converged solution would not necessarily match the interpolated
energy after a postponed cube update. This would lead to a violation
of energy conservation. This problem does not arise with the internal
energy in the solution of the transport equation because only the
energy exchange rate, \( e_{i'}^{\rm ext} \), from Eq. (\ref{eq_yeupdate_fd})
is transferred to the hydrodynamics equations in step (3). The extrapolated
energy, \( e_{i'}^{*} \), is discarded after step (1).

The Jacobian of the hydrodynamics equations is much smaller. But it
has a less regular structure due to the strong nonlinearities in the
equations of general relativistic hydrodynamics. Furthermore, the
adaptive grid couples changes of all variables to the grid motion,
which itself depends on the gradients of the variables. The spatial
smoothing operator in the adaptive grid equation couples variables
from five adjacent zones. In order to eliminate the considerable risk
of producing an erroneous Jacobian in lengthy manual derivations,
the hydrodynamics code {\sc agile} constructs the Jacobian automatically
by taking numerical derivatives \citep{Liebendoerfer_Rosswog_Thielemann_02}.
We start with the evaluation of the residuum vector \( R=R(y^{n},\tilde{y}^{n+1};dt) \)
based on an initial guess \( \tilde{y}^{n+1} \) and define, based
on machine precision, a small number \( \varepsilon  \) for the calculation
of the numerical derivatives, \citep{Press_et_al_92}. Next, we sort
the components of the state vector, denoted with label \( i \), into
\( J \) distinct groups \( g_{j} \) according to the rule that none
of the equations \( R \) may depend on more than one state vector
component out of the same group. We try to choose these groups such
that the number of distinct groups is minimized. We select a group
\( g_{j} \) and create a variation \( \tilde{y}^{var,n+1} \), \begin{eqnarray}
\tilde{y}[i]^{var,n+1} & = & \tilde{y}[i]^{n+1}+\varepsilon y[i]^{n,{\textrm{scl}}}\nonumber \\
\Delta [i] & = & \frac{\tilde{y}[i]^{var,n+1}-\tilde{y}[i]^{n+1}}{y[i]^{n,{\textrm{scl}}}},
\end{eqnarray}
 for all components \( i\in g_{j} \). The residuum vector \( R^{var}=R(y^{n},\tilde{y}^{var,n+1};dt) \)
is then evaluated based on the varied guess. From the two residuals
we can extract the components of the Jacobian \begin{equation}
A[k,i]=y[i]^{n,{\textrm{scl}}}\frac{\partial R[k](y^{n},\tilde{y}^{n+1};dt)}{\partial y[i]^{n+1}}=\frac{R[k]^{var}-R[k]}{\Delta [i]}
\end{equation}
 for all \( i\in g_{j} \). The Jacobian is complete when this procedure
has been performed for all groups \( J \). Finally, we scale the
rows of \( A \) and the right hand side residuum vector by the maximum
component in the row and solve the linear system \begin{equation}
\sum _{i}A[k,i]\left( \Delta y[i]/y[i]^{n,{\textrm{scl}}}\right) =-R[k].
\end{equation}
 to get the corrections \( \Delta y \) for the update of the guess.
Again, we use direct Gauss elimination for band-diagonal matrices.
Before the next iteration, all zeros in the Jacobian are detected
and the sparsity structure is refined. This allows further reduction
of the number of required groups for all following time steps. The
evaluation of \( R^{var}=R(y^{n},\tilde{y}^{var,n+1};dt) \) for a
group \( j \) is completely independent from the corresponding calculation
for a different group \( j'\neq j \). In order to compose a complete
Jacobian, the system of equations has to be calculated once with the
actual guess, and \( J \) times in parallel with varied guesses.
The hydrodynamics solution usually converges in \( 3 \) iterations.

The actual time step is set such that the hydrodynamics variables
don't change by more than a percent per time step. For the neutrino
distribution functions, it has been most efficient to apply the weak
condition \( \left( \rho F-\bar{\rho }\bar{F}\right) /\left( \bar{\rho }\bar{F}+0.1\right) <0.1 \) combined
with a general upper limit of the time step to \( 0.1 \) ms. Both
radiation transport and hydrodynamics proceed with the minimum time
step compatible with any of these conditions. A simulation from \( 100 \)
ms before bounce to about \( 600 \) ms after bounce takes about \( 12000 \)
time steps. The time steps are smallest around bounce, where they
show typical values of \( 10^{-4} \) ms.

\section{Code Verification}\label{section_code_verification}

In this section we present a test of the terms and finite difference
representations introduced since \citep{Mezzacappa_Bruenn_93a}. First,
we investigate the diffusion limit in our finite difference representation
and demonstrate that a small diffusive flux is accurate in the presence
of a large, nearly isotropic radiation field. We also check the other
extreme, the evolution of the radiation moments in a free streaming
situation in spherically symmetric geometry. Another test in stationary
space-time investigates the implementation of gravitational terms,
such as number luminosity conservation, gravitational frequency shift,
and gravitational bending. The frequency shift and angular aberration
of the radiation field are probed at the shock front, where we relate
the discontinuity in radiation quantities in the comoving frame to
the smooth radiation field in the view of stationary observers. Then,
we investigate the resolution dependence of our results and perform
a detailed energy and lepton number conservation analysis to check
the overall consistency of our code. Finally, in the intended application
of stellar core collapse and postbounce evolution, we compare our
results with the independently developed multi-group flux-limited
diffusion ({\sc mgfld}) code of \citet{Bruenn_DeNisco_Mezzacappa_01}.
The latter is based on a sophisticated but approximative treatment
of radiative transfer in spherical symmetry and uses a different hydrodynamics
code.

\subsection{Diffusion limit}

The neutrinos in the interior region of the protoneutron star are
almost completely trapped. Because of the high electron chemical potential,
they equilibrate with matter at high root mean square energies. The
neutrino number density multiplied by the speed of light exceeds the
neutrino number flux by orders of magnitude. It is therefore essential
that the finite difference representation of the Boltzmann equation
does not allow errors in the large particle density and pressure to
swamp the markedly smaller particle flux. A derivation of the analytical
diffusion equation from the Boltzmann equation is the first step in
the derivation of a multi-group flux-limited diffusion scheme \citep{Bruenn_85}.
We perform here exactly the same procedure, but apply all operations
to the finite difference representation of the Boltzmann equation.
For simplicity, we assume a stationary background. The particle distribution
function is split into an isotropic and a flux component,\[
F_{i',j',k'}=\psi _{i',k'}^{0}+\mu _{j'}\psi _{i',k'}^{1}.\]
 We further assume that the flux component is stationary, i.e., that
its time dependence in the radiation momentum equation is negligible.
We substitute this approximation for the neutrino distribution function
into the finite difference representation of the stationary-state
Boltzmann equation, \( C_{t}+D_{a}+D_{\mu }+D_{E}=C_{c} \), and obtain
the radiation momentum equation by the application of the operator
\( \left( 3/2\right) \sum _{j}\mu _{j'}w_{j'} \) and the use of the
identity \( \left( 3/2\right) \sum _{j}\mu ^{2}_{j'}w_{j'}=1 \):\[
R^{0}_{a}+R_{a}^{1}+R_{\mu }^{0}+R_{\mu }^{1}+R^{0}_{E}=R_{c}\]
with\begin{eqnarray}
R^{0}_{a} & = & \frac{2\pi }{\alpha _{i'}da_{i'}}\left[ r^{2}_{i+1}\left( \alpha _{i'}\rho _{i'}\psi _{i',k'}^{0}+\alpha _{i'+1}\rho _{i'+1}\psi _{i'+1,k'}^{0}\right) \right. \nonumber \\
 & - & \left. r^{2}_{i}\left( \alpha _{i'-1}\rho _{i'-1}\psi _{i'-1,k'}^{0}+\alpha _{i'}\rho _{i'}\psi _{i',k'}^{0}\right) \right] \nonumber \\
R_{a}^{1} & = & \frac{-3\pi }{2\alpha _{i'}da_{i'}}\left[ r_{i+1}^{2}\left( 1-2\beta _{i+1,k'}\right) \left( \alpha _{i'}\rho _{i'}\psi _{i',k'}^{1}-\alpha _{i'+1}\rho _{i'+1}\psi _{i'+1,k'}^{1}\right) \right. \nonumber \\
 & - & \left. r_{i}^{2}\left( 1-2\beta _{i,k'}\right) \left( \alpha _{i'-1}\rho _{i'-1}\psi _{i'-1,k'}^{1}-\alpha _{i'}\rho _{i'}\psi _{i',k'}^{1}\right) \right] \nonumber \\
R^{0}_{\mu } & = & -2\Upsilon _{i+1}\psi _{i',k'}^{0}\nonumber \\
R_{\mu }^{1} & = & 3\Upsilon _{i+1}\sum _{j}\mu _{j'}\left[ \left( 1-\gamma _{i',k'}\right) \zeta _{j+1}\left( \mu _{j'+1}-\mu _{j'}\right) +\gamma _{i',k'}\zeta _{j}\left( \mu _{j'}-\mu _{j'-1}\right) \right] \psi ^{1}_{i',k'}\nonumber \\
R^{0}_{E} & = & -\frac{G^{E}_{i+1}}{E^{2}_{k'}dE_{k'}}\left( \frac{dE_{k'-dk}}{E_{k'}-E_{k'-dk}}E_{k'-dk}^{3}\psi _{i',k'-dk}^{0}-\frac{dE_{k'}}{E_{k'+dk}-E_{k'}}E_{k'}^{3}\psi _{i',k'}^{0}\right) \nonumber \\
R^{1}_{c} & = & -\chi _{i',k'}(\rho _{i'},T_{i'}^{*},Y_{e,i'}^{*})\psi _{i',k'}^{1}.\label{eq_diffusive_radiation_moment_fd} 
\end{eqnarray}

The simple expression \( R_{\mu }^{0} \) results from the specific
property (\ref{eq_def_angular_diff_coff}) of the angular difference
coefficients. The radial dependence of \( \beta _{i,k'} \) in Eq.
(\ref{eq_transport_coefficients_fd}) can lead to an asymmetric treatment
of ingoing and outgoing particles, producing the numerical term \( R_{a}^{1} \).
Also \( R_{\mu }^{1} \) arises as an unwanted term with no analytical
correspondence. It is therefore essential to set the transport coefficients
in the diffusive regime precisely to \( \beta _{i,k'}=1/2 \). With
the choice in Eq. (\ref{eq_transport_coefficients_fd}) we immediately
find \( R_{a}^{1}=0 \). Also \( R_{\mu }^{1} \) vanishes because
the term under the sum over \( j \) becomes antisymmetric with \( \gamma _{i',k'}=\beta _{i',k'}=1/2 \).
For the next step, we need a more concise notation for the difference
and average of a zone edge variable, \( V \):\begin{eqnarray*}
d\left\langle V\right\rangle _{i'} & = & V_{i+1}-V_{i}\\
\left\langle V\right\rangle _{i'} & = & \frac{1}{2}\left( V_{i+1}+V_{i}\right) .
\end{eqnarray*}
For zone center variables, we use the analogous definitions, e.g.
\( d\left\langle \alpha \rho \psi _{k'}^{0}\right\rangle _{i}=\alpha _{i'}\rho _{i'}\psi _{i',k'}^{0}-\alpha _{i'-1}\rho _{i'-1}\psi _{i'-1,k'}^{0} \).
With these, we may rewrite the terms \( R_{a}^{0} \) and \( R_{\mu }^{0} \)
as\begin{eqnarray}
R_{a}^{0} & = & \frac{d\left\langle 4\pi r^{2}\right\rangle _{i'}}{da_{i'}}\rho _{i'}\psi _{i',k'}^{0}+\frac{4\pi }{da_{i'}}\frac{1}{2\alpha _{i'}}\left[ r_{i+1}^{2}d\left\langle \alpha \rho \psi _{k'}^{0}\right\rangle _{i+1}+r_{i}^{2}d\left\langle \alpha \rho \psi _{k'}^{0}\right\rangle _{i}\right] \nonumber \\
R_{\mu }^{0} & = & -\left( \Gamma _{i+1}\frac{3d\left\langle r^{2}\right\rangle _{i'}}{2d\left\langle r^{3}\right\rangle _{i'}}-G_{i+1}^{\mu }\right) 2\psi _{i',k'}^{0}.\label{eq_ra0_rmu0_cancellation} 
\end{eqnarray}
If the hydrodynamics equations relate the rest mass element to the
volume element by \( \Gamma _{i+1}da_{i'}=\left( 4\pi /3\right) \rho _{i'}d\left\langle r^{3}\right\rangle _{i'} \),
the first term in \( R_{a}^{0} \) cancels with the first term in
\( R_{\mu }^{0} \). Finally we try to extract the lapse function
\( \alpha  \) from the gradient in the remainder of \( R_{a}^{0} \),\[
\frac{4\pi }{da_{i'}}\frac{1}{\alpha _{i'}}\left\langle r^{2}d\left\langle \alpha \rho \psi _{k'}^{0}\right\rangle \right\rangle _{i'}=\frac{1}{da_{i'}\alpha _{i'}}\left\langle 4\pi r^{2}\left\langle \rho \psi _{k'}^{0}\right\rangle d\alpha \right\rangle _{i'}+\frac{1}{da_{i'}\alpha _{i'}}\left\langle 4\pi r^{2}\alpha d\left\langle \rho \psi _{k'}^{0}\right\rangle \right\rangle _{i'},\]
 and find for the diffusion limit in the finite difference representation\[
R^{\prime 0}_{a}+R_{\mu }^{\prime 0}+R^{0}_{E}=R_{c}\]
with\begin{eqnarray}
R_{a}^{\prime 0} & = & \frac{1}{\alpha _{i'}da_{i'}}\left\langle 4\pi r^{2}\alpha d\left\langle \rho \psi _{k'}^{0}\right\rangle \right\rangle _{i'}\nonumber \\
R_{\mu }^{\prime 0} & = & \frac{1}{\alpha _{i'}da_{i'}}\left\langle 4\pi r^{2}\left\langle \rho \psi _{k'}^{0}\right\rangle d\alpha \right\rangle _{i'}+G_{i+1}^{\mu }2\psi _{i',k'}^{0}\nonumber \\
R^{0}_{E} & = & -\frac{G^{E}_{i+1}}{E^{2}_{k'}dE_{k'}}\left( \frac{dE_{k'-dk}}{E_{k'}-E_{k'-dk}}E_{k'-dk}^{3}\psi _{i',k'-dk}^{0}-\frac{dE_{k'}}{E_{k'+dk}-E_{k'}}E_{k'}^{3}\psi _{i',k'}^{0}\right) \nonumber \\
R^{1}_{c} & = & -\chi _{i',k'}(\rho _{i'},T_{i'}^{*},Y_{e,i'}^{*})\psi _{i',k'}^{1}.\label{eq_diffusion_limit_fd} 
\end{eqnarray}
 If we compare this to the analytic radiation momentum equation in
the diffusion limit,\begin{equation}
\label{eq_diffusion_limit_analytical}
4\pi r^{2}\frac{\partial }{\partial a}\left( \rho K^{N}(E)\right) +4\pi r^{2}\frac{1}{\alpha }\frac{\partial \alpha }{\partial a}\rho J^{N}(E)-\frac{\Gamma }{\alpha }\frac{\partial \alpha }{\partial r}\frac{\partial }{E^{2}\partial E}\left( E^{3}K^{N}(E)\right) =-\chi H^{N}(E),
\end{equation}
where we have set \( J-3K=0 \), we can identify the physical meaning
of all remaining terms. The diffusive flux \( \psi ^{1} \) appearing
in \( R^{1}_{c} \) is determined by the opacity \( \chi  \) and
operators acting on the particle density. The term in \( R^{\prime 0}_{a} \)
describes the flux generated by a gradient in the particle density.
It is the only term that survives in the Newtonian limit. In full
general relativity, we have an additional gravitational force acting
on the radiation particles. It contributes to the particle flux and
is described in the term \( R_{\mu }^{\prime 0} \). Its analytical
analogue is \( R_{\mu }^{\prime 0}=3\left( \Gamma /\alpha \right) \left( \partial \alpha /\partial r\right) \psi ^{0} \).
The factor of three stems from our choice \( 3\psi ^{0}=3K=J \). 

\begin{figure}
{\centering \resizebox*{0.7\textwidth}{!}{\includegraphics{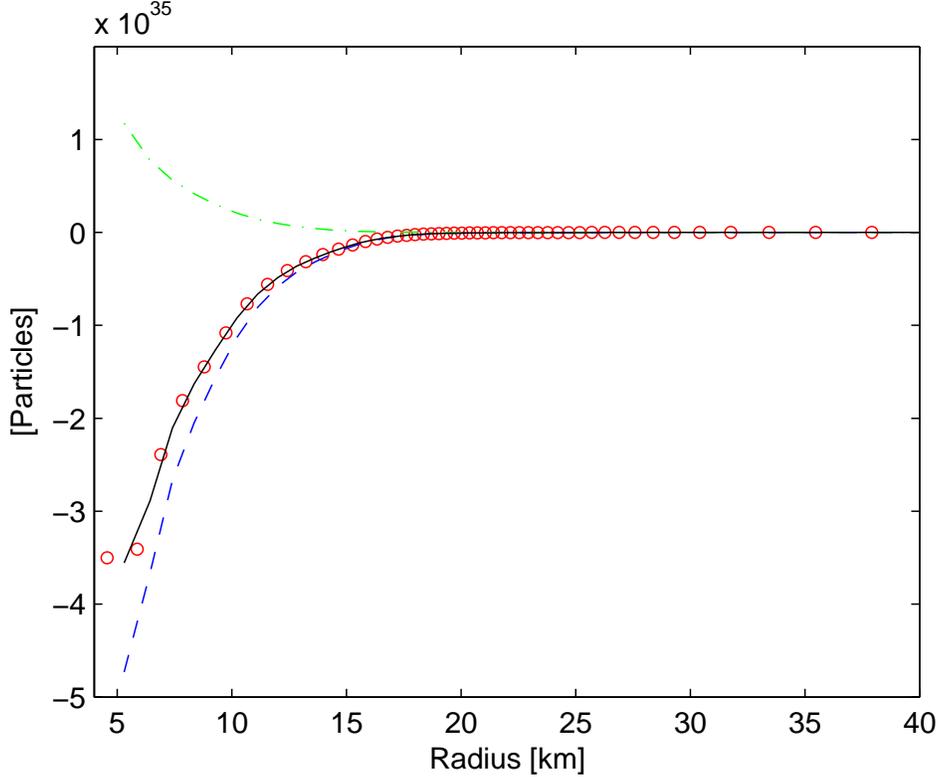}} \par}

\caption{The sum of the terms \protect\( R_{a}^{\prime 0}\protect \) and
\protect\( R_{\mu }^{\prime 0}\protect \) in Eq. (\ref{eq_diffusion_limit_fd})
determines the diffusive flux in the discrete Boltzmann equation.
We check these terms in a stationary-state situation at \protect\( 400\protect \)
ms after bounce in the \protect\( 40\protect \) M\protect\( _{\odot }\protect \)
model. Plotted are the contributions from \protect\( R_{a}^{\prime 0}\protect \)
(dashed line) and \protect\( R_{\mu }^{\prime 0}\protect \) (dash-dotted
line) as determined in Eq. (\ref{eq_gravitational_diffusion_fd})
from our finite differencing of the Boltzmann equation. Their sum
(solid line) is compared with a straightforward finite difference
representation of the same quantity (circles) according to Eq. (\ref{eq_natural_diffusion_source_fd})
(The circle at \protect\( 5\protect \) km radius is the zone edge
of our innermost zone, the other quantities are evaluated on zone
centers). The agreement demonstrates that we obtain an accurate diffusion
flux in the numerical solution of the Boltzmann equation.\label{fig_diffusioncheck.ps}}
\end{figure}

The term \( R_{E}^{0} \) corrects for the fact that the gradient
in \( R_{a}^{\prime 0} \) has been taken at constant local particle
energy instead of constant particle energy at infinity. The terms
\( R_{a}^{\prime 0} \) and \( R_{c}^{1} \) emerge in a perfect way
from our finite difference representation because of the special measures
taken in the finite difference representation of the propagation terms
\( D_{a} \) and \( D_{\mu } \) in \citep{Mezzacappa_Bruenn_93a}.
However, the requirement of global energy conservation ties the finite
difference representation of this basic choice to many other terms,
as outlined in table (\ref{eq_cancellation_table}). These constraints
impose a rather obscure finite differencing in the diffusive limit
for the general relativistic corrections \( R_{\mu }^{\prime 0} \)
and \( R_{E}^{0} \). However, each term is recognizable as a finite
difference representation of a physical term. If we integrate the
corrections over energy, the contributions from \( R_{E}^{0} \) are
guaranteed to vanish. We therefore focus on the remaining correction
\( R_{\mu }^{\prime 0} \). In order to ensure a vanishing flux in
equilibrium, it remains to be shown that the results from our finite
difference representation do not significantly deviate from a natural
finite difference representation of Eq. (\ref{eq_diffusion_limit_analytical}),
which becomes after an integration over energy,\begin{equation}
\label{eq_diffusive_equilibrium_2}
\frac{4\pi r^{2}}{\alpha ^{3}}\frac{\partial }{\partial a}\left( \alpha ^{3}\rho K^{N}\right) =-\chi H^{N}.
\end{equation}
To this purpose, we collect from Eq. (\ref{eq_diffusion_limit_fd})
the finite difference representations\begin{eqnarray}
\frac{\rho _{i'}}{\Gamma _{i+1}}dr_{i'}R_{a}^{\prime 0} & = & \frac{3}{4\pi \left( r_{i+1}^{2}+r_{i+1}r_{i}+r_{i}^{3}\right) \alpha _{i'}}\left\langle 4\pi r^{2}\alpha d\left\langle \rho \psi ^{0}\right\rangle \right\rangle _{i'}\nonumber \\
\frac{\rho _{i'}}{\Gamma _{i+1}}dr_{i'}R_{\mu }^{\prime 0} & = & \frac{3}{4\pi \left( r_{i+1}^{2}+r_{i+1}r_{i}+r_{i}^{3}\right) \alpha _{i'}}\left\langle 4\pi r^{2}\left\langle \rho \psi ^{0}\right\rangle d\alpha \right\rangle _{i'}\nonumber \\
 & + & \left( r_{i+1}-r_{i}\right) G_{i+1}^{\mu }2\frac{\rho _{i'}\psi _{i'}^{0}}{\Gamma _{i+1}}\label{eq_gravitational_diffusion_fd} 
\end{eqnarray}
and compare their sum with the natural finite difference representation
of Eq. (\ref{eq_diffusive_equilibrium_2}),\begin{equation}
\label{eq_natural_diffusion_source_fd}
\frac{1}{\alpha _{i}^{3}}d\left\langle \alpha ^{3}\rho \psi _{k'}^{0}\right\rangle _{i}.
\end{equation}
 The comparison is performed in a time slice in the evolution of the
\( 40 \) M\( _{\odot } \) progenitor of \citep{Woosley_Weaver_95}
at \( 400 \) ms after bounce. We force the hydrodynamics to be static
and evaluate the finite difference expressions (\ref{eq_gravitational_diffusion_fd})
and (\ref{eq_natural_diffusion_source_fd}) in the high density region,
where the diffusion limit is valid and gravitational terms sizeable.
We find in Figure (\ref{fig_diffusioncheck.ps})
that the terms stemming from \( R_{a}^{\prime 0} \) (dashed line)
and \( R_{\mu }^{\prime 0} \) (dash-dotted line) compose a sum (solid
line) that agrees well with the more accurate finite difference representation
given in Eq. (\ref{eq_natural_diffusion_source_fd}). Thus, we conclude
that the opacities safely determine the diffusive flux in our finite
difference representation of the Boltzmann equation.

\subsection{Redshift, gravitational bending, and the evolution of angular moments}\label{subsection_anglecheck}

In this subsection we test free streaming in spherically symmetric
geometry. This is the opposite limit to the diffusion investigated
in the previous subsection. For the time being, we stay with our stationary
time slice at \( 400 \) ms after bounce, but focus on radii larger
than \( 40 \) km. Outside this radius, we switch all interactions
between the radiation field and matter off. This facilitates the comparison
of the neutrino density, neutrino number flux, and neutrino luminosity
with the analytical behavior. Unlike in the Newtonian limit, they
are subject to time lapse, gravitational frequency shift, and gravitational
bending. 

\begin{figure}
{\centering \resizebox*{0.7\textwidth}{!}{\includegraphics{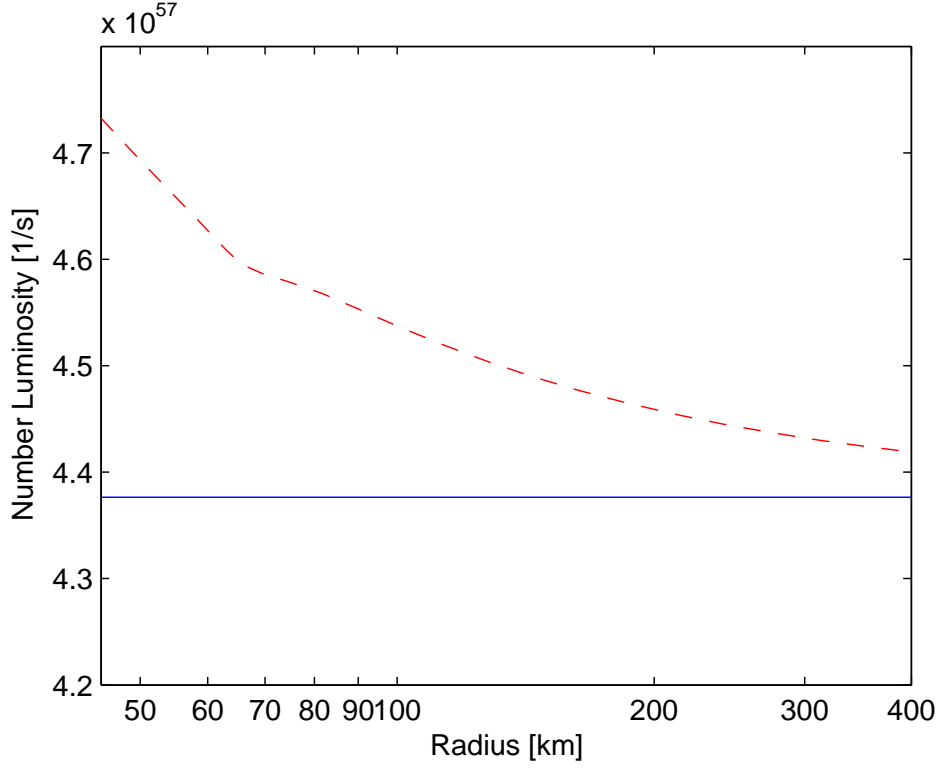}} \par}

\caption{Shown are the radial variations in the locally observed neutrino
number luminosity \protect\( 4\pi r^{2}\rho H^{N}\protect \) (dashed
line) and the conserved number luminosity \protect\( 4\pi r^{2}\alpha \rho H^{N}\protect \)
(solid line) in a stationary-state situation at \protect\( 400\protect \)
ms after bounce in the \protect\( 40\protect \) M\protect\( _{\odot }\protect \)
model. Eq. (\ref{eq_check_number_conservation}) is fullfilled by
construction.\label{fig_fluxcheck.ps}}
\end{figure}

\begin{figure}
{\centering \resizebox*{0.8\textwidth}{!}{\includegraphics{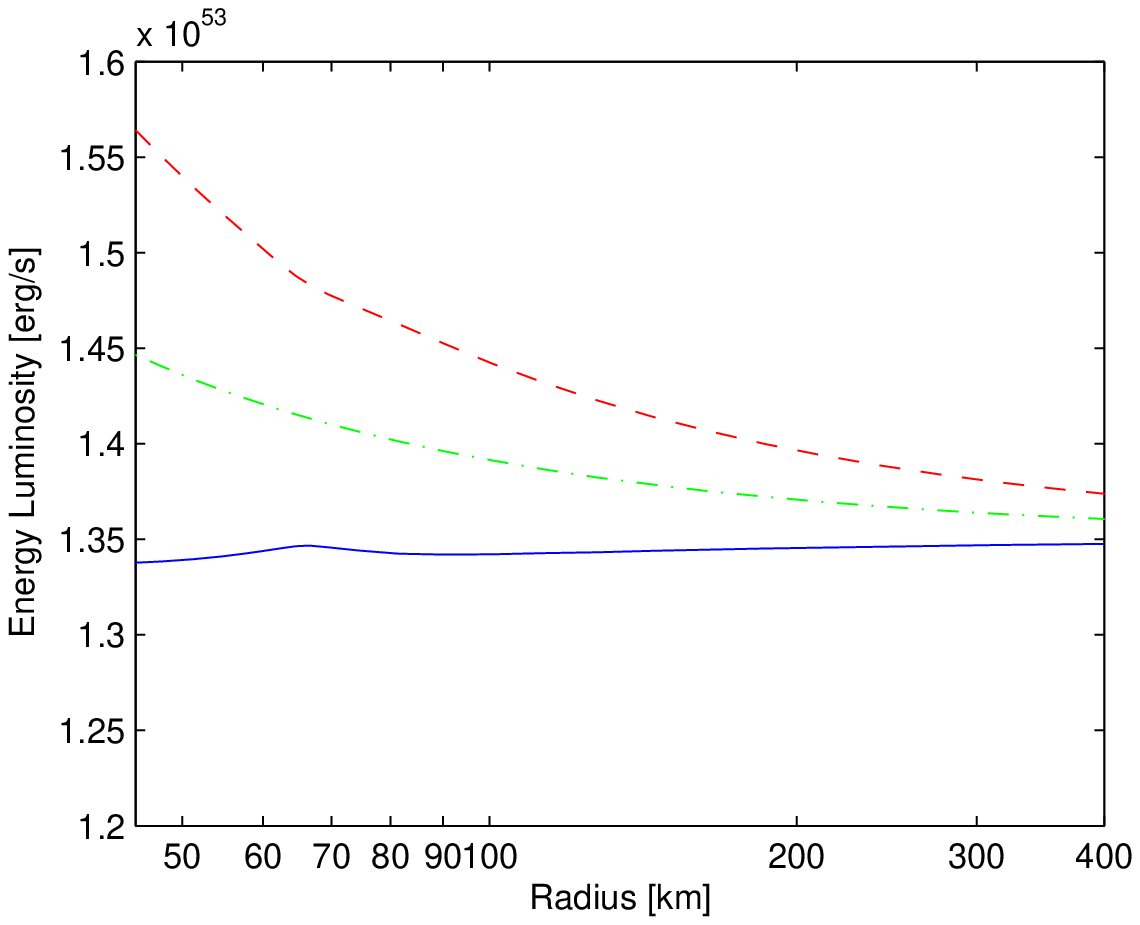}} \par}

\caption{Shown are the energy luminosities \protect\( 4\pi r^{2}\rho H\protect \)
(dashed line), \protect\( 4\pi r^{2}\alpha \rho H\protect \) (dash-dotted
line), and \protect\( 4\pi r^{2}\alpha ^{2}\rho H\protect \) (solid
line) as functions of radius in a stationary-state situation at \protect\( 400\protect \)
ms after bounce in the \protect\( 40\protect \) M\protect\( _{\odot }\protect \)
model. Deviations from Eq. (\ref{eq_check_redshift}) do not exceed
\protect\( 1.5\%\protect \) in the third expression for the luminosity,
which is expected to be conserved. The other two expressions illustrate
that the gravitational effects contribute substantially in this test.\label{fig_lumincheck.ps}}
\end{figure}

For a stationary neutrino flux in the free streaming limit, Eq. (\ref{eq_neutrino_number_conservation})
expresses particle number conservation\begin{equation}
\label{eq_check_number_conservation}
\frac{\partial }{\partial a}\left[ 4\pi r^{2}\alpha \rho H^{N}\right] =0.
\end{equation}
Figure (\ref{fig_fluxcheck.ps})
shows the radial dependence of the locally observed neutrino number
luminosity \( 4\pi r^{2}\rho H^{N} \) (dashed line) and the conserved
number luminosity \( 4\pi r^{2}\alpha \rho H^{N} \) (solid line).
We find that the latter is constant to machine precision as required
by Eq. (\ref{eq_check_number_conservation}). This is the direct result
of our finite differencing of the term \( D_{a} \), because we have
placed the lapse function \( \alpha  \) inside the space derivative.
The lapse function in the bracket converts the locally advected particle
number per proper time to the particle advection per global coordinate
time. There is a similar equation for the conservation of the energy
luminosity. Eq. (\ref{eq_radiation_internal_energy}) for the particle
energy carries a second lapse function inside the spatial derivative
from the gravitational frequency shift,\begin{equation}
\label{eq_check_redshift}
\frac{\partial }{\partial a}\left[ 4\pi r^{2}\alpha ^{2}\rho H\right] =0.
\end{equation}
Figure (\ref{fig_lumincheck.ps})
 shows the radial dependence of the locally observed luminosity \( 4\pi r^{2}\rho H \)
(dashed line), the luminosity with the time lapse correction only,
\( 4\pi r^{2}\alpha \rho H \) (dash-dotted line), and the conserved
luminosity, \( 4\pi r^{2}\alpha ^{2}\rho H \), with the gravitational
redshift included. We see again, that the latter is fairly constant
in the interaction free region. Unlike the number luminosity, the
conservation of energy luminosity is not automatically fulfilled by
the finite difference representation, and small deviations can be
observed. As the local mean energies of the moving particles are determined
by the ratio of energy and number flux, we may conclude from the fulfillment
of Eqs. (\ref{eq_check_number_conservation}) and (\ref{eq_check_redshift})
that, in this regime, the gravitational energy shift of the particles
is accurately implemented.

\begin{figure}
{\centering \resizebox*{0.8\textwidth}{!}{\includegraphics{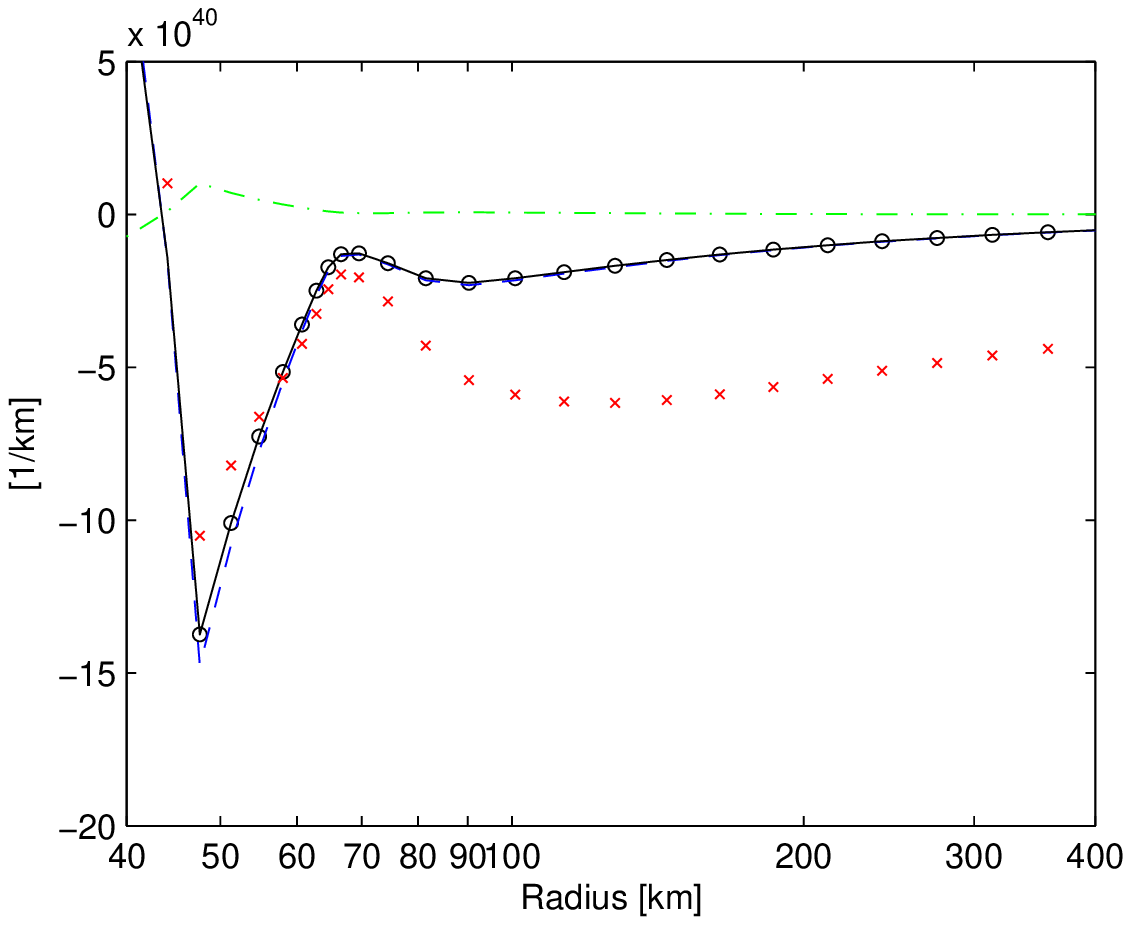}} \par}

\caption{This figure shows radial increments of the particle number density
(actually plotted is the difference in the particle density across
a zone, multiplied by \protect\( 4\pi r^{2}\protect \)). The particle
number luminosity is constant in the free streaming limit and the
changes in the particle density depends on the angular distribution
of the radiation field. The dashed line shows the changes induced
by the increasing distance to the neutrino source in Newtonian geometry.
The dash-dotted line shows the contribution from gravitational bending.
The sum (solid line) represents the right hand side of Eq. (\ref{eq_anglecheck_fd}).
It exactly matches the left hand side (circles). This indicates a
correct derivation and implementation. However, the crosses mark a
more accurate solution to the right hand side of Eq. (\ref{eq_anglecheck_analytic})
than the finite differencing imposed by the discrete Boltzmann equation.
It has been evaluated according to Eq. (\ref{eq_anglecheck_natural_fd}).
The discrepancy is moderate at flux factors \protect\( h<0.75\protect \)
and large far from the source where \protect\( h\protect \) should
converge to \protect\( 1\protect \).\label{fig_anglecheck.ps}}
\end{figure}

In order to fully constrain the redshift, gravitational bending, and
evolution of the angular moments, we need to test the radial dependence
of a third quantity. Most appropriate is the particle number density.
However, some special care is necessary to clearly separate the following
three effects on the angular neutrino distribution: (i) gravitational
bending, (ii) the changing opening angle as a function of the geometric
distance from the source, (iii) the numerical bending caused by limited
angular resolution in our finite difference method. We find an analytical
expression for the radial dependence of the particle density by applying
the operator \( \int dEd\mu E^{2}\mu ^{-1} \) to the interaction-free
stationary state limit of the Boltzmann equation (\ref{eq_relativistic_boltzmann}),\begin{eqnarray}
\frac{\partial }{\alpha \partial a}\left( 4\pi r^{2}\alpha \rho J^{N}\right) +\Gamma \left( \frac{1}{r}-\frac{1}{\alpha }\frac{\partial \alpha }{\partial r}\right)  & \times  & \nonumber \\
\left( 4\int F(\mu =0)E^{2}dE-\int \left[ F-F(\mu =0)\right] E^{2}dE\frac{1-\mu ^{2}}{\mu ^{2}}d\mu \right)  & = & 0.\label{eq_anglecheck_analytic} 
\end{eqnarray}
As before, we may compare this to our finite difference representation.
The application of the operator \( \sum dE_{k'}w_{j'}E_{k'}^{2}\mu _{j'}^{-1} \)
to the finite difference representation of the Boltzmann equation
leads to\begin{eqnarray}
\left( 4\pi r_{i+1}^{2}\alpha _{i+1}\rho _{i+1}J^{N}_{i+1}-4\pi r_{i}^{2}\alpha _{i}\rho _{i}J_{i}^{N}\right)  & = & \nonumber \\
-\alpha _{i'}\Upsilon _{i+1}\sum F_{i',j,k'}da_{i'}E_{k'}^{2}dE_{k'}\frac{\zeta _{j}}{\mu _{j'}\mu _{j'-1}}\left( \mu _{j'}-\mu _{j'-1}\right) . &  & \label{eq_anglecheck_fd} 
\end{eqnarray}
Here, \( \alpha _{i}\rho _{i}J_{i}^{N} \) is interpolated according
to Eq. (\ref{eq_Fi_interpolation}) and \( F_{i',j,k'} \) according
to Eq. (\ref{eq_Fj_interpolation}). Figure (\ref{fig_anglecheck.ps}) 
compares the left hand side of Eq. (\ref{eq_anglecheck_fd}) (circles)
to the right hand side (solid line), evaluated in our \( 400 \) ms
post bounce time slice. The match is to machine precision; it demonstrates
that we made no errors in the derivation of Eq. (\ref{eq_anglecheck_fd})
or its implementation. We split this quantity into the contribution
from geometric changes according to the opening angle of the source,
proportional to \( \Gamma /r \), (dashed line) and the gravitational
bending, proportional to \( \left( \Gamma /\alpha \right) \left( \partial \alpha /\partial r\right)  \),
(dash-dotted line). Because the latter is much smaller, we exclude
the gravitational bending contribution to \( \Upsilon  \) as a primary
source of uncertainties. We can then focus on the comparison of the
analytical integral in Eq. (\ref{eq_anglecheck_analytic}) with its
finite difference representation in Eq. (\ref{eq_anglecheck_fd}).
We start with a more accurate numerical evaluation of the integral
in Eq. (\ref{eq_anglecheck_analytic}) in order to compare with Eq.
(\ref{eq_anglecheck_fd}) in our time slice. The application of a
Gaussian quadrature in Eq. (\ref{eq_anglecheck_analytic}) converts
the integral to the sum \begin{equation}
\label{eq_anglecheck_natural_fd}
-\alpha _{i'}\Upsilon _{i+1}\left( 4\sum F(0)_{i',k'}E_{k'}^{2}dE_{k'}-\sum \left[ F_{i',j',k'}-F(0)_{i',k'}\right] da_{i'}E_{k'}^{2}dE_{k'}\frac{1-\mu _{j'}^{2}}{\mu _{j'}^{2}}w_{j'}\right) 
\end{equation}
The result of Eq. (\ref{eq_anglecheck_natural_fd}) depends quite
sensitively on the choice of the distribution function in the tangential
direction, \( F\left( 0\right) _{i',k'} \). We determine it here
by the maximum entropy model for Maxwell-Boltzmann statistics in angle
space, \begin{equation}
\label{eq_analytical_angular_f}
F_{j'}=A\exp \left( B\mu _{j'}\right) ,
\end{equation}
 (see e.g. \citep{Smit_VanDenHorn_Bludman_00} for an overview on maximum
entropy closures). The zeroth and second moments of Eq. (\ref{eq_analytical_angular_f})
define the flux factor, \( h=H/J \), as a function of the parameters
\( A \) and \( B \),%
\footnote{Note that our definition of moments in Eq. (\ref{eq_number_moments_definition})
is twice as large as in standard references because this leads to
more consistency between our code and its description.
} \begin{eqnarray}
J & = & \frac{2A}{B}\sinh (B)\nonumber \\
h & = & \coth (B)-\frac{1}{B},\label{eq_max_entropy_relations} 
\end{eqnarray}
We derive the free parameters \( A \) and \( B \) by the inversion
of Eq. (\ref{eq_max_entropy_relations}) from the zeroth and first
angular moments of the numerically obtained neutrino distribution
function and define \( F(0)_{i',k'}=A_{i',k'} \). The result is shown
with cross markers in Fig. (\ref{fig_anglecheck.ps}). We observe
comparable values between the two finite difference representations
(\ref{eq_anglecheck_fd}) and (\ref{eq_anglecheck_natural_fd}) at
smaller radius, where the flux factor is smaller (\( \sim 0.75 \)).
Large differences are found at larger radii, where the flux factor
would be expected to be close to one. I.e., the primary source of
inaccuracy in the particle density stems from the fact that the sum
over angles on the right hand side of the finite difference Eq. (\ref{eq_anglecheck_fd})
is a poor representation of the angle integral in the analytic Eq.
(\ref{eq_anglecheck_analytic}).

In order to study this discrepancy in more detail, we construct a
series of analytical particle distribution functions according to
the maximum entropy model in Eq. (\ref{eq_analytical_angular_f})
with \( B=\left\{ 0,1,2,5,10\right\}  \). This corresponds to flux
factors of \( h=\left\{ 0,0.31,0.54,0.8,0.9\right\}  \). Then, we
evaluate the integral on the right hand side of Eq. (\ref{eq_anglecheck_fd})
in different ways. First, as given by the finite differencing in Eq.
(\ref{eq_anglecheck_fd}) itself, then, with the more natural Gauss
quadrature in Eq. (\ref{eq_anglecheck_natural_fd}), and finally by
an analytic integration. The analytic integration yields\begin{eqnarray}
\frac{1}{J}\int _{-1}^{1}\frac{1}{\mu }\frac{\partial }{\partial \mu }\left[ \left( 1-\mu ^{2}\right) A\exp \left( B\mu \right) \right] d\mu  & = & \nonumber \\
-2-Bh+\frac{B^{2}}{\sinh (B)}\left( B+\frac{B^{3}}{3\cdot 3!}+\frac{B^{5}}{5\cdot 5!}+\frac{B^{7}}{7\cdot 7!}+\ldots \right) . &  & \label{eq_analytical_angle_integral} 
\end{eqnarray}
The result for different flux factors and angular resolutions is shown
in Table (\ref{tab_fluxfactor_evolution}).
\begin{table}
\caption{Integral on the right
hand side of Eq. (\ref{eq_anglecheck_analytic}).}
\label{tab_fluxfactor_evolution}
\begin{center}
\begin{tabular}{|c|c|c|c|c|c|c|}
\hline 
&
B&
flux factor
\protect\footnote{This table investigates the accuracy of the integral on the right
hand side of Eq. (\ref{eq_anglecheck_analytic}) in a parameter study
based on analytic distribution functions with different flux factors.
The result of the numerical integrations are given for discretizations
with \protect\( j_{\rm max}=6\protect \), \protect\( j_{\rm max}=12\protect \),
and \protect\( j_{\rm max}=48\protect \) angular bins. The uppermost
block shows the evaluation according to Eq. (\ref{eq_anglecheck_fd}).
It emerges from the chosen finite differencing of the term \protect\( D_{\mu }\protect \)
in Eq. (\ref{eq_dmu_fd}) and converges only slowly to the exact solution.
The center block shows the same evaluation without upwind differencing
for the advected \protect\( F\protect \) (i.e. with \protect\( \gamma _{i',k'}=0.5\protect \)).
The convergence is much better in this case. The lowermost block shows
the evaluation based on the more natural Gauss quadrature (\ref{eq_anglecheck_natural_fd}),
which converges very rapidly to the exact result.}&
j\( _{\rm max} \)=6&
j\( _{\rm max} \)=12&
j\( _{\rm max} \)=48&
analytical\\
\hline
Eq. (\ref{eq_anglecheck_fd})&
0&
0&
-2.00&
-2.00&
-2.00&
-2.00\\
\hline 
&
1&
0.31&
-1.10&
-1.23&
-1.36&
-1.41\\
\hline 
&
2&
0.54&
-0.21&
-0.22&
-0.28&
-0.32\\
\hline 
&
5&
0.80&
0.28&
0.47&
0.68&
0.77\\
\hline 
&
10&
0.90&
0.06&
0.14&
0.26&
0.31\\
\hline
\hline 
Eq. (\ref{eq_anglecheck_fd}),&
0&
0&
-2.00&
-2.00&
-2.00&
-2.00\\
\hline 
but no&
1&
0.31&
-1.48&
-1.43&
-1.41&
-1.41\\
\hline 
upwind&
2&
0.54&
-0.47&
-0.36&
-0.32&
-0.32\\
\hline 
diff.&
5&
0.80&
0.66&
0.74&
0.77&
0.77\\
\hline 
&
10&
0.90&
0.27&
0.30&
0.31&
0.31\\
\hline
\hline 
Eq. (\ref{eq_anglecheck_natural_fd})&
0&
0&
-2.00&
-2.00&
-2.00&
-2.00\\
\hline 
&
1&
0.31&
-1.41&
-1.41&
-1.41&
-1.41\\
\hline 
&
2&
0.54&
-0.32&
-0.32&
-0.32&
-0.32\\
\hline 
&
5&
0.80&
0.77&
0.77&
0.77&
0.77\\
\hline 
&
10&
0.90&
0.32&
0.31&
0.31&
0.31\\
\hline
\end{tabular}
\end{center}
\end{table}
We find quite poor convergence for the evaluation of the integral
as it emerges from the finite differencing in the Boltzmann equation.
The comparison with the center block, where no upwind differencing
was used for the advection terms, demonstrates that the advective
terms play a major r\^ole in the accuracy of the representation of
this integral. A forward peaked radiation field shows a steep gradient
in the angular distribution. The choice of upwind differencing always
underestimates the advective flux at the edge of the angular zone.
This is necessary for the stability of the scheme for arbitrary large
time steps, but can lead to a substantial underestimation of the integral
in Eq. (\ref{eq_anglecheck_analytic}). The most populated forward
bin, for example, never contributes to the integral. The distribution
function asymptotically approaches a highly populated forward bin
while all other angular bins are emptied. The maximum achievable flux
factor is smaller than one. Although higher angular resolution improves
the situation, it does not eliminate this systematic effect. The choice
of a higher order advection scheme in angular space might be promising.
More rigorous, however, appears the implementation of an adaptive
angular grid in the transparent regime \citep{Yamada_Janka_Suzuki_99}.
It would be designed to minimize the advective flow in angle space
such that the undesired effects are eliminated at the root. This difficulty
with angular advection does not occur in methods that solve a model
Boltzmann equation along particle characteristics to close the system
of radiation moments equations with a variable Eddington factor \citep{Burrows_et_al_00,Rampp_Janka_02}.
An independent and much smaller source of inaccuracy stems from the
fact that the integral in Eq. (\ref{eq_anglecheck_fd}) is not written
as a sum over a function evaluated at Gaussian quadrature points and
multiplied with Gaussian weights. Hence, it does not take profit from
the fast convergence of Gaussian quadrature. This can be seen if the
result is compared with the evaluation based on the Gaussian quadrature
as shown in the third block in Table (\ref{tab_fluxfactor_evolution}).

In our supernova application, however, stationary-state convergence
tests have shown that \( 6 \) angle Gaussian quadrature produces
physically reasonable results \citep{Messer_et_al_98,Messer_00}. We
confirm this in section \ref{subsection_resolution} in a comparison
with a dynamical run discretized with \( 12 \) angular bins. In the
regions of free streaming, where numerical angular diffusion becomes
apparent, the neutrino field is already decoupled from the matter
and does not influence the dynamical evolution anymore.

\subsection{Observer corrections}

\begin{figure}
{\centering \resizebox*{0.8\textwidth}{!}{\includegraphics{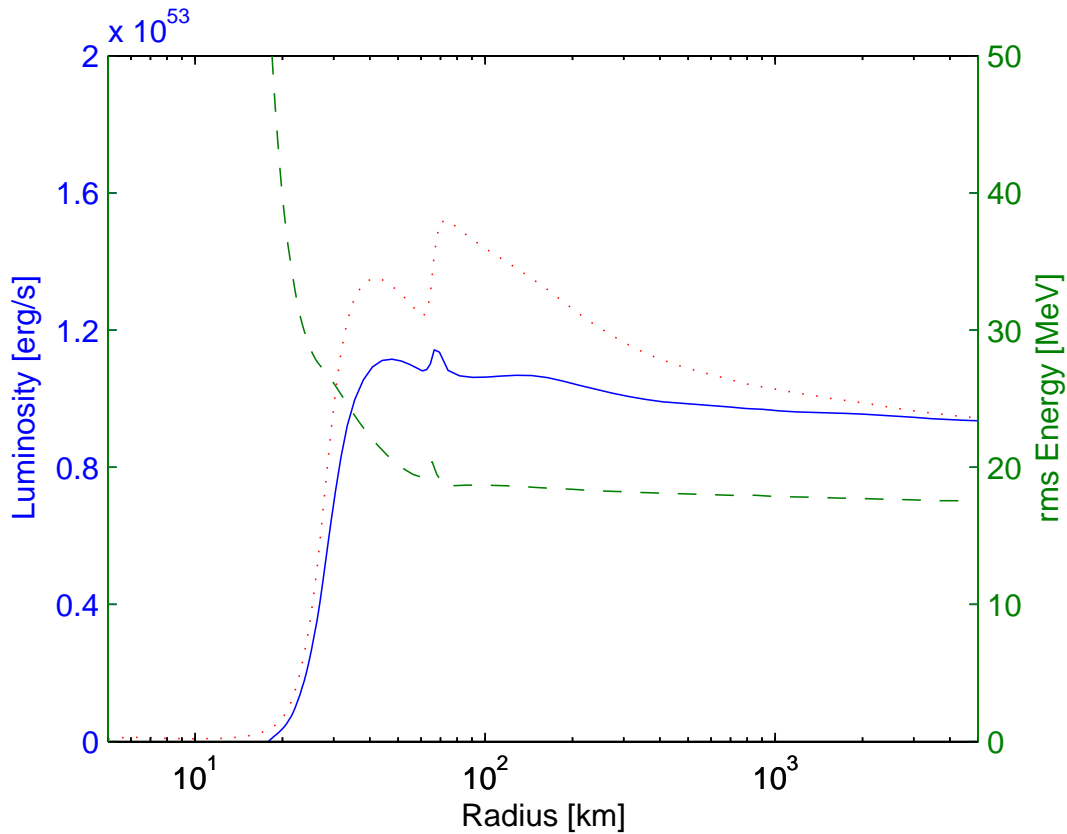}} \par}

\caption{Shown are the electron neutrino luminosity (solid line) and rms energy
(dashed) line after the transformation from the comoving frame to
the rest frame of a distant observer in the evolution of the \protect\( 40\protect \)
M\protect\( _{\odot }\protect \) model at \protect\( 400\protect \)
ms after bounce. Aside of a local bump, the two profiles show the
expected continuous transition at the shock front. Also shown is the
luminosity in the comoving frame (dotted line) to illustrate the effect
of the transformation.\label{fig_observer.ps}}
\end{figure}

For the check of the observer corrections, we go back to the original
solution at \( 400 \) ms after bounce. Both, the neutrino luminosities
and rms energies show a discontinuity across the shock front because
of the Doppler effect and angular aberration. In the laboratory frame
of a distant observer, however, the radiation field should not be
affected by the shock. We may therefore test the implementation of
the observer corrections by transforming the electron neutrino luminosity
and rms energy from our comoving frame to the rest frame. In \citep{Liebendoerfer_Mezzacappa_Thielemann_01}
we determined for the particle propagation angle, \( \mu ^{S} \),
and particle energy, \( E^{S} \), in Schwarzschild coordinates the
relationship\begin{eqnarray}
\left( 1-\mu ^{S2}\right) \left( \Gamma ^{S}\right) ^{-2} & = & \left( 1-\mu ^{2}\right) \left( \Gamma +u\mu \right) ^{-2}\nonumber \\
\Gamma ^{S}E^{S} & = & \left( \Gamma +u\mu \right) E,\label{eq_schwarzschild_transformation} 
\end{eqnarray}
where \( \Gamma ^{S}=\left( 1-2m/r\right) ^{1/2} \). Assuming free
outwards streaming with transport coefficients \( \beta =1 \) in
the interesting region around the shock, we evaluate for the luminosity,
\( L^{S} \), and rms energy, \( \left\langle E^{S}\right\rangle  \),
the finite difference expression\begin{eqnarray}
L_{i}^{S} & = & 4\pi r_{i}^{2}\frac{2\pi }{\left( hc\right) ^{3}}\sum _{j,k}\rho _{i'-1}F_{i'-1,j,k}E_{i,j',k'}^{S}E_{k'}^{2}dE_{k'}\mu ^{S}_{i',j'}cw_{j'}\nonumber \\
\left\langle E^{S}\right\rangle _{i'} & = & \left( \sum _{j,k}\left( E^{S}_{i,j',k'}\right) ^{2}F_{i',j',k'}E_{k'}^{2}dE_{k'}w_{j'}\right) ^{\frac{1}{2}}\left( \sum _{j,k}F_{i',j',k'}E_{k'}^{2}dE_{k'}w_{j'}\right) ^{-\frac{1}{2}}.\label{eq_schwarzschild_transformation_fd} 
\end{eqnarray}
The result is shown in Figure (\ref{fig_observer.ps}).
The solid line is the luminosity in the rest frame according to Eq.
(\ref{eq_schwarzschild_transformation_fd}). The dotted line shows
the luminosity in the comoving frame for comparison. We find that
the luminosity profile shows a numerical local distortion on top of
the shock front. However, if we compare the left hand side values
with the right hand side values, we find the expected disappearance
of the discontinuity, indicating that the observer corrections in
the comoving frame are well represented. The same behavior can be
seen in the rms energies.

\subsection{Radiation pulse propagation}\label{subsection_pulse_propagation}

\begin{figure}
{\centering \resizebox*{0.8\textwidth}{!}{\includegraphics{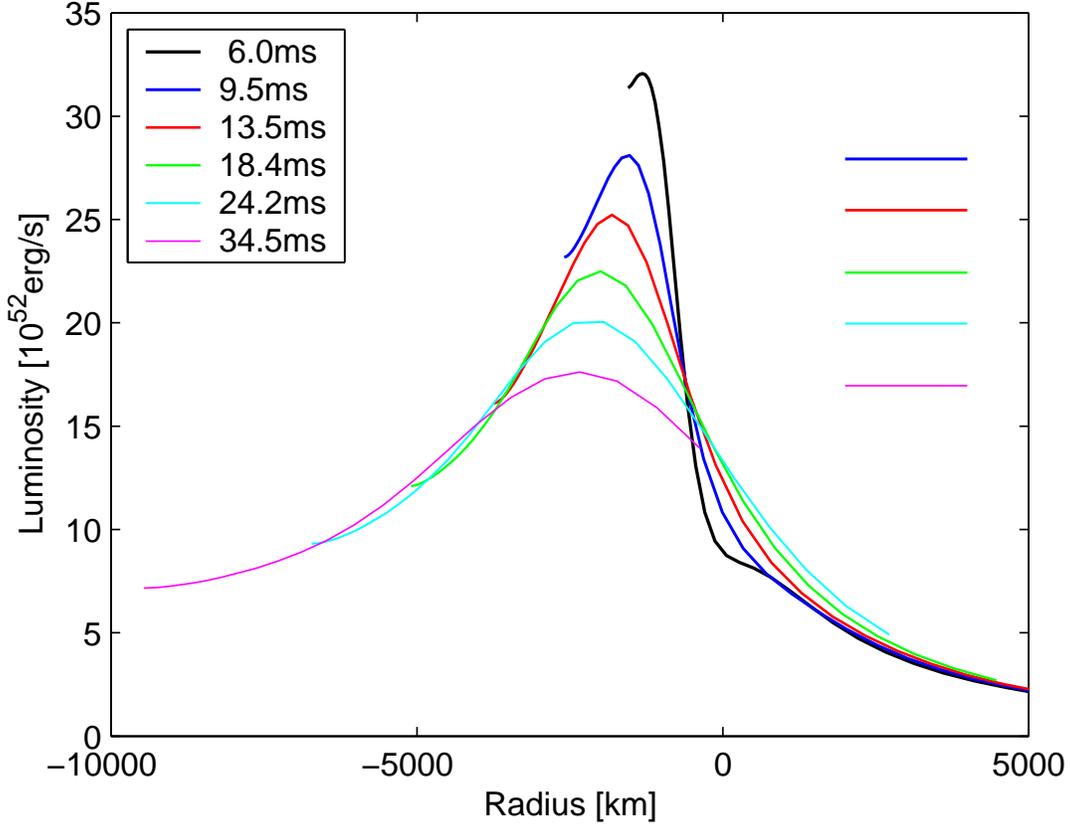}} \par}

\caption{Shown is the radial profile of the electron neutrino luminosity at
different times when the neutrino burst propagates through the star
(\protect\( 40\protect \) M\protect\( _{\odot }\protect \) model).
The radial coordinate of each time slice has been shifted to the left
by the distance a signal with light speed would have traveled since
bounce. This makes the pulse profile roughly stationary in position.
The levels on the right hand side in the graph give estimates of the
peak height for the next time slice. The evaluation is based on the
estimated numerical diffusion from first order upwind differencing
in the free streaming regime in the transport equation. The expected
peak heights accurately explain the visible decay in the luminosity
pulse.\label{fig_pulse.ps}}
\end{figure}

We have now performed many checks of the particle transport in stationary-state
situations. In this subsection, we check the dynamics of the radiation
field itself. The neutrino transport in the supernova is most dynamic
when the shock crosses the neutrinosphere. As soon as the shock has
propagated to densities where the opacities are low enough that neutrinos
can escape from the hot material behind the shock, immediate electron
capture will occur to refill the emptied phase space with new electron
neutrinos. A neutrino burst carries an energy of order \( 10^{51} \)
erg throughout the star towards the distant observer. Although the
neutrino burst is not completely interaction free, it is an excellent
example for a time-dependent radiation field where pulse propagation
can be studied. We do this by plotting the luminosity profiles at
selected times after the launch of the neutrino burst. We shift the
radial coordinate, \( r \), in each time slice by the amount a free
massless neutrino would have propagated during the time \( t \) since
bounce, \( r'=r-ct \). With an ideal numerical solution to free propagation
(and assuming a point source), we would expect congruent pulse shapes
from each time slice at the same positions \( r' \). Figure (\ref{fig_pulse.ps})
 shows the neutrino burst with the described radius adjustment in
the evolution of the \( 40 \) M\( _{\odot } \) progenitor. We first
observe, that the position of the pulse is fairly stationary. Therefore,
the pulse indeed travels with speed of light through the star. However,
we also observe that the shape of the pulse broadens considerably
with ongoing time. From the check of the neutrino number luminosity
conservation in Fig. (\ref{fig_fluxcheck.ps}), we already know that
the number and energy of the emitted neutrinos is conserved during
their propagation to larger radii. The broadening is likely to occur
by artificial diffusion in our numerical finite differencing scheme.
Indeed, from considerations in \citep{Liebendoerfer_Rosswog_Thielemann_02}
we remember that first order upwind differencing introduces a numerical
diffusivity proportional to the advection speed and zone width,\begin{equation}
\label{eq_numerical_diffusivity}
D_{i'}=cdr_{i'}.
\end{equation}
We quantify the influence of this effect to the evolution of our neutrino
pulse by approximating the diffusion between the time steps with \( \Delta L\simeq D\left( \partial ^{2}L/\partial r^{2}\right) \Delta t \).
We evaluate the diffusivity (\ref{eq_numerical_diffusivity}) and
the second derivative of the luminosity at the pulse peak in the first
time slice. The time span \( \Delta t \) to the next time slice gives
us an estimate \( \Delta L \) for the change of the peak luminosity.
We subtract \( \Delta L \) from the pulse height and draw a horizontal
line on the right hand side in Fig. (\ref{fig_pulse.ps}) at this
estimated peak level for the next time slice. The close to perfect
agreement with the numerically evolved pulse height leads to the conclusion
that the diffusivity (\ref{eq_numerical_diffusivity}) introduced
with the low order advection scheme fully accounts for the observed
pulse broadening. The resolution study in the next section illustrates
that this is the dominant effect.

\subsection{Resolution}\label{subsection_resolution}

\begin{figure}
{\centering \resizebox*{0.95\textwidth}{!}{\includegraphics{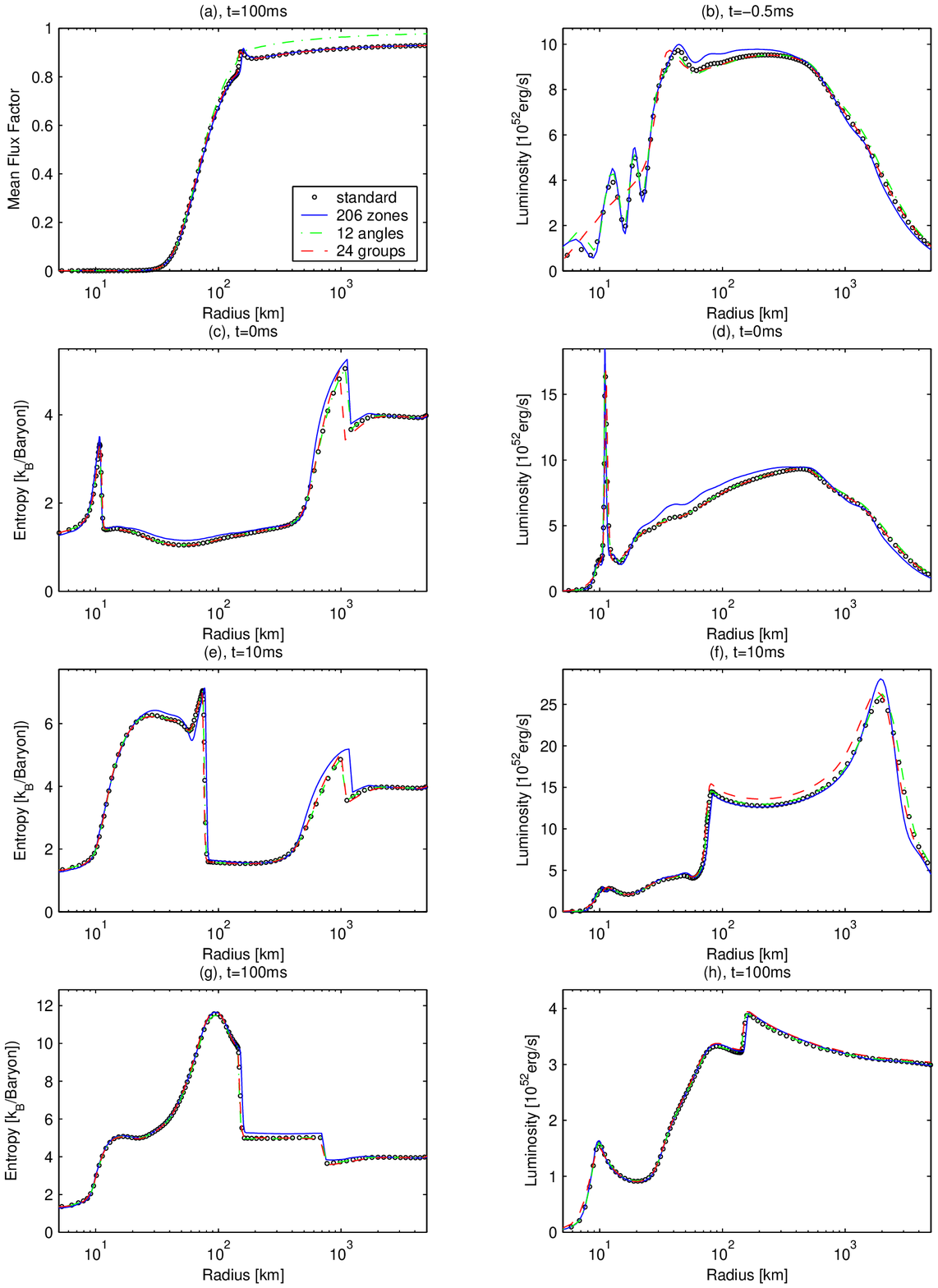}} \par}

\caption{Resolution dependence of the results. The standard run (circles)
was calculated with \protect\( 103\protect \) adaptive zones, \protect\( 6\protect \)
angular bins, and \protect\( 12\protect \) energy groups. We compare
with runs with \protect\( 206\protect \) zones (solid line), \protect\( 12\protect \)
angular bins (dash-dotted line), and \protect\( 24\protect \) energy
groups (dashed line).\label{fig_resolution}}
\end{figure}

In spite of our ambition to produce accurate results already at low
resolution, the reader has been left wondering about the actual resolution
dependence of our data. In this subsection, we demonstrate reasonable
convergence of our results by separately doubling the resolution in
each phase space dimension in otherwise identical simulations. Historically,
our standard resolution with \( 103 \) adaptive zones, \( 6 \) angular
bins, and \( 12 \) energy groups formed as the minimum resolution
where we felt safe about our physical conclusions. In Fig. (\ref{fig_resolution})
we compare the results with a simulation that uses \( 206 \) adaptive
zones, one that uses \( 12 \) angular bins, and one that uses \( 24 \)
energy groups. We focus on the quantities and instances where we find
the largest deviations. We start with two limitations that are well-known.
Graph (a) shows the mean flux factor at \( 100 \) ms after bounce
in the neutrino heating phase. It can clearly be seen, that the run
with \( 12 \) angular bins determines the flux factor with more accuracy
in the free streaming limit \citep{Messer_et_al_98,Yamada_Janka_Suzuki_99}.
The angular resolution determines how close to the radial direction
the most forward peaked angular bins are. The closer they are the
larger is the asymptotic limit of the flux factor. In graph (b), we
show another previously documented effect \citep{Mezzacappa_Bruenn_93b}.
In the last half of a millisecond before bounce, insufficient resolution
of the energy phase space leads to a poor representation of the Fermi
energy in the degenerate neutrino gas when the trapped neutrinos are
compressed to high densities. The consequence are rapid and transient
local displacements of neutrinos. At poor energy resolution, the Fermi
energies rather increase in steps than continuously, and the neutrino
fluxes try to balance the numerical variations in the Fermi energies
between adjacent zones. Therefore, the effect is best seen in the
electron neutrino luminosities as shown in graph (b). While the conserved
lepton number cannot show numerical variations, the electron fraction
and neutrino abundances also reflect the transient steps in the Fermi
energies. Variations in these quantities are of order \( 5\% \).
However, because the neutrinos are trapped and because our scheme
is conservative, this transient wiggles do not lead to any differences
that survive bounce.

This is shown in the two graphs (c) and (d) at bounce where the entropy
and luminosity profile, respectively, have again converged with respect
to energy resolution. At this stage, we detect an influence of the
spatial resolution. During collapse, there is no special region with
a high concentration of grid points. At bounce, however, the grid
points speed inwards to resolve the newborn shock wave. We detect
an effect of this rapid grid displacement in the entropy and electron
neutrino luminosity profiles to the extent shown in graphs (c) and
(d). Differences in the entropies can also be seen at the interface
between the silicon layer and nuclear statistical equilibrium at a
radius \( \sim 1000 \) km. The difference is of no concern because
it stems from the granular triggering of zone conversions from silicon
to NSE. Each run determines autonomously when output files are dumped.
For a given time, we then compare the output with the closest available
output files in other runs. Hence, it can happen that the conversion
of a zone took already place in one run, but not in another one. The
location of the transition to NSE shows an uncertainty of at least
one zone width. The first \( 10 \) ms after bounce are probably the
most dynamical time in the simulations. At this time we find again
the most prominent resolution dependencies in the entropy and luminosity
profiles. It is still a dependency on the space resolution alone.
The entropy in the high energy resolution run carries the slight enhancement
from before, overlapped by a slightly narrower and deeper cooling
at the launch of the neutrino burst. The higher luminosity peak in
graph (f) at higher spatial resolution comes not unexpectedly, as
we have seen in subsection \ref{subsection_pulse_propagation} that
the decay of the outwards propagating neutrino burst depends on the
zone width. We can also demonstrate by this resolution study, that
the described pulse spreading does only marginally depend on the angular
resolution. Note that the apparent difference in the run with higher
energy resolution is not a real difference, it stems from an insufficient
time match between the output files and shows the rapidly decaying
luminosity profile at a slightly earlier time. 

Finally, we compare the runs with different resolutions during the
important neutrino heating phase, e.g. at \( 100 \) ms after bounce.
Graphs (g) and (h) show again the entropy and luminosity profiles.
We are happy to report that the simulations are converged at this
stage. With the exception of the slightly higher entropy in the outer
layers of the high space resolution run, none of the previously discussed
differences has survived to this time. Moreover, the quality of agreement
presented in these graphs is similar to what we find in the velocity,
logarithmic density, electron fraction, and rms neutrino energy profiles
at any time during the simulations. We conclude that convergence issues
are far from affecting our physical conclusions from the simulations.
Only if one asks for high precision numbers in specific quantities,
we may, with descending importance, recommend an increase of space-,
energy-, and angle-resolution.

\subsection{Energy conservation}\label{subsection_energy_conservation}

\begin{figure}
{\centering \resizebox*{0.8\textwidth}{!}{\includegraphics{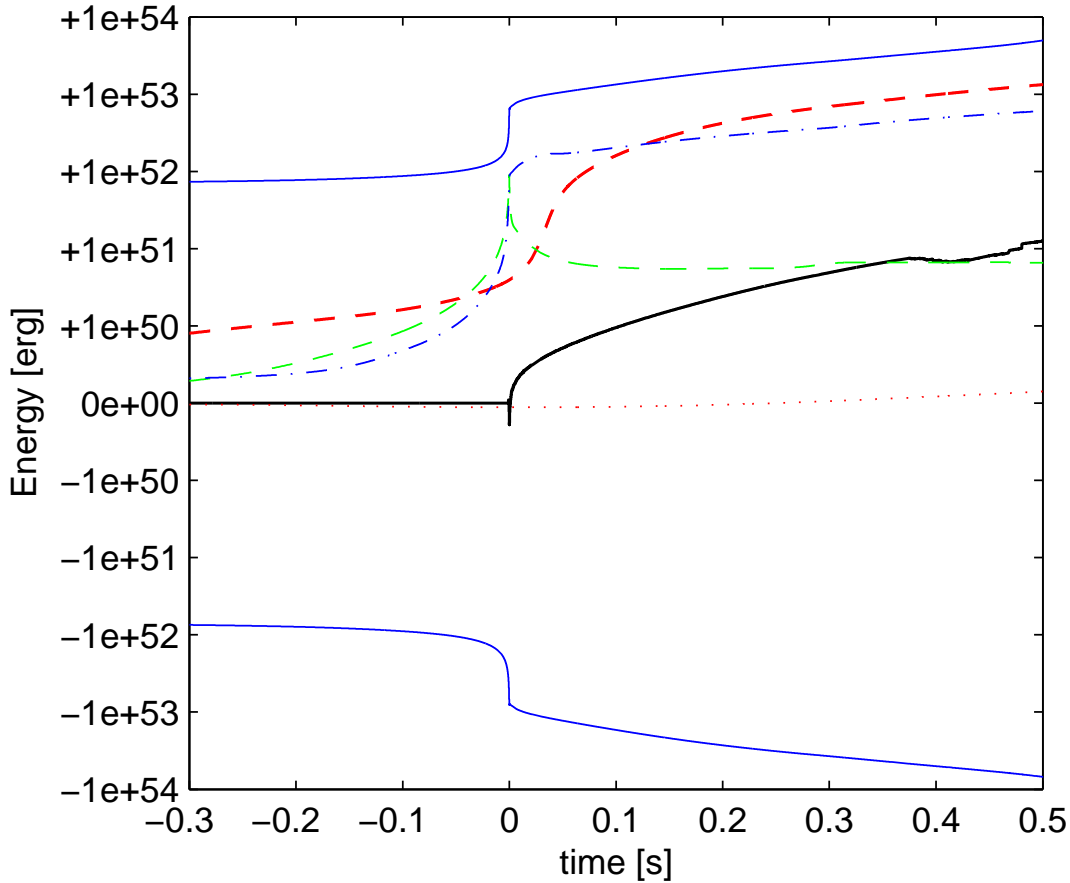}} \par}

\caption{The different components of the energy budget are shown. The internal
energy of the fluid is represented with a thin solid line in the upper
half of the figure. We set the free constant in the internal energy
such that we start with zero total energy (thick solid line). Most
of the internal energy is balanced by the gravitational energy (thin
solid line in the lower half of the figure). We show also the kinetic
energy of the fluid (thin dashed line) and the surface work exerted
at the border of the computational domain (dotted line). The dash-dotted
line represents the energy of the neutrinos in the computational domain
and the dashed thick line represents the accumulated energy of the
escaped neutrinos. The total energy is nicely conserved during core
collapse and exhibits a systematic increase of order \protect\( 10\%\protect \)
of an explosion energy during the crucial phase around \protect\( 100\protect \)
ms after bounce.\label{fig_energy1}}
\end{figure}

After having put much effort into the consistent finite differencing
of the transport equation in favor of an accurate evolution of expectation
values, we investigate in this section the energy conservation properties
in the most challenging run started from the massive \( 40 \) M\( _{\odot } \)
progenitor star. Our scheme preserves lepton numbers to machine precision
by construction. In Fig. (\ref{fig_energy1})
we show how the total energy in the simulation divides up into different
energy forms during the simulation. The gravitational energy (thin
solid line in the lower part of the figure) and the internal energy
of the fluid (thin solid line in the upper part of the figure) form
the largest contributions. For this figure, we have set the undetermined
constant in the internal energy such that the total energy vanishes
at the start of the simulation. The energy stored in neutrinos (dash-dotted
line) grows with respect to the internal fluid energy until bounce.
Afterwards, it evolves almost proportionally to the internal fluid
energy. The kinetic energy (thin dashed line) also grows during collapse.
It peaks at bounce and settles on a lower level afterwards, slightly
decaying during the stationary postbounce phase because of the decreasing
density of infalling material. The dotted line at the center of the
figure is the work exerted on the surface of the computational domain.
The thick dashed line is the accumulated energy emitted by neutrinos.
Its steep increase around bounce is delayed with respect to the other
energy contributions because of the delay in the neutrino burst and
the propagation time to the surface of the computational domain at
\( 10^{4} \) km radius. The thick solid line represents the evolution
of the total energy in our simulation. It is very accurately conserved
during the core collapse phase. It shows a perturbation of order \( 5\times 10^{49} \)
erg in the most dynamic phase around bounce when the grid points rush
to the center to resolve the shock front. It systematically increases
afterwards and reaches the order of an explosion energy at the end
of the simulation when the neutron star collapses to a black hole.
We further investigate this energy violation in the next two figures.

\begin{figure}
{\centering \resizebox*{0.8\textwidth}{!}{\includegraphics{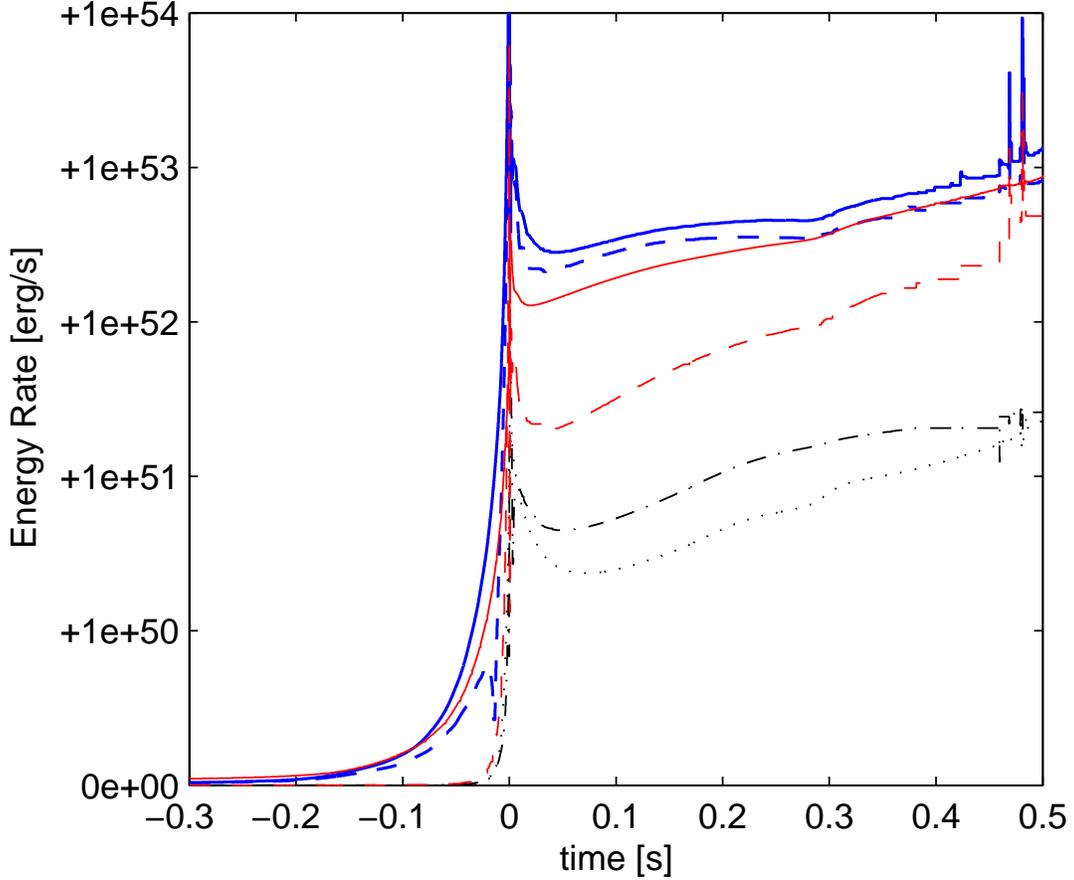}} \par}

\caption{For the six chains of terms that are supposed to cancel in Eq. (\ref{eq_cancellation_table})
we show the size of all positive contributions to measure the importance
of accurate cancellations. The implementation of the {\cal O}\protect\( \left( v/c\right) \protect \)
cancellations \protect\( (D_{\mu }^{12}O_{E}^{34})\protect \) and
\protect\( (D_{a}^{4}O_{E}^{1})\protect \) are most important (thick
solid line and thick dashed line, respectively). Of similar importance
is a matching in the general relativistic term \protect\( (C_{t}^{4}D_{a}^{2}D_{E}^{1}O_{\mu }^{1})-4\pi r(1+e+p/\rho )H\protect \)
(thin solid line). About one order of magnitude less important is
the matching \protect\( (C_{t}^{2}D_{\mu }^{34}D_{E}^{2})\protect \)
(thin dashed line). Of lowest importance is the matching among observer
corrections themselves in \protect\( (O_{E}^{2}O_{\mu }^{2})\protect \)
and \protect\( (O_{E}^{56}O_{\mu }^{34})\protect \) (dash-dotted
and dotted lines, respectively). The figure illustrates the steep
increase in the challenge to conserve energy after bounce.\label{fig_energy2}}
\end{figure}

We would obtain energy conservation to machine precision if we could
enforce perfect cancellation in the six chains discussed after Eq.
(\ref{eq_cancellation_table}), i.e. \( (D_{a}^{4}O_{E}^{1}) \),
\( (D_{\mu }^{12}O_{E}^{34}) \), \( (O_{E}^{2}O_{\mu }^{2}) \),
\( (O_{E}^{56}O_{\mu }^{34}) \), \( (C_{t}^{2}D_{\mu }^{34}D_{E}^{2}) \),
and \( (C_{t}^{4}D_{a}^{2}D_{E}^{1}O_{\mu }^{1})-4\pi r^{2}(1+e+p/\rho )H \).
The hydrodynamics scheme is designed to absorb the energy exchange
by the collision integral to machine precision and conserve energy
perfectly, even on the adaptive grid. By selecting only the positive
contributions in the canceling terms we obtain a measure of the importance
for an accurate cancellation. After an integration over the computational
domain, Fig. (\ref{fig_energy2})
illustrates this measure for the six above-mentioned expressions with
a thick dashed line, a thick solid line, a dash-dotted line, a dotted
line, a thin dashed line, and a thin solid line, respectively. After
bounce, the maximum individual contribution to the energy conservation
equation in Eq. (\ref{eq_radiation_energy_conservation}) reaches
a typical level of \( 5\times 10^{52} \) erg/s. Fig. (\ref{fig_energy2})
makes immediately evident that maintaining accurate energy conservation
is more challenging after bounce than before bounce. Moreover, we
note that the {\cal O}\( (v/c) \) terms \( (D_{a}^{4}O_{E}^{1}) \)
and \( (D_{\mu }^{12}O_{E}^{34}) \) are much larger than the higher
order terms \( (O_{E}^{2}O_{\mu }^{2}) \) and \( (O_{E}^{56}O_{\mu }^{34}) \).

\begin{figure}
{\centering \resizebox*{0.8\textwidth}{!}{\includegraphics{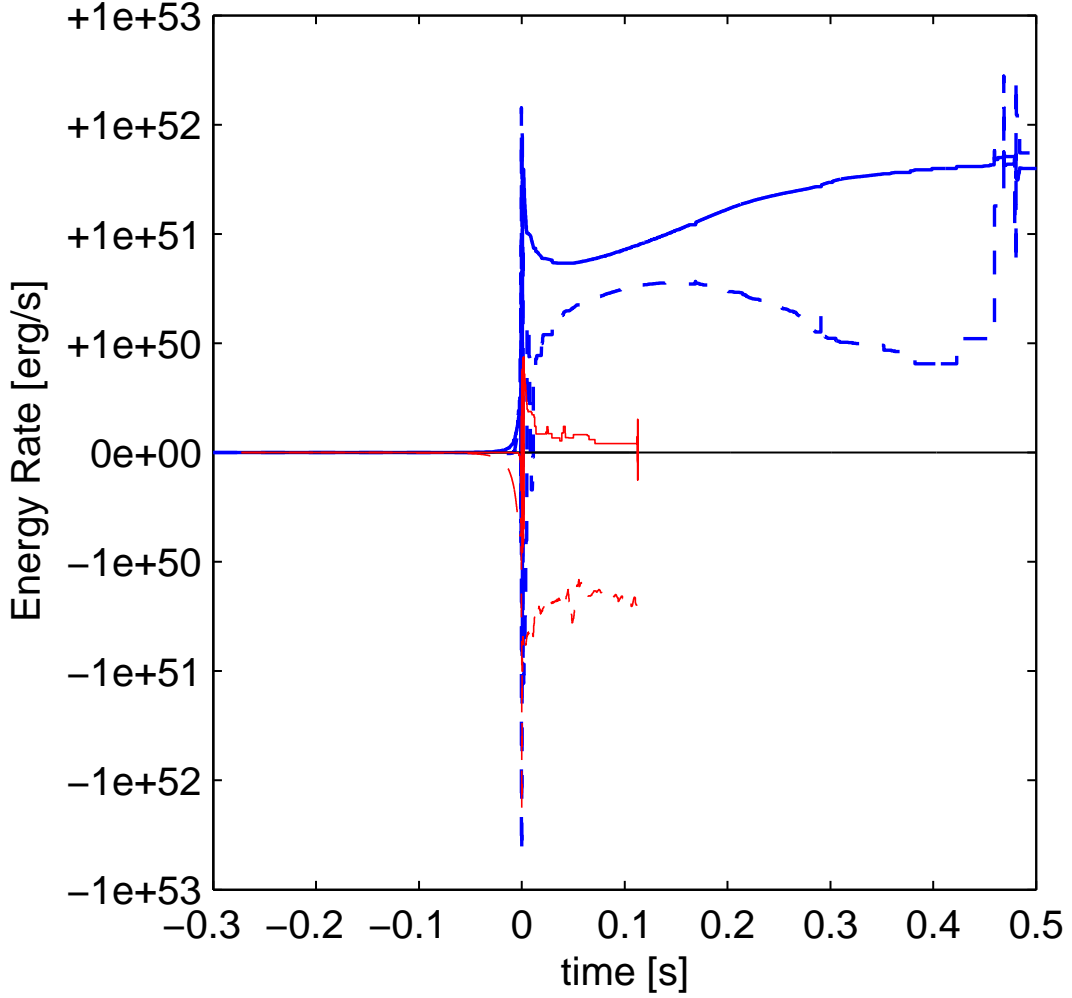}} \par}

\caption{The thick solid line and thick dashed line represent the error in
the matching of the terms \protect\( (C_{t}^{4}D_{a}^{2}D_{E}^{1}O_{\mu }^{1})-4\pi r(1+e+p/\rho )H\protect \)
and \protect\( (C_{t}^{2}D_{\mu }^{34}D_{E}^{2})\protect \) respectively.
All other matches are to machine precision. Therefore, the time integration
of these energy rates must reproduce the drift in the total energy
represented by a thick solid line in Fig. (\ref{fig_energy1}). The
two monitored expressions do not allow a distinction of errors induced
by approximations in the matching procedure in Eqs. (\ref{eq_misner_dGdt_fd})
and (\ref{eq_Ge_fd}) from errors induced by the application of the
adaptive grid to radiation quantities in the comoving frame (see discussion
following Eq. (\ref{eq_fbar_reynolds})). In an attempt to disentangle
these two contributions to the violation of energy conservation, we
compare the cancellation in the terms \protect\( (C_{t}^{4}D_{a}^{2}D_{E}^{1}O_{\mu }^{1})-4\pi r(1+e+p/\rho )H\protect \)
and \protect\( (C_{t}^{2}D_{\mu }^{34}D_{E}^{2})\protect \) with
a Lagrangian run with \protect\( 400\protect \) zones and \protect\( 2\protect \)
angular bins (thin solid line and thin dashed line, respectively).
The energy violation of the former term is greatly reduced on a fixed
grid.\label{fig_energy3}}
\end{figure}

Nevertheless, these terms are perfectly matched in our implementation.
Violation of energy conservation therefore stems from the terms \( (C_{t}^{2}D_{\mu }^{34}D_{E}^{2}) \)
(the matching is tuned for an isotropic neutrino distribution) and
\( (C_{t}^{4}D_{a}^{2}D_{E}^{1}O_{\mu }^{1})-4\pi r(1+e+p/\rho )H \)
(the matching is based on an energy flux averaging) as discussed in
the context of Eqs. (\ref{eq_Gmu_fd}) and (\ref{eq_Ge_fd}) respectively.
In section \ref{subsection_adaptive_grid} we have discussed energy
conservation violations by the adaptive grid corrections when they
are applied to the radiation quantities. Where do they enter the conservation
check? If we evaluate Eq. (\ref{eq_radiation_energy_conservation})
on the adaptive grid according to the recipe in Eq. (\ref{eq_reynolds_theorem_fd}),
we note that the integration of the energy over the whole star reduces
the adaptive grid corrections to surface terms. These surface terms
vanish because the grid velocity is zero at the center and the surface
of the star. If we compare with the more detailed Eq. (\ref{eq_cancellation_table})
we find that this time derivative corresponds exactly to the term
\( C_{t}^{1}+C_{t}^{3} \). The energy violations by the adaptive
grid show only up in the terms \( C_{t}^{2} \) and \( C_{t}^{4} \).
If these terms are evaluated with all the grid corrections in the
time evolution of \( \Gamma  \), \( u \), \( J \), and \( H \),
they will numerically differ from the terms \( C_{t}^{2} \) and \( C_{t}^{4} \)
we have used in Eqs. (\ref{eq_misner_dGdt_fd}) and (\ref{eq_Ge_fd})
for the approximate matching. We check the influence of the adaptive
grid corrections by a comparison with a run using a pure Lagrangian
grid, where no adaptive grid corrections can compromise energy conservation.
However, in order to get enough resolution in the run with the fixed
grid, we had to run with \( 400 \) spatial zones instead of the \( 103 \)
we used with the adaptive grid. This run is extremely slow. On the
one hand, the solution vector is four times larger. But much more
important is that every single zone has to change its value from the
preshock conditions to the postshock conditions in the allowed \( 1\% \)-change
steps. This requires an almost \( 10 \) times smaller time step than
with the adaptive grid, where a zone can follow the shock. In the
latter case, the conditions in the zone changes on a much longer time
scale determined by the drift between the zone speed and the shock
propagation. In order to let a run reach the interesting phase around
\( 100 \) ms after bounce in reasonable time, we had to reduce the
angular resolution to only two angular bins. Both measures, the increase
of spatial and time resolution and the decrease of angular resolution
can in principle affect energy conservation. Nevertheless, we hope
to get the correct impression of the influence of the adaptive grid
on the energy conservation. The two terms \( C_{t}^{2} \) and \( C_{t}^{4} \)
of the Lagrangian run are also shown in Fig. (\ref{fig_energy3})
(thin lines). We find that the order of magnitude of energy violation
in the cancellation \( (C_{t}^{2}D_{\mu }^{34}D_{E}^{2}) \) does
not significantly change on the fixed grid. However, the cancellation
in the term \( (C_{t}^{4}D_{a}^{2}D_{E}^{1}O_{\mu }^{1})-4\pi r(1+e+p/\rho )H \)
is greatly reduced on the fixed grid. This suggests that the adaptive
grid corrections of the radiation quantities are the dominant remaining
sources of energy violation, about five times larger than the mismatch
in \( (C_{t}^{2}D_{\mu }^{34}D_{E}^{2}) \).

The absolute necessity of accurate energy conservation may be discussed.
But certainly, it provided an invaluable check for the congruence
between the programmers intention and the actual implementation of
the many intricate finite difference expressions in our code.

\subsection{Comparison with Multi-Group Flux-Limited Diffusion}

Following our tradition \citep{Mezzacappa_Bruenn_93a}, we conclude
this paper with a comparison with simulations using the Multi-Group
Flux-Limited Diffusion ({\sc mgfld}) approximation. We note that,
excepting the pioneering Boltzmann solver of \citep{Wilson_71}, codes
with {\sc mgfld} currently provide the only alternative numerical
data for the evolution of a supernova in full general relativity \citep{Myra_et_al_87,Schinder_Bludman_89,Bruenn_DeNisco_Mezzacappa_01}.
We compare the evolution of a \( 13 \) M\( _{\odot } \) progenitor
\citep{Nomoto_Hashimoto_88}. This is a stellar model with a small
iron core which---in the hope of seeing an explosion in numerical
simulations---has been used throughout the supernova literature. Questions
concerning the accuracy of the {\sc mgfld} approximation in the dynamic
semi-transparent region between the neutrino sphere and the heating
region have been posed and answered in several studies with different
flux limiters (in the supernova context e.g. \citep{Janka_92,Messer_et_al_98,Yamada_Janka_Suzuki_99}).
For general relativistic {\sc mgfld} simulations \citep{Bruenn_DeNisco_Mezzacappa_01}
Bruenn developed a new flux limiter that consists of two parts \citep{Bruenn_02}:
The first part is a specific implementation of the usual scheme for
interpolating between the optically thick diffusion regime and the
optically thin free streaming regime, namely\[
{\cal F}_{\rm intrp}\left( E\right) =\left( 1+\frac{1}{3}\lambda _{t}\left( E\right) \frac{\left| \partial \psi ^{0}\left( E\right) /\partial r\right| }{\psi ^{0}\left( E\right) }\right) ^{-1},\]
where \( \lambda _{t}\left( E\right)  \) is the total energy-dependent
transport mean free path. This ensures that\[
\psi ^{1}\left( E\right) =-\frac{\lambda _{t}\left( E\right) }{3}\frac{\partial \psi ^{0}\left( E\right) }{\partial r}\]
in the diffusion limit, and that \( \psi ^{1}\left( E\right) =\psi ^{0}\left( E\right)  \)
in the free streaming limit. This part alone suffers from the generic
problem of too rapid a transition to the free streaming limit when
matter goes from optically thick to optically thin abruptly. To avoid
this problem, a second piece of the flux limiter is constructed. It
basically prevents the neutrino angular distribution from becoming
more forward peaked than the geometrical limit. At radius \( r \),
the second part depends on the radius of the neutrinosphere, \( R_{\nu }\left( E\right)  \),
and is given by\[
{\cal F}_{\rm geom}\left( E\right) =\left\{ \begin{array}{cc}
\frac{\frac{1}{2}\left( 1+\mu _{0}\left( E\right) \right) \psi ^{0}\left( E\right) }{\frac{1}{3}\lambda _{t}\left( E\right) \left| \partial \psi ^{0}\left( E\right) /\partial r\right| } & {\rm if}\quad r>R_{\nu }\left( E\right) \\
1 & {\rm if}\quad r\leq R_{\nu }\left( E\right) 
\end{array}\right. \]
where\begin{eqnarray*}
\mu _{0}\left( E\right)  & = & \frac{\mu \left( E\right) +v}{1-\mu \left( E\right) v},\\
\mu \left( E\right)  & = & \sqrt{1-\left( \frac{R_{\nu }\left( E\right) }{r}\right) ^{2}}{\cal G},
\end{eqnarray*}
and\[
{\cal G}=\left( 1-\frac{2m}{r}\right) ^{1/2}\left( 1-\frac{2m}{R_{\nu }\left( E\right) }\right) ^{-1/2}.\]
 The quantity \( \mu  \) is the cosine of the angle from the limb
of the neutrinosphere to the point at \( r \), corrected for gravitational
bending, and \( \mu _{0} \) is that same angle as seen in the fluid
frame. The net diffusivity in the diffusion equation is then set by
the minimum of the two parts of the flux limiter,\[
{\cal D}\left( E\right) =\frac{\lambda _{t}\left( E\right) }{3}\min \left( {\cal F}_{\rm intrp}\left( E\right) ,{\cal F}_{\rm geom}\left( E\right) \right) .\]
In the following, we compare Bruenn's {\sc mgfld} simulation with
this flux limiter to the solution we have obtained by the complete
solution of the transport equation. Beside of the different methods
implemented for the radiation transport, the two codes also use different
schemes to solve the hydrodynamics equations. While the code described
in this paper adapts to spatial resolution requirements by continuously
displacing zones, the {\sc mgfld} code adjusts the resolution by occasionally
inserting or removing zones. It uses of order \( 150 \) zones at
bounce and \( 250 \) zones half a second after bounce, while {\sc agile-boltztran}
works with \( 103 \) zones throughout. The energy resolution in {\sc mgfld}
is given by \( 20 \) groups which adapt to the general relativistic
redshift. In the Boltzmann solver, we used \( 12 \) fixed energy
groups.

\begin{figure}
{\centering \resizebox*{0.95\textwidth}{!}{\includegraphics{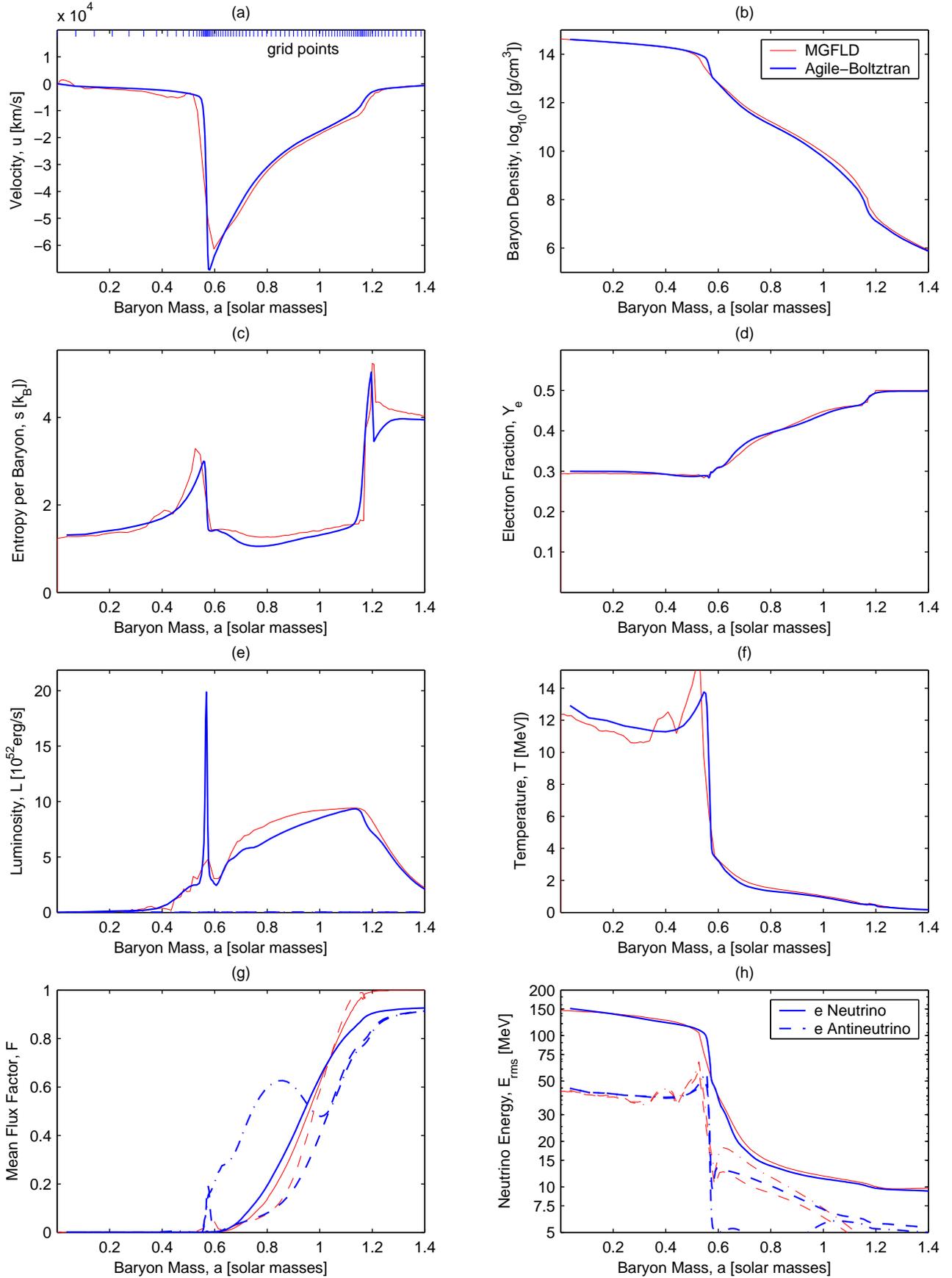}} \par}

\caption{Comparison with {\sc mgfld}, \protect\( 13\protect \) M\protect\( _{\odot }\protect \)
model at bounce.\label{fig_bounce.ps}}
\end{figure}

We start with the comparison of a time slice at bounce in Figure (\ref{fig_bounce.ps}).
The ordinates of graphs (a-h) display the enclosed mass. At the end
of core collapse, we find differences in the density profiles in graph
(b) of \( 3\% \) at the center, and less than \( 1\% \) in the inner
core. Differences of up to \( 80\% \) around the shock front are
due to a different resolution of the shock. Outside of the shock,
we find a \( \sim 30\% \) lower density in {\sc agile-boltztran}.
The shock position, however, is in reasonable agreement as shown in
the velocity profiles in graph (a). This is a consequence of the agreement
in the thermodynamical state within the homologous inner core up to
the sonic point at an enclosed mass of \( \sim 0.54 \) M\( _{\odot } \).
Differences in the infall velocities are of order \( 6\% \). In the
diffusive domain, we find agreement in the entropies of order \( 5\% \)
(graph (c)). Around an enclosed mass of \( 0.8 \) M\( _{\odot } \),
however, the deviations are systematic and of order \( 10\% \). At
least part of it is due to a narrow choice of the spatial resolution
in {\sc agile-boltztran} as discussed in section \ref{subsection_resolution}.
Immediately connected to the entropy profile is the temperature profile
in graph (f). The temperature is very sensitive to entropy differences
in the degenerate high density material. The electron fractions in
graph (d) differ typically by \( 2\% \). The maximum deviation of
\( 4\% \) is found where the entropy deviation is largest. At lower
densities than \( \sim 10^{10} \) g/cm\( ^{3} \), the dominant source
of neutrinos is provided by electron capture on heavy nuclei. As the
mass fraction of heavy nuclei is not very temperature dependent, the
entropy difference only affects the deleptonization by the dependence
of the electron capture rates. {\sc agile-boltztran} finds a slightly
lower electron fraction in this regime. At densities exceeding \( \sim 10^{10} \)
g/cm\( ^{3} \), however, standard input physics used to prematurely
switch off electron captures on heavy nuclei in the oversimplified
independent particle model \citep{Langanke_et_al_03} such that electron
captures on free protons became dominant in the simulations compared
here. The lower entropy in the simulation with {\sc agile-boltztran} leads
to a smaller mass fraction of free protons and reduced deleptonization
with respect to the {\sc mgfld} simulation. The slightly larger electron
abundance between \( 0.65 \) and \( 0.8 \) M\( _{\odot } \) is
a consequence. The agreement in the luminosities in graph (e) is of
order \( 15\% \). The solution in the diffusive domain is smoother
in {\sc agile-boltztran} because the expanding zones of the adaptive
grid equilibrate local discontinuities introduced by the equation
of state. In contrast to these random fluctuations inside the shock
front, we find a systematic deviation of \( 15\% \) in the luminosities
outside of the shock front. This difference might also be a consequence
of the lower entropy in {\sc agile-boltztran}. The neutrino rms energies
(graph (h)) in the inner core are determined by thermal equilibrium.
There, the agreement is to \( 3\% \). Outside of the shock front,
the deviation is initially more around \( 6\% \), before it becomes
better again towards very large radii. The mean flux factor, the ratio
of neutrino flux over neutrino density, \( {\rm F}=H/\left( cJ\right)  \),
in graph (g) shows differences of order \( 20\% \) at this time.
The smaller flux factor of the {\sc mgfld} neutrinos at an enclosed
mass of about \( 0.8 \) M\( _{\odot } \) is probably due to increased
isotropic neutrino emission at the higher entropy. It leads to slightly
higher luminosities and higher rms energies, and is consistent with
the smaller electron fraction. We should note here that it is very
typical for such comparisons that each code presents an itself consistent
picture of the dynamics such that it is sometimes almost impossible
to isolate a single cause for a difference in the strongly coupled
variables. The smaller flux factor in {\sc agile-boltztran} outside
\( 1.1 \) M\( _{\odot } \) stems from the limited angular resolution
discussed in section (\ref{subsection_anglecheck}). It is systematic
and visible in all the following figures. Its impact on the dynamics,
however, is negligible because of the small coupling between the neutrino
flux and matter at large radii.

\begin{figure}
{\centering \resizebox*{0.95\textwidth}{!}{\includegraphics{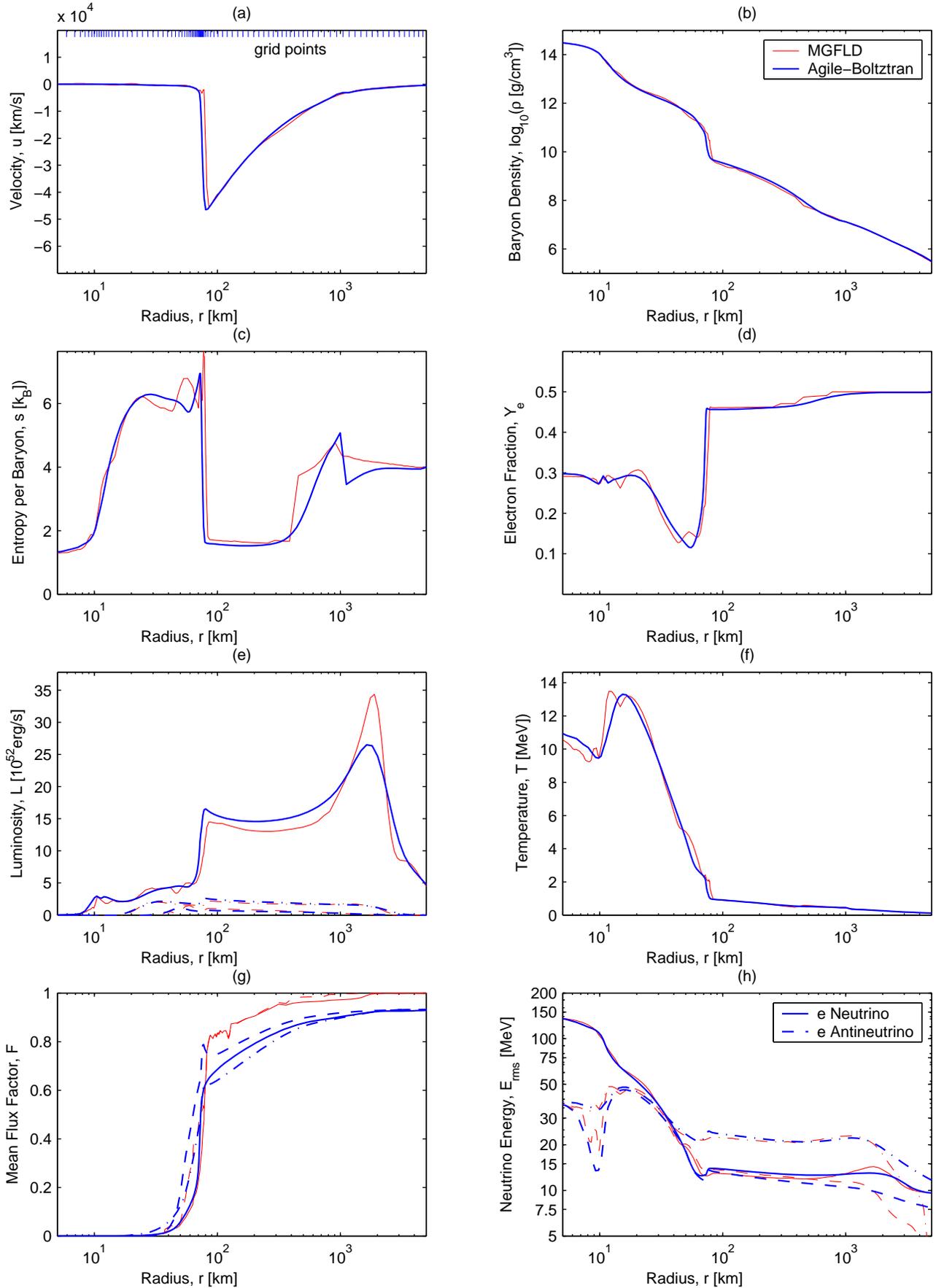}} \par}

\caption{Comparison with {\sc mgfld}, \protect\( 13\protect \) M\protect\( _{\odot }\protect \)
model, \protect\( 10\protect \) ms after bounce.\label{fig_burst.ps}}
\end{figure}

Figure (\ref{fig_burst.ps})
shows a time slice just after the launch of the neutrino burst. This
is the most dynamical phase for the radiation transport. The neutrino
burst in graph (e) is evident in both codes as a propagating peak
in the luminosity profile. The peak at \( \sim 2000 \) km radius
is broader and by \( 25\% \) smaller in {\sc agile-boltztran} because
of the numerical diffusion we have analyzed in section \ref{subsection_pulse_propagation}.
The {\sc mgfld} burst changes its shape more slowly because the {\sc mgfld}
code reverts, in the transparent regime, to centered difference advection
(rather than first order upwind) which is second order accurate in
space (on a uniform spatial grid). The luminosity in the region between
the shock and the luminosity peak decays very rapidly. Differences
of order \( 15\% \) are due to a small time mismatch between the
two solutions. Inside the shock front, the luminosity is still much
noisier in the {\sc mgfld} solution. This is also true for the entropy
profile in graph (c), where local differences of \( 20\% \) behind
the shock or \( 50\% \) at \( 400 \) km radius, disturb the otherwise
nice agreement of order \( 5\% \). Note however, that the temperature
profiles in graph (f) is less sensitive to this entropy variation.
The temperature difference at \( 400 \) km radius is \( 12\% \).
The agreement in the infall velocities in graph (a) has improved to
generally \( 2\% \) agreement with the exception of \( 9\% \) around
\( 400 \) km radius, i.e. the region with the entropy differences.
The entropy deviations inside the shock front and at \( 400 \) km
radius cause density differences of \( 25\% \) in these regions.
The central density agrees to \( 1\% \) and the agreement in the
very distant layers is of order \( 3\% \). The electron fraction
profile is affected by the numerical noise as well. Deviations, however,
are only of order \( 2\% \), with local exceptions showing differences
of \( 10\% \). the rms energies in graph (h) typically agree to \( 5\% \).
The differences in the flux factors in graph (g) show systematic deviations
of \( 10\% \) outside of \( 100 \) km radius and random fluctuations
of order \( 20\% \) interior to it. The latter are probably a consequence
of the variations in the density and entropy profiles.

\begin{figure}
{\centering \resizebox*{0.95\textwidth}{!}{\includegraphics{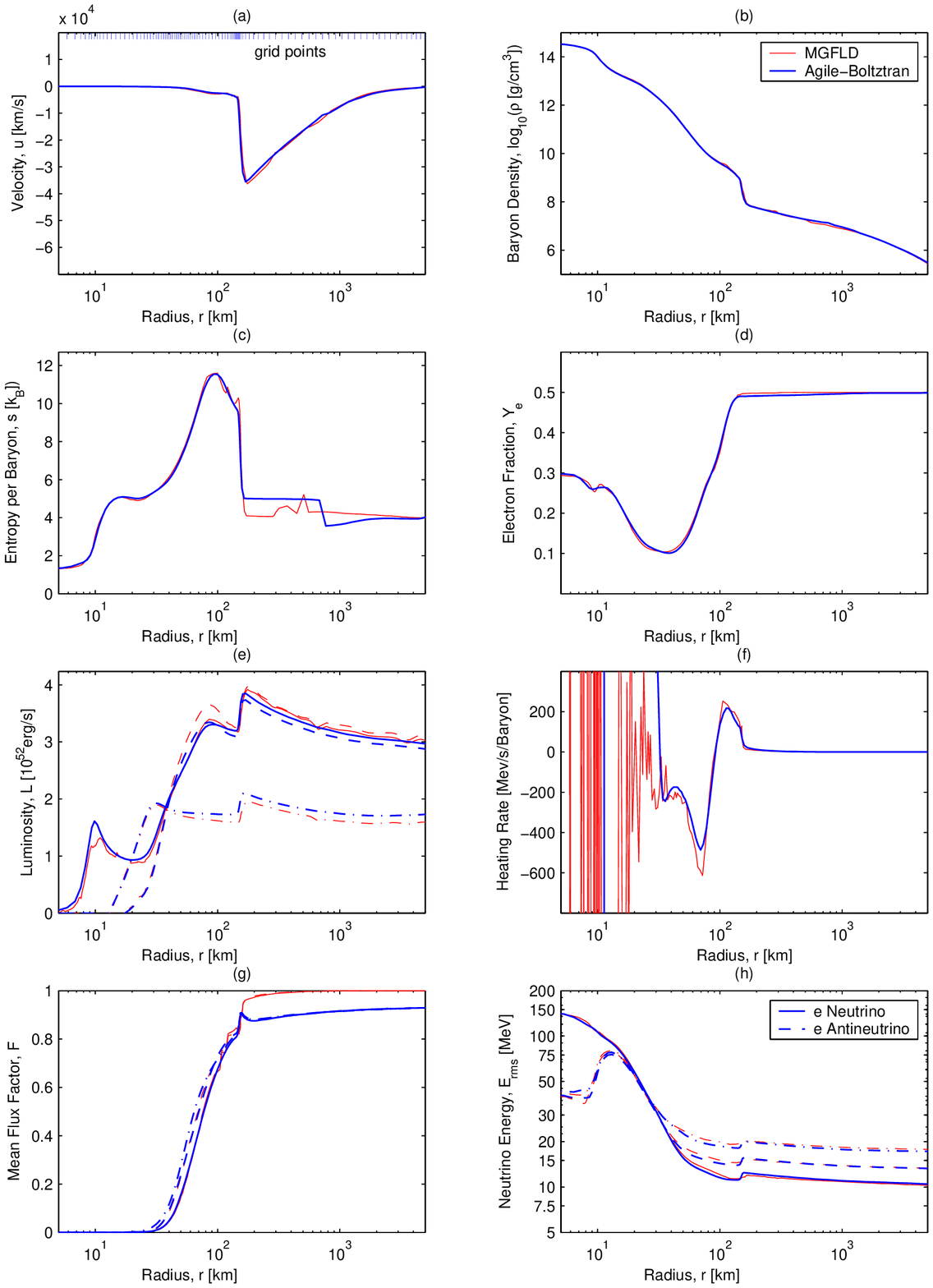}} \par}

\caption{Comparison with {\sc mgfld}, \protect\( 13\protect \) M\protect\( _{\odot }\protect \)
model, \protect\( 100\protect \) ms after bounce.\label{fig_heat.ps}}
\end{figure}

Most important for the success or failure of the supernova explosion
in our spherically symmetric simulations is the time \( \sim 100 \)
ms after bounce. Though, we have to state clearly that the general
relativistic runs are not failing marginally. They are failing with
officious insistence, and differences of the size we have described
above are far from changing this, as we will show with the next two
time slices. Figure (\ref{fig_heat.ps})
shows the comparison of the two simulations at \( 100 \) ms after
bounce, when the efficiency of the neutrino heating is close to maximum.
The agreement of the data in this phase is remarkable. In this quasi-stationary
phase, the position of the stalled shock is in accurate agreement
as one can see in graph (a). We note however, that the important infall
velocity in the heating region (e.g. at \( 100 \) km radius) is \( 10\% \)
larger in {\sc mgfld} than in {\sc agile-boltztran}. Since this infall
velocity determines the time the infalling matter spends in the heating
region before being accreted onto the protoneutron star, its value
is as relevant as the heating rate. Better visible in the graph are
the differences in the infall velocity which have only a size of \( 3\% \).
Again, they reflect differences in the entropy profile in graph (c).
Outside the shock the entropy deviations of order \( 20\% \) seem
large. However, they correspond to temperature variations of only
\( 10\% \) and are explicable by the assumption of instantaneous
silicon burning in {\sc agile-boltztran}. {\sc mgfld} uses a \( 9 \)
species nuclear network to burn nuclei to iron until the temperature
exceeds \( 5.106\times 10^{9} \) K, at which point the zone is flashed
to NSE. Inside the shock front, the agreement in the entropies is
good to \( 2\% \) (except for some local fluctuations in the {\sc mgfld}
solution in the heating region). As the rise in entropy between the
shock front and the location at \( 100 \) km radius is entirely due
to neutrino heating, this is an encouraging result. In this time slice,
we have replaced the temperature graph (f) with a graph of the more
important heating rates. Plotted is the sum of the heating rates by
electron neutrino absorption and electron antineutrino absorption.
The rate is negative in the cooling region, representing cooling rates
by the inverse processes, electron capture and positron capture. We
find \( 7\% \) difference in the heating region and \( 10\% \) difference
in the cooling region. The values in the {\sc mgfld} simulation are
larger. This is perfectly consistent with the larger electron antineutrino
luminosity in graph (e) and the larger infall velocity behind the
shock in graph (a). The electron fraction profiles in graph (d) agree
to \( 2\% \) accuracy with the exception of two local \( 6\% \)
deviations at \( 25 \) km and \( 45 \) km radius, where the electron
fraction is as small as \( \sim 0.1 \). This close agreement in the
entropy and electron fraction leads to a similar thermodynamical state
of the fluid in the inner core. Good agreement in the nearly hydrostatic
density profile in graph (b) is the consequence. Differences are \( 0.5\% \)
at the center and on average around \( 5\% \) up to a radius of \( 400 \)
km. Further out, differences stemming from the silicon/iron layer
interface are visible. The good agreement in the density profiles
facilitates the comparison of other density-dependent quantities.
E.g. the neutrino rms energies in graph (h) agree to \( 2\% \) in
the innermost core, to \( 6\% \) in the heating regions, and to \( 3\% \)
outside of the shock front in the neutrino signal. In the heating
region, the rms energies are lower in {\sc agile-boltztran}. At larger
radii, the electron neutrino energies are lower in {\sc mgfld}, while
the \( \mu  \)- and \( \tau  \)-neutrino energies are lower in {\sc agile-boltztran}.
The electron antineutrino energies do not show a discernible tendency.
The electron neutrino luminosities in graph (e) are \( 5\% \) larger
in {\sc agile-boltztran} in the diffusive domain and \( 2\% \) larger
in {\sc mgfld} in the heating region. The electron antineutrino luminosity
is about \( 6\% \) larger in {\sc agile-boltztran} in both domains.
In the \( \mu  \)- and \( \tau  \)-neutrino luminosities, deviations
start at \( 20 \) km radius with \( 2\% \)and linearly increase
to \( 10\% \) at the shock position. With respect to the neutrino
signal, the electron flavor luminosities are larger in the simulation
with {\sc mgfld} and the \( \mu  \)- and \( \tau  \)-luminosities
are larger in the simulation with {\sc agile-boltztran}. This is just
inverse to the relation between the rms energies. While, for example,
an increase in the rms energy contributes only linearly towards a
luminosity increase, the neutrino flux, however, is determined by
the quadratically decreasing mean free path. Where the flux factors
in graph (g) are below \( 0.5 \), they are typically \( 5\% \) lower
in {\sc mgfld}. Between this domain and the shock front (including
the heating region) the flux factors show on average no discernible
deviation, but fluctuations of \( \pm 5\% \) are evident. The flux
factors in {\sc agile-boltztran} disagree with the asymptotic limit,
\( F=1 \), by \( 8\% \) at distant radii due to the limitation in
angular resolution as discussed in subsections \ref{subsection_anglecheck}
and \ref{subsection_resolution}. 

\begin{figure}
{\centering \resizebox*{0.95\textwidth}{!}{\includegraphics{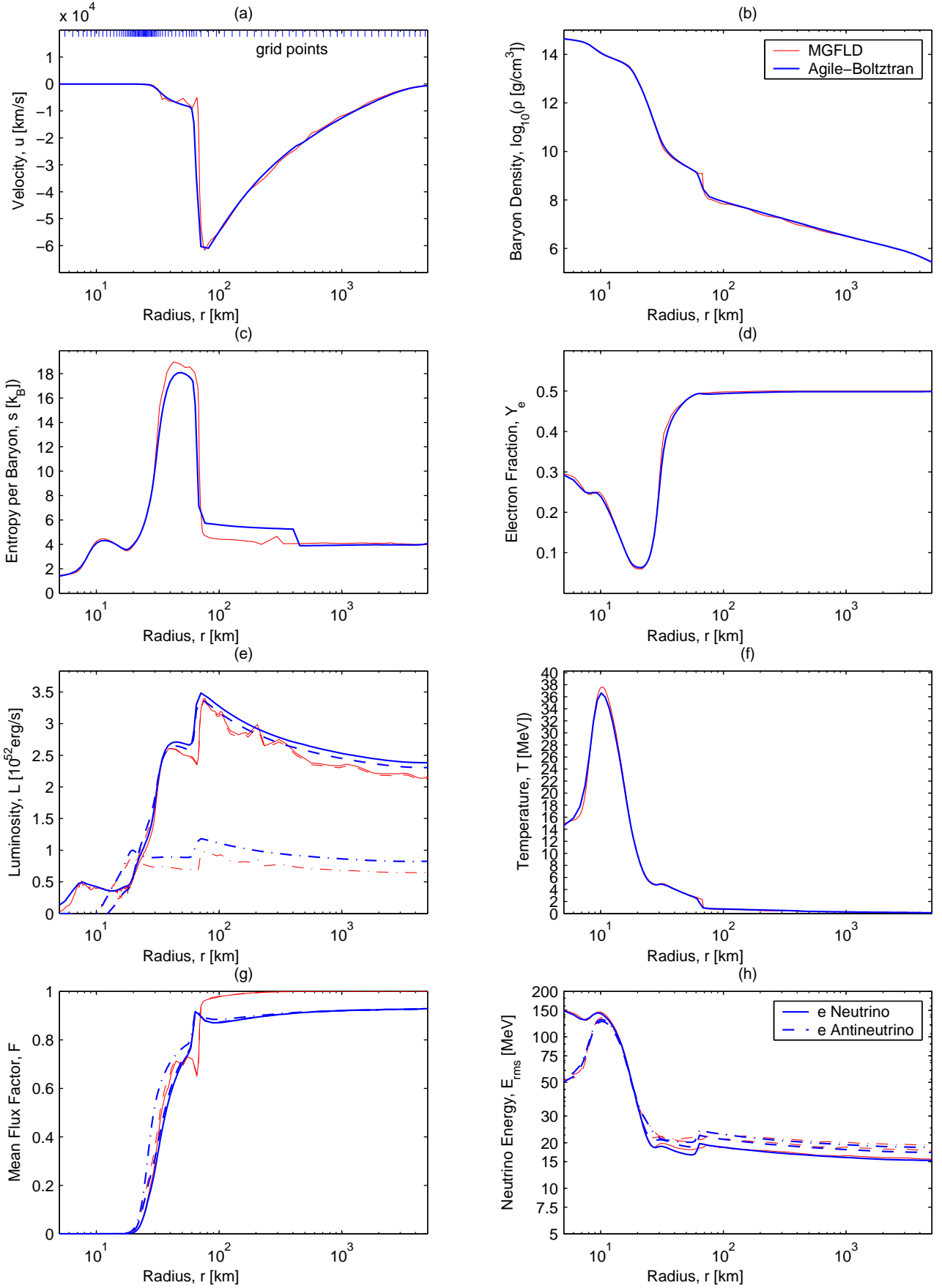}} \par}

\caption{Comparison with {\sc mgfld}, \protect\( 13\protect \) M\protect\( _{\odot }\protect \)
model, \protect\( 400\protect \) ms after bounce.\label{fig_moreretract.ps}}
\end{figure}

Similar agreements and differences are found in later stages of the
quasistationary phase, e.g. at \( 400 \) ms after bounce (Fig. (\ref{fig_moreretract.ps})).
 We find again hotter neutrinos between the neutrino sphere and the
shock position in the {\sc mgfld} simulation and a larger diffusive
flux at very high densities in {\sc agile-boltztran}. On a longer
time scale, the latter leads to a marginally larger deleptonization,
visible in the innermost \( 7 \) km of the electron fraction profile.
The electron fraction in {\sc agile-boltztran} is now \( 2\% \) smaller,
whereas it was \( 2\% \) larger early after bounce. A new difference
of \( 6\% \) appears in the peak of the entropy profile, which might
be due to differences in the hydrodynamics scheme for the solution
of the shock jump conditions. We also point to the fact that the resolution
of the shock in this stage is quite poor. The thin marks at the top
of graph (a) represent the grid point locations in the {\sc agile-boltztran}
simulation. Most of the grid points cluster around the forming density
cliff between \( 20 \) km and \( 30 \) km radius. A high resolution
in this domain is important for the radiation transport because most
neutrinos are emitted from the cliff. However, this leaves less grid
points to focus on the shock front. Graph (a) shows only a small increase
of the grid point concentration at the shock front. The infall velocities
inside the shock in graph (a) have increased to \( 10^{4} \) km/s.
At this speed, the infalling material is only very shortly exposed
to neutrino heating. This is consistent with the observation that
the entropy increase from the shock to the peak entropy is only marginal
in both runs. Without unaccounted effects, a shock revival is impossible
under these conditions. The most significant new differences appear
in the luminosity and rms energy profiles. {\sc mgfld} has now consistently
smaller luminosities and higher rms energies. The differences are
\( 20\% \) in the luminosities and unchanged \( 6\% \) in the rms
energies in the heating region. Also interesting to note is that the
rms energies agree perfectly right outside the shock at \( 100 \)
km radius, and develop deviations of \( 3\%-4\% \) at far distances.
There might be small differences in the implementation of the observer
corrections. The flux factors in graph (g) are more difficult to compare,
because the neutrinos no longer decouple at exactly the same radius.
The flux factor in {\sc agile-boltztran} is still numerically limited
in its asymptote, while the flux factor in {\sc mgfld} exhibits somewhat
peculiar fluctuations in the heating region.

Overall, in our general relativistic simulations of stellar core collapse
and postbounce evolution in spherically symmetric space-time, we find
quite exactly the same physical evolution in both runs. One implementing
the full Boltzmann transport equation, and the other based on the
multi-group flux-limited diffusion approximation with a new sophisticated
flux limiter. This is especially true in all phases of stationary
radiation transport. There are modest quantitative differences in
these simulations of failed supernovae. However, except for some obvious
cases, like e.g. the limited asymptotic flux factor in {\sc agile-boltztran},
or the numerical fluctuations that the equation of state induces to
purely Lagrangian hydrodynamics in {\sc mgfld}, these quantitative
differences cannot uniquely be traced back to principal weaknesses
or strengths of either method. They may rather be due to method-specific
details in the implementation of two very different methods to solve
a complex time-dependent physical problem. The advantage of the Boltzmann
solution is that it is complete, and, beside of numerical resolution,
does not need to justify or verify any basic assumptions. The advantage
of {\sc mgfld} is the transparency of the approach, and that it produces
accurate results with much higher efficiency in the application we
have tested here.

\section{Conclusion}

{\sc agile-boltztran} directly solves the general relativistic Boltzmann
transport equation for the specific neutrino distribution function
in spherically symmetric space-time. In combination with general relativistic
hydrodynamics, we present a consistent finite differencing for fermion
radiation hydrodynamics in spherical symmetry. We test our code in
the context of stellar core collapse and postbounce evolution of a
smaller \( 13 \) M\( _{\odot } \), and a more massive \( 40 \)
M\( _{\odot } \) progenitor star. They embrace the progenitor mass
range expected to produce explosions after core collapse. But no explosions
are obtained in our simulations. We analyze the regions of neutrino
emission by a statistical description allowing the detailed presentation
of the specific production reactions and energies of escaping neutrinos.

The Boltzmann transport equation plays a fundamental role in kinetic
gas theory. One may derive from it the diffusion equation, or---if
one drives the limit further to vanishing transport---the hydrodynamics
equations. Thermodynamical quantities are given by the expectation
values of various operators on the fundamental particle distribution
functions. It is quite unique to numerically solve a single equation
for the fundamental particle distribution functions in an object of
astronomical size! However, this basic approach becomes intricate
in the detailed finite difference representation. In theory, the solution
of the Boltzmann equation guarantees the correct evolution of expectation
values in the energy conservation equation or the diffusion limit.
Also the equation of state for radiation, \( p\propto \rho /3, \)
will emerge correctly. This is not automatically the case in a numerically
solved Boltzmann equation. Especially in the astrophysical application,
a distant observer will rather observe macroscopic phenomena than
the microscopic local state of the matter and the radiation field.
Therefore we compose the finite difference representation of the microscopic
physics such that important macroscopic quantities are accurately
met. This optimizes the accuracy of the simulations at moderate resolution
settings. However, it would be more straightforward to comply with
this ambition if there would not be an additional complication. Physically,
the Boltzmann transport equation is a trivial equation, describing
free particle propagation on geodesics between independent collisions
(even more so in spherical symmetry, where only two independent spatial
degrees of freedom have to be considered). With respect to the collisions,
the supernova has been investigated as a site of interesting nuclear
physics and neutrino-matter interactions. The cross sections depend
on the local neutrino energies and scattering angles. They are most
conveniently evaluated in the fluid rest frame and most easily described
in combination with comoving coordinates. The latter are also favored
by the initial core collapse where the relevant computational domain
collapses with the matter from several thousands to some hundreds
of kilometer radius, before it hopefully expands again with the explosion.
The comoving coordinates convert the simple statement of the Boltzmann
equation, that the derivative of the invariant particle distribution
function along the phase flow vanishes between collisions, into a
complicated sum of partial derivatives along the comoving coordinates.
The finite differencing of all comoving frame correction terms have
to be adjusted in a mutually dependent way to correctly transform
to conservation equations for the observer in a laboratory frame.
We construct number conservation to be guaranteed. With respect to
energy conservation, we match the largest {\cal O}\( \left( v/c\right)  \)
terms to machine precision, while higher order gravitational terms
are matched approximately, guided by specific limiting cases. The
adaptive grid applied to the neutrino distribution function in the
comoving frame introduces additional deviations from perfect energy
conservation in the laboratory frame. In the worst case, i.e. in the
evolution of the very massive \( 40 \) M\( _{\odot } \) progenitor
star, all these deviations together reach the order of a supernova
explosion energy (\( \sim 10^{51} \) erg) at the time the protoneutron
star collapses to a black hole. However, when the conditions are most
favorable for an explosion (we should rather say least unfavorable),
the numerical energy gain has not yet exceeded \( 10^{50} \) erg
and we are confident that the remaining energy drift does not influence
our physical conclusions. We have also tested the implementation of
the gravitational redshift and bending, the angular advection in the
free streaming regime, and the propagation of the neutrino burst.
Quite generally, we find that the choice of first order upwind differencing
is the limiting factor for the accuracy in our numerical solution.
In space, it causes artificial pulse spreading. In the advection terms
of the adaptive grid, it causes artificial diffusion. In momentum
space, it causes angular diffusion in the evolution of the particle
distribution function. Although these limitations can be reduced by
the choice of higher resolution, more elaborate advection schemes
could generally be beneficial for the accuracy of the simulations.
With an investigation of the resolution dependence of the results
in each phase space dimension, however, we demonstrate that our physical
results are sufficiently converged in the supernova application. The
undesired effects of low order upwind differencing predominantly affect
the regions far from the neutrinospheres, where the radiation field
is decoupled from the matter such that an adverse influence on the
dynamics of the model can be excluded. We derived the moments entering
the energy conservation equation. Moreover, we algebraically derived
the finite difference representation of the diffusive limit, the nonrelativistic
limit, and the radial dependence of the radiation quantities in a
stationary state free streaming radiation field. These algebraic gymnastics
led to a better understanding of the internal mechanisms in our finite
difference representation and, by comparison with the code results,
enhanced the confidence that the implementation exactly corresponds
to the programmer's intention. In order to perform simulations over
a second or more, the radiation transport and hydrodynamics have been
implemented with implicit finite differencing. This allows reasonably
sized time steps during the neutrino heating phase. Finally we compared
the evolution of a \( 13 \) M\( _{\odot } \) star with an independently
implemented general relativistic supernova code which applies the
multi-group flux-limited diffusion approximation with a recently developed
flux limiter \citep{Bruenn_02}. We find significant agreement that
is especially impressive in the neutrino heating phase. Other phases
show regions with moderate deviations, but those are far from affecting
any physical conclusions.

With the exponentially growing power of computer hardware, computational
astrophysics may reach a comparable status in astronomical, nuclear
and particle physics research as terrestrial experiments have received
in the previous century. We may sort scientific progress into three
phases: First, the astronomical observation or a new theoretical concept
lead to a primary idea about the basic physics involved in an event.
At this stage, order of magnitude estimates are made and compared
to the qualitative observational data. The second step involves plausibility
studies of the suggested scenario, which may include detailed numerical
simulations and comparisons to observations. Often, however, approximations
have to be used that are dictated by technical limitations in the
newly developing field. Only in the simplest cases it can be shown
that all assumptions are based on undeniably justifiable physical
considerations. This exciting phase of discovery and controversy should
be followed by more rigorous numerical simulations in a third step,
when technical limitations fade and remaining approximations become
quantifiable. This phase has to be accompanied by code documentation
that allows independent researchers to analyze and eventually reproduce
the numerical results, similar to the way the value of experimental
data is enhanced by a detailed description of the reproduceable experiment.
The recent emergence of functional Boltzmann solvers leads the radiation
hydrodynamics in spherically symmetric supernova models into this
third phase. The third phase, by principle, should be accompanied
by solid agreement in the numerical solution found with independent
implementations of the same physical ingredients. In order to perform
such comparisons, we appreciate the situation that a basic set of
standard nuclear and weak interaction physics has been defined and
used for over \( 15 \) years (e.g. \citep{Tubbs_Schramm_75,Schinder_Shapiro_82,Bruenn_85}).
In a comparison with the second phase---which used the former standard
scheme for neutrino transport in supernovae---multi-group flux-limited
diffusion, we hope to increase the confidence in our results. On the
other hand, we support the validity of the {\sc mgfld} approximation
in spherical symmetry if the flux limiter is chosen carefully. As
satisfying as these close results are, they should not be mistaken
as an example of convergence to agreement in the sense of the third
phase. The third phase is only reached if similar agreement is found
between independent groups and other documented codes that solve physically
complete neutrino transport equations with general relativistic effects,
e.g. the variable Eddington factor method implemented by \citet{Rampp_Janka_02}.
We hope that spherically symmetric simulations with neutrino radiation
hydrodynamics have reached phase three and that accurate and detailed
neutrino information continues to be useful for the exploration and
improvement of the local microscopic input physics.

\section*{Acknowledgment}

We thank Raph Hix for fruitful discussions and comments on the manuscript.
We enjoyed discussions with Markus Rampp, Thomas Janka, Gabriel Martinez-Pinedo,
and Karlheinz Langanke. We acknowledge support from the National Science
Foundation under contract AST-9877130, the Oak Ridge National Laboratory,
managed by UT-Batelle, LLC, for the U.S. Department of Energy under
contract DE-AC05-00OR22725, the Swiss National Science Foundation
under contract 20-61822.00, the NSF under contract 96-18423, the NASA
under contract NAG5-3903, the Joint Institute for Heavy Ion Research,
a DoE PECASE Grant, and the DoE HENP Scientific Discovery through
Advanced Computing Program. Our regular simulations were carried out
on the National Energy Research Supercomputer Center Cray SV-1, the
high resolution runs and auxiliary checks on the CITA Intel Itanium
I.

\appendix

\section{{\cal O}\protect\( (v/c)\protect \) limit of the finite difference
representation}

The {\cal O}\( (v/c) \) Boltzmann equation has been derived by Castor
\citep{Castor_72}. One can also obtain it by the elimination of higher
order terms from Eq. (\ref{eq_relativistic_boltzmann}). It is enough
to set \( \alpha =\Gamma =const.=1 \) and to replace \( u \) by
the nonrelativistic velocity \( v \). The conservation properties
of the {\cal O}\( (v/c) \) Boltzmann equation become apparent when
we take its energy and angular moments \citep{Mihalas_Mihalas_84}:
\begin{eqnarray}
\frac{\partial J}{\partial t} & + & \frac{\partial }{\partial a}\left[ 4\pi r^{2}\rho H\right] +\frac{v}{r}\left( J-K\right) -\left( \frac{\partial \ln \rho }{\partial t}+\frac{2v}{r}\right) K\nonumber \\
 & - & \int \frac{j}{\rho }E^{3}dEd\mu +\int \chi FE^{3}dEd\mu =0,\nonumber \\
\frac{\partial H}{\partial t} & + & \frac{\partial }{\partial a}\left[ 4\pi r^{2}\rho K\right] -\frac{1}{r}\left( J-K\right) -\left( \frac{\partial \ln \rho }{\partial t}+\frac{2v}{r}\right) H\nonumber \\
 & + & \int \chi FE^{3}dE\mu d\mu =0.\label{eq_ovc_radiation_moment_evolution} 
\end{eqnarray}
 As we construct the specific radiation energy in the laboratory frame,
\( J+vH \), we keep \emph{all} terms that arise from the {\cal O}\( (v/c) \)
Boltzmann equation and obtain, in analogy to Eq. (\ref{eq_radiation_energy_conservation}),
an almost conservative evolution equation for the radiation moments,
\begin{eqnarray}
0 & = & \frac{\partial }{\partial t}\left( J+vH\right) +\frac{\partial }{\partial a}\left[ 4\pi r^{2}\rho \left( vK+H\right) \right] \nonumber \\
 & - & \int \frac{j}{\rho }E^{3}dEd\mu +\int \chi FE^{3}dEd\mu +v\int \chi FE^{3}dE\mu d\mu \nonumber \\
 & - & \frac{1}{4\pi r^{2}\rho }\frac{\partial }{\partial t}\left( 4\pi r^{2}\rho v\right) H.\label{eq_ovc_radiation_energy_conservation} 
\end{eqnarray}
 Although terms of higher order than \( \left( v/c\right)  \), especially
the gravitational terms, require consideration in realistic supernova
simulations, the {\cal O}\( (v/c) \) limit may be useful for various
explorative studies. We give below a finite difference representation
of the {\cal O}\( (v/c) \) Boltzmann equation by simply substituting
\( \alpha =\Gamma =1 \) in the corresponding general relativistic
expressions:\begin{eqnarray}
C_{t} & = & \frac{F_{i',j',k'}-\overline{F}_{i',j',k'}\frac{\overline{da}_{i'}}{da_{i'}}}{dt}+\frac{1}{da_{i'}}\left[ u_{i+1}^{{\rm rel}}F_{i+1,j',k'}^{*}-u_{i}^{{\rm rel}}F_{i,j',k'}^{*}\right] \label{eq_ovc_ct_fd} \\
D_{a} & = & \frac{\mu _{j'}}{da_{i'}}\left[ 4\pi r^{2}_{i+1}\rho _{i+1}F_{i+1,j',k'}-4\pi r^{2}_{i}\rho _{i}F_{i,j',k'}\right] \label{eq_ovc_da_fd} \\
D_{\mu } & = & \frac{3\left[ r^{2}_{i+1}-r^{2}_{i}\right] }{2\left[ r^{3}_{i+1}-r^{3}_{i}\right] }\frac{1}{w_{j'}}\left( \zeta _{j+1}F_{i',j+1,k'}-\zeta _{j}F_{i',j,k'}\right) \label{eq_ovc_dmu_fd} \\
O_{E} & = & \frac{1}{E^{2}_{k'}dE_{k'}}\left[ \left( \mu _{j'}^{2}A_{i',k'-dk}-B_{i',j'}\right) \frac{dE_{k'-dk}}{E_{k'}-E_{k'-dk}}E_{k'-dk}^{3}F_{i',j',k'-dk}\right. \nonumber \\
 & - & \left. \left( \mu _{j'}^{2}A_{i',k'}-B_{i',j'}\right) \frac{dE_{k'}}{E_{k'+dk}-E_{k'}}E_{k'}^{3}F_{i',j',k'}\right] \label{eq_ovc_oe_fd} \\
O_{\mu } & = & \frac{1}{w_{j'}}\left[ \left( A_{i',k'}+B_{i',j'-dj}/\zeta _{j'-dj}\right) \frac{w_{j'-dj}}{\mu _{j'}-\mu _{j'-dj}}\zeta _{j'-dj}\mu _{j'-dj}F_{i',j'-dj,k'}\right. \nonumber \\
 & - & \left. \left( A_{i',k'}+B_{i',j'}/\zeta _{j'}\right) \frac{w_{j'}}{\mu _{j'+dj}-\mu _{j'}}\zeta _{j'}\mu _{j'}F_{i',j',k'}\right] \label{eq_ovc_omu_fd} \\
C_{c} & = & \frac{j_{i',k'}(\rho _{i'},T_{i'}^{*},Y_{e,i'}^{*})}{\rho _{i'}}-\chi _{i',k'}(\rho _{i'},T_{i'}^{*},Y_{e,i'}^{*})F_{i',j',k'}.\label{eq_ovc_cc_fd} 
\end{eqnarray}
The expressions basically reduce to the finite difference representation
described in \citep{Mezzacappa_Bruenn_93a}, except for the adaptive
grid extension, the new code flow outlined in section \ref{subsection_code_flow},
the improved choice of transport coefficients \( \beta _{i,k'} \)
and \( \gamma _{i',k'} \) in Eq. (\ref{eq_transport_coefficients_fd}),
the new discretization of angular aberration, and the matched finite
difference representation of \( A_{i',k'} \) and \( B_{i',j',k'} \)
,\begin{eqnarray}
A_{i',k'} & = & \frac{4\pi \rho _{i'}}{da_{i'}}\left( r^{2}_{i+1}\left( v_{i+2}-v_{i+1}\right) \beta _{i+1,k'}+r^{2}_{i}\left( v_{i+1}-v_{i}\right) \left( 1-\beta _{i,k'}\right) \right) \label{eq_ovc_A_fd} \\
B_{i',j',k'} & = & \frac{3}{2}\frac{r^{2}_{i+1}-r^{2}_{i}}{r^{3}_{i+1}-r^{3}_{i}}\frac{v_{i+1}}{w_{j'}}\left[ \gamma _{i',k'}\zeta _{j+1}\left( \mu _{j'+1}-\mu _{j'}\right) +\left( 1-\gamma _{i',k'}\right) \zeta _{j}\left( \mu _{j'}-\mu _{j'-1}\right) \right] .\label{eq_ovc_B_fd} 
\end{eqnarray}
The angular difference coefficients, \( \zeta _{j} \), are defined
in Eq. (\ref{eq_def_angular_diff_coff}).

\section{Attenuation factors for the presentation of interaction rates}\label{appendix_interaction_rates}

Whenever one composes a graph for the discussion of weak interaction
rates in the supernova environment, one has to circumvent the inherently
large scale differences of the rates at different locations in the
star. In a logarithmic presentation, many details are hidden and differences
between absorption and emission are difficult to appreciate. One can
focus on the neutrinospheres and investigate the rates of interest
under the corresponding conditions. However, the definition of the
neutrinosphere is only based on the opacity and does not account for
large emissivities outside the neutrino sphere. One may miss important
sources of the total neutrino luminosity and fail in the explanation
of the spectra. Moreover, it is common to average the extremely energy-dependent
location where a given optical depth is reached to one single neutrinosphere---an
even more problematic concept. In this appendix, we motivate a convenient
presentation of interaction rates according to their relevance to
the total luminosities. The approach aims to produce intuitively accessible
figures with information about where the neutrinos come from, which
reactions contribute to the total luminosity, how they locally compare
to other reactions, and how the neutrino spectra are formed.

\begin{figure}
{\centering \resizebox*{0.8\textwidth}{!}{\includegraphics{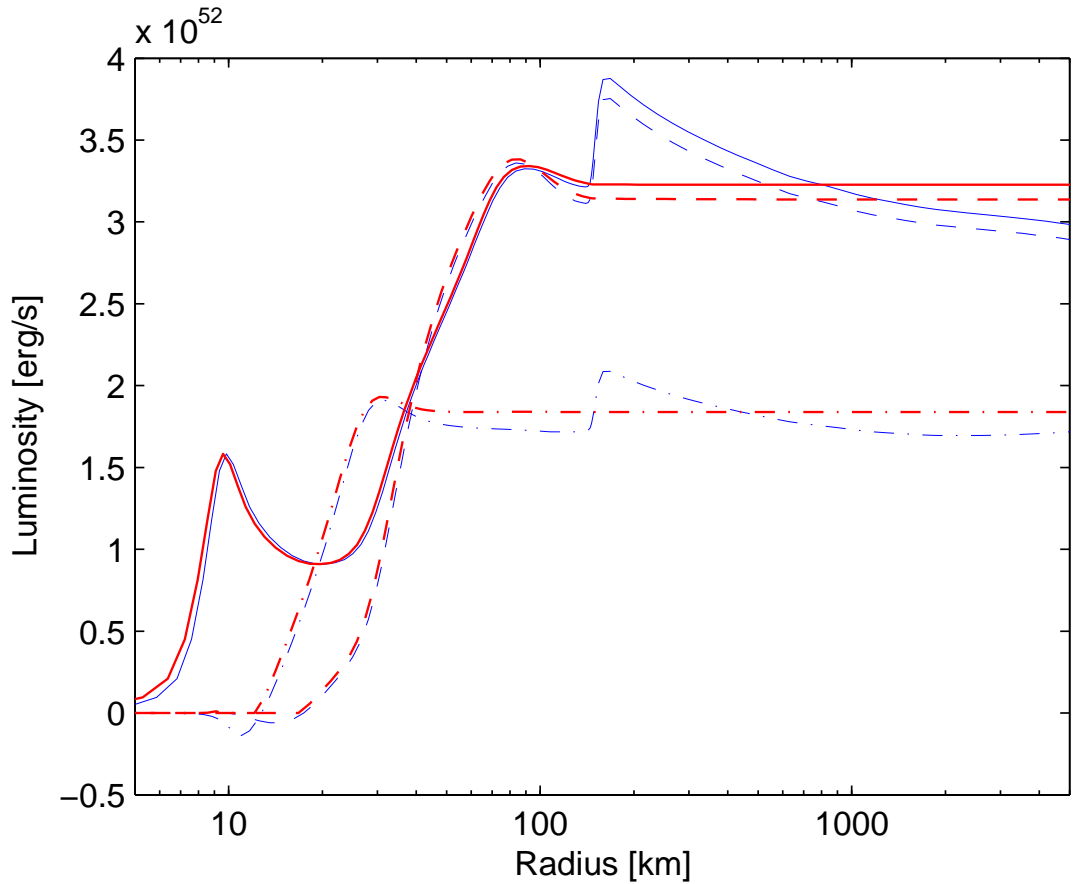}} \par}

\caption{The thin lines show the luminosity profiles in the evolution of the
\protect\( 13\protect \) M\protect\( _{\odot }\protect \) progenitor
star at \protect\( 100\protect \) ms after bounce. The solid line
represents the electron neutrino luminosity, the dashed line the electron
antineutrino luminosity, and the dash-dotted line the \protect\( \mu \protect \)-
and \protect\( \tau \protect \)-neutrino luminosity. The thick lines
show the luminosity profiles evaluated according to Eq. (\ref{eq_luminosity_composition}),
i.e. based on attenuated local emissivities. The reconstruction obviously
misses the observer corrections in regions with low interaction rates.
Otherwise, it reproduces the original luminosities sufficiently well
to be accurate in the analysis of the formation of the neutrino spectra.\label{fig_attenuation.ps}}
\end{figure}

We derive auxiliary attenuation factors in terms of a staggered grid
with zone edge indices \( i \) and zone center indices \( i'=i+1/2 \).
We start with a conserved luminosity \( L_{i} \) with respect to
spheres around the symmetry center. We may choose the neutrino number
luminosity in units of particles per second or the neutrino energy
luminosity in ergs per second. Each zone may have a number or energy
source, \( em_{i'} \), and a number or energy sink, \( ab_{i'} \),
in number or ergs per gram and second. The luminosity is then recursively
defined by its central value \( L_{0}=0 \) and\begin{equation}
\label{eq_recursive_flux}
L_{i+1}=L_{i}+\left( em_{i'}-ab_{i'}\right) da_{i'},
\end{equation}
where \( da_{i'} \) denotes the rest mass contained in the mass shell
\( i' \). The entering luminosity, \( L_{i} \), and the source,
\( em_{i'}da_{i'} \), are subject to absorption in this shell. We
define an attenuation factor, \( x_{i'}\leq 1 \), which accounts
for the reduction of these quantities,\begin{equation}
\label{eq_for_x_definition}
L_{i+1}=x_{i'}\left( L_{i}+em_{i'}da_{i'}\right) .
\end{equation}
From Eq. (\ref{eq_recursive_flux}) and (\ref{eq_for_x_definition}),
we can readily isolate \( x_{i'} \):\[
x_{i'}=\frac{L_{i+1}}{L_{i}+em_{i'}da_{i'}}=\frac{L_{i+1}}{L_{i+1}+ab_{i'}da_{i'}}.\]
In the rare cases of a negative \( x_{i'} \), we set \( x_{i'} \)
to zero. If we calculate a mean free path, \( \lambda _{i'} \), from
all reactions that may act as a sink, the absorption rate can also
be derived from the local specific number or energy density, \( J_{i'} \),
according to \( ab_{i'}=cJ_{i'}/\lambda _{i'} \). The attenuation
factor \( x_{i'} \) can then be expressed by physical quantities
that are well accessible in a numerical evolution of radiation hydrodynamics,\begin{equation}
\label{eq_x_definition}
x_{i'}=\frac{L_{i+1}}{L_{i+1}+\frac{cJ_{i'}}{\lambda _{i'}}da_{i'}}.
\end{equation}
On the other hand, we derive from Eq. (\ref{eq_for_x_definition})
by two recursive self-substitutions,\[
L_{i+1}=x_{i'}\left( x_{i'-1}\left( x_{i'-2}\left( L_{i-2}+em_{i'-2}da_{i'-2}\right) +em_{i'-1}da_{i'-1}\right) +em_{i'}da_{i'}\right) .\]
The continuation to \( L_{0}=0 \) and a rearrangement of the terms
leads to\begin{equation}
\label{eq_luminosity_composition}
L_{n+1}=\sum _{i=1}^{n}\left( \prod _{l=i}^{n}x_{l'}\right) em_{i'}da_{i'}.
\end{equation}
 This equation describes how the total number or energy luminosity
at a radius \( r_{n+1} \) is composed by contributions of distributed
emissivities in the star. The attenuation coefficients\begin{equation}
\label{eq_attenuation_coefficients}
\xi _{n+1,i'}=\prod _{l=i}^{n}x_{l'}=\prod _{l=i}^{n}\frac{L_{l+1}}{L_{l+1}+\frac{cJ_{l'}}{\lambda _{l'}}da_{l'}}
\end{equation}
suppress irrelevant sources that are subject to large reabsorption.
It is instructive to go a step further. We introduce the flux factor
\( h_{i'}=L_{i+1}/\left( 4\pi r^{2}c\rho J_{i'}\right)  \) and rewrite
Eq. (\ref{eq_attenuation_coefficients}) as\[
\xi _{n+1,i'}=\left( 1+\frac{dr_{i'}}{h_{i'}\lambda _{i'}}\right) ^{-1}\xi _{n+1,i'+1}\]
in order to extract the logarithmic derivative of \( \xi  \),\[
\frac{\xi _{n+1,i'+1}-\xi _{n+1,i'}}{\xi _{n+1,i'}}=\frac{dr_{i'}}{h_{i'}\lambda _{i'}}.\]
This is a finite difference representation of the equation\[
\frac{d\ln \xi }{dr}=\frac{1}{h\lambda }\]
 with the solution \( \xi =\exp \left( -\int \left( h\lambda \right) ^{-1}dr\right)  \).
It is the familiar attenuation \( \exp \left( -\tau \right)  \),
if the radius-dependent flux factors are properly accounted for in
the evaluation of the optical depth, \( \tau =\int \left( h\lambda \right) ^{-1}dr \).
However, we evaluate the attenuation coefficients according to Eq.
(\ref{eq_attenuation_coefficients}) where the finite differencing
is consistent with the conservation laws in our implementation of
the Boltzmann equation. As a consistency check, we compare in Fig.
(\ref{fig_attenuation.ps})
the original luminosities \( L_{n+1} \) from the simulation with
the luminosities reconstructed according to Eq. (\ref{eq_luminosity_composition}).

We can use the attenuation coefficients in many convenient ways. A
graph showing\[
g_{n+1}\left( r_{i'}\right) =\xi _{n+1,i'}em_{i'}\frac{da_{i'}}{dr_{i'}}\]
as a function of radius, \( r_{i'} \), visualizes the contribution
of each region in the star towards the total luminosity (represented
by the area under the graph) at radius \( r_{n+1} \). In a comparison
of different neutrino species, the graphs illustrate the decoupling
at different radii. If the gravitational well is not too deep, the
neutrino energy is approximately a constant of motion and the neutrino
in different energy groups can be treated like different species.
Moreover, instead of considering total emissivities and opacities,
one can disentangle them into different reactions which we will enumerate
with a superscript \( \ell  \). A figure with graphs showing\begin{equation}
\label{eq_emission_rate}
g_{n+1}^{\ell }(r_{i'})=\xi _{n+1,i'}em_{i'}^{\ell }\frac{da_{i'}}{dr_{i'}}
\end{equation}
 as a function of radius \( r_{i'} \) would visualize the contribution
of reaction \( \ell  \) to the total luminosity at radius \( r_{n+1} \)
by the enclosed area under the line \( g_{n+1}^{\ell } \). The graph
is automatically scaled such that the most important reactions for
the total luminosity are presented most prominently. If a reaction
conserves the analyzed quantity we are free to include or not include
it in the evaluation of the mean free path in Eq. (\ref{eq_attenuation_coefficients})
and the emissivities in Eq. (\ref{eq_luminosity_composition}). Scattering,
for example, does not change the neutrino number. If we include scattering
as a reaction in the analysis of the number luminosity, the attenuation
coefficients would indicate the probability to escape from a given
location without any further scattering. If we do not include scattering,
the attenuation coefficients indicate the (in many cases larger) probability
to escape from a given location without being absorbed on the way
out. The number luminosity is the same in both cases because scattering
conserves the number of propagating neutrinos. If we work with the
energy luminosity instead of the number luminosity, we have to include
neutrino-electron scattering because this reaction affects the energy
of escaping neutrinos after their production. We simply decompose
a neutrino-electron scattering reaction into a neutrino absorption
at the incoming neutrino energy and a neutrino production at the outgoing
neutrino energy. These terms are then included in the opacities and
emissivities in Eqs. (\ref{eq_attenuation_coefficients}) and (\ref{eq_luminosity_composition}).
In this case, the attenuation coefficients indicate the probability
to escape from a given location without an energy-changing reaction.
We found this choice to be the most interesting for the analysis of
the formation of the neutrino spectra. The omitted isoenergetic scattering
reactions are reflected in the transport spheres which are easily
displayed as a complement. Finally, we mention the potential use of
the attenuation coefficients for statistical evaluations, e.g. for
the average radius of neutrino emission,\begin{equation}
\label{eq_rms_radius}
\left\langle r\right\rangle =\frac{\sum _{i=1}^{n}r_{i'}\xi _{n+1,i'}em_{i'}da_{i'}}{\sum _{i=1}^{n}\xi _{n+1,i'}em_{i'}da_{i'}}.
\end{equation}
In contrast to the classical definition of the neutrinosphere, the
quantity \( \left\langle r\right\rangle  \) accounts for regions
with high emissivities in transparent regimes. Many other statistical
informations at the origin of the neutrino emission may be obtained
analogously.

\end{document}